\documentclass[aps,prd,reprint,11pt,amsmath,amssymb,nofootinbib,groupedaddress,superscriptaddress,floatfix,onecolumn,tightenlines,eqsecnum]{revtex4-2}

\usepackage{macros}

\graphicspath{{./plots/}}

\begin{document}

\title{Searching for coupled, hyperlight scalars across cosmic history}

\author{Masha Baryakhtar}\email[]{mbaryakh@uw.edu}
\affiliation{Department of Physics, University of Washington, Seattle, Washington, 98195, USA}
\author{Olivier Simon}\email[]{osimon@princeton.edu}
\affiliation{Princeton Center for Theoretical Science, Princeton University, Princeton, New Jersey, 08544, USA}
\affiliation{Department of Physics, Princeton University, Princeton, New Jersey, 08544, USA}
\author{Zachary J. Weiner}\email[]{zweiner@perimeterinstitute.ca}
\affiliation{Department of Physics, University of Washington, Seattle, Washington, 98195, USA}
\affiliation{Perimeter Institute for Theoretical Physics, Waterloo, Ontario N2L 2Y5, Canada}

\date{\today}
\begin{abstract}
Cosmological scalar fields coupled to the Standard Model drive temporal variations in the fundamental constants that grow with redshift, positioning the early Universe as a powerful tool to study such models.
We investigate the dynamics and phenomenology of coupled scalars from the early Universe to the present to consistently leverage the myriad searches for time-varying constants and the cosmological signatures of scalars' gravitational effects.
We compute the in-medium contribution from Standard Model particles to the scalar's dynamics and identify only a limited range of couplings for which the scalar has an observable impact on the fundamental constants without either evolving before recombination or gravitating non-negligibly.
We then extend existing laboratory and astrophysical bounds to the hyperlight scalar regime.
We present joint limits from the early and late Universe, specializing to hyperlight, quadratically
coupled scalars that modulate the mass of the electron or the strength of electromagnetism and make
up a subcomponent of the dark matter today.
Our dedicated analysis of observations of the cosmic microwave background, baryon acoustic
oscillations, and type Ia supernovae provides the most stringent constraints on quadratically
coupled scalars with masses from $10^{-28.5}$ to $\sim 10^{-31}~\mathrm{eV}$, below which quasar absorption spectra yield stronger bounds.
These results jointly limit hyperlight scalars that comprise a few percent of the current dark matter
density to near- or subgravitational couplings to electrons or photons.
\end{abstract}

\maketitle
\makeatletter
\def\l@subsubsection#1#2{}
\makeatother
\tableofcontents

\section{Introduction}
\label{sec:introduction}

Particle physics often progresses by postulating that the parameters of a theory, hitherto considered fundamental constants, are dynamical functions of an underlying field.
The Standard Model Higgs mechanism provides the most successful such example~\cite{Englert:1964et,Higgs:1964pj,Guralnik:1964eu}, while others are proposed solutions to extant issues, such as the Peccei-Quinn mechanism to address the strong \emph{CP} problem~\cite{Peccei:1977hh,Wilczek:1977pj,Weinberg:1977ma}.
In this work we consider extensions to the Standard Model that introduce new scalar fields whose values control one or more fundamental ``constants,'' which are then meaningfully and observably \emph{non}constant throughout spacetime when the scalar has nontrivial dynamics.

Variations of fundamental constants can be mechanized by new scalar fields that arise, for example, in scalar-tensor theories of gravity~\cite{Brans:1961sx,Damour:1995kt,Fujii:2003pa} and in theories with compactified extra dimensions (like string theory) in the form of the so-called dilaton or moduli fields~\cite{Brans:1961sx,Scherk:1974ca,Green:2012pqa,Taylor:1988nw,Cho:1987xy,Damour:1994ya,Kaplan:2000hh,Gasperini:2001pc,Damour:2002mi,Dimopoulos:1996kp}.
``Bottom-up'' perspectives use principles of effective field theory to build Lagrangians for
dynamical variations in fundamental constants by promoting Standard Model (SM) parameters to depend
on new scalars constrained by symmetry principles~\cite{Terazawa:1981ga, Bekenstein:1982eu, Damour:2010rm, Damour:2010rp}.
These scalars result in three (related) types of observational signatures that have spurred vibrant experimental programs: modulations of fundamental constants with space, time, and environment~\cite{Brans:1961sx,Bekenstein:1982eu,Damour:1994ya,Green:2012pqa,Olive:2007aj,Arvanitaki:2014faa},
observable violations of the weak equivalence principle~\cite{Taylor:1988nw,Kaplan:2000hh,Damour:2010rp,Damour:2010rm,Graham:2015ifn,Hees:2018fpg}, and deviations from the $1/r$ law of gravitational interactions~\cite{Dimopoulos:1996kp, Arkani-Hamed:1998jmv, Adelberger:2003zx}.

Scalar fields have cosmological consequences as well---even simply through
gravity if they are sufficiently abundant.
A cosmological density of the new scalar may be produced in the very early Universe,
for instance through the misalignment mechanism~\cite{Preskill:1982cy, Abbott:1982af, Dine:1982ah, Turner:1985si, Linde:1987bx, Cho:1987xy, Hu:2000ke, Piazza:2010ye, Hui:2016ltb}.
Depending on its initial conditions and interactions, the scalar makes up at least a
subcomponent of the cosmological dark sector today as matter, radiation, or something more exotic.
In general, scalar fields redshift with the expansion of the Universe and their observational
effects are more pronounced earlier in cosmological history---more so if the leading dependence of
fundamental constants on the scalar field amplitude is quadratic~\cite{Olive:2007aj,
Hees:2018fpg, Bouley:2022eer, Banerjee:2022sqg} or higher rather than linear.
In an era of precision cosmology, early-Universe observations are particularly apt probes of dynamical
scalar fields and may explore parameter space complementary to that from traditional particle
physics tests.

Primordial nucleosynthesis and recombination are the two earliest processes with observable relics
that are sensitive to SM parameters; measurements of nuclear abundances and the cosmic microwave
background (CMB) therefore test for possible evolution of fundamental constants at temperatures
around an MeV and eV, respectively.
Limits on the variation in the fine-structure constant~\cite{Hannestad:1998xp, Kaplinghat:1998ry,
Avelino:2000ea, Battye:2000ds, Avelino:2001nr, Landau:2001st, Martins:2003pe, Rocha:2003gc,
Stefanescu:2007aa, Nakashima:2008cb, Menegoni:2009rg, Menegoni:2012tq} and the electron
mass~\cite{Kujat:1999rk, Ichikawa:2006nm, Landau:2008re, Scoccola:2008jw, Nakashima:2009cs,
Landau:2010zs, Scoccola:2012ny, Schoneberg:2024ynd} at recombination have been placed largely
without regard for the microphysics involved.
On the other hand, nucleosynthesis provides some of the leading constraints on quadratically coupled
scalar dark matter~\cite{Sibiryakov:2020eir, Bouley:2022eer}.
Unlike the yields of primordial elements, which depend only on the conditions while they formed, CMB
anisotropies are sensitive not just to physics at the time of recombination but also to intervening
dynamics of the Universe up to the present day.
Therefore, a consistent CMB analysis must take into account the joint effect of a new scalar's cosmological abundance as well as
its microphysical couplings to matter~\cite{Baryakhtar:2024rky}.
Grounding an analysis with a concrete model enables a consistent study of the gravitational
interactions of the scalar field, the impact of SM matter on the cosmological evolution of the
field, and complementary constraints of fundamental constants across epochs.

In this work, we explore the parameter space of a hyperlight
scalar field responsible for variations in the mass of the electron $m_e$ and the
electromagnetic fine-structure constant $\alpha$.
We place constraints on the scalar's cosmological abundance and microphysical coupling strengths with cosmological datasets, utilizing our prior work in which a light scalar field is consistently implemented~\cite{Baryakhtar:2024rky}, as well as with complementary late-time
astrophysical and terrestrial probes.
We target regimes that mimic the phenomenological models of prior study in which $\alpha$ and $m_e$ took on different but time-independent values during
recombination, which imposes nontrivial constraints on viable models.
In particular, we show that such scalar fields may neither be too heavy nor too strongly coupled to
SM matter, for in either case the scalar would become dynamical before photon last scattering.
But neither can the scalar be too light, for it must have redshifted via its cosmological dynamics
sufficiently to satisfy stringent constraints on fundamental constant variations nearer to the
present day.
We therefore focus on massive, hyperlight subcomponents of dark matter with particle mass
$10^{-32}~\eV \lesssim m_\varphi \lesssim 10^{-28}~\eV$.
Constraints from late-time searches have to be reinterpreted for such low masses;
we find that cosmological constraints are the most stringent down to
$10^{-31}$-$10^{-30}~\eV$, where they are superseded by measurements of quasar spectra.

This work is divided as follows.
In \cref{sec:models}, we review the effective field theory of dynamical modulations in
fundamental constants by a scalar field, with a focus on $\alpha$ and $m_e$.
We also discuss the influence of the cosmological abundance of SM particles on the scalar's
effective potential and dynamics.
\Cref{sec:late-probes} recasts searches for varying constants and new scalar fields
coupled to the SM from quasar absorption lines, equivalence principle violation, and precision
metrology.
Finally, in \cref{sec:early-probes}, we detail the physical impact of varying fundamental constants on
nucleosynthesis and review the constraints using CMB and other cosmological observables derived in
Ref.~\cite{Baryakhtar:2024rky}.
We combine the results of the preceding sections in \cref{sec:discussion} to set limits on a new
scalar's coupling to the photon or the electron across the hyperlight parameter space, and we
conclude in \cref{sec:conclusions}.

We use natural units in which $\hbar = c = 1$ and define the reduced Planck mass
$\Mpl \equiv 1 / \sqrt{8 \pi G} = 2.435 \times 10^{18}~\GeV$.
We employ the Einstein summation convention for spacetime indices; repeated spatial indices (Latin
characters) are implicitly summed regardless of their placement.
We also use upright boldface to denote spatial vectors.
Unless otherwise specified, we fix a homogeneous Friedmann-Lema\^itre-Robertson-Walker
spacetime with metric
\begin{align}\label{eqn:flrw-metric}
  \ud s^2
  &= \ud t^2
    - a(t)^2 \delta_{i j} \ud x^i \ud x^j,
\end{align}
with $a(t)$ the scale factor.
Dots denote derivatives with respect to cosmic time $t$, and the Hubble rate is $H \equiv \dot{a} / a$.

\section{Scalar field models for varying fundamental constants}\label{sec:models}

In this section, we review the field theoretic description of the modulation of fundamental
constants by a new scalar field.
After some general considerations, we narrow our scope to the mass of the electron and the
electromagnetic fine-structure constant.
We summarize the fundamental couplings of such fields to matter and radiation (\cref{sec: Varying
constants with new scalars}) as well as their effective coupling to composite matter
(\cref{sec:Effective Lagrangian for phi-matter coupling}).
We also detail how the cosmological abundance of SM particles contributes to the scalar's effective
potential (\cref{sec:matter-potentials}) and comment on variations in $\alpha$ and $m_e$ due to such
in-medium effects (\cref{sec:sm-sources-of-variation}).
Finally, we describe the cosmological dynamics of coupled scalars in
\cref{sec:cosmological-dynamics}, considering in turn the case in which the scalar's dynamics in an
expanding Universe are wholly dictated by its bare potential and where SM matter makes a
nonnegligible contribution to its effective potential.

\subsection{Varying constants with new scalars}
\label{sec: Varying constants with new scalars}

We set the stage by defining notation for a general description of theories featuring new scalars coupled to the SM.
We consider a new, real scalar field that couples to the Standard Model in a manner that effectively
generates spacetime dependence of some subset of SM parameters.
We take the simplest case of a massive field $\varphi$ with negligible self-interactions,
\begin{align}
    \label{eq:general-lagrangian}
    \mathcal{L}
    &= \frac{1}{8 \pi G} \left( \partial_\mu \varphi \partial^\mu \varphi - m_\varphi^2 \varphi^2 \right)
        + \mathcal{L}_\mathrm{SM}[\varphi, \ldots];
\end{align}
the ellipsis denotes other SM fields.
The scalar's couplings each take the generic form of a function of $\varphi$
multiplying a dimension-four operator already present in the SM Lagrangian
$\mathcal{L}_\mathrm{SM}$.
The canonical, dimensionful field is $\phi \equiv \varphi / \sqrt{4 \pi G} = \sqrt{2} \Mpl \varphi$, while $\varphi$'s kinetic term is analogous to that of gravity.
The normalization of $\varphi$ is useful for parametrizing its strength relative to dimensionless
metric perturbations in cosmological and fifth-force contexts, as the SM contribution to the
Klein-Gordon equation for $\varphi$ resembles the gravitational Poisson equation:
\begin{align}
    \ddot{\varphi}(t, \three{x})
        + 3 H \dot{\varphi}(t, \three{x})
        - \frac{1}{a(t)^2} \partial_i \partial_i \varphi(t, \three{x})
        + m_\varphi^2 \varphi(t, \three{x})
    &= 4 \pi G \frac{\partial \mathcal{L}_\mathrm{SM}}{\partial \varphi}.
    \label{eq:scalar-eom}
\end{align}
If the bare mass term dominates both Hubble friction and the interaction term on the right-hand side
of \cref{eq:scalar-eom}, the scalar behaves as (a component of) dark matter.

We prescribe the spacetime dependence of a fundamental constant $\lambda$, inherited from a slowly
varying background field $\varphi(t, \three{x})$, via coupling functions $g_\lambda$ defined by
\begin{eq}
    g'_\lambda(\varphi)
    \equiv \frac{1}{\lambda(\varphi)}
    \dd{\lambda(\varphi)}{\varphi}.
    \label{eq:coupling-function-def}
\end{eq}
Integrating \cref{eq:coupling-function-def} and taking $g_\lambda$ to vanish at the vacuum
expectation value (VEV) $\left\langle \varphi \right\rangle$ gives
\begin{align}
    \label{eq:kappa_as_function_of_phi}
    \lambda(\varphi)
    &= \lambda(\left\langle \varphi \right\rangle) e^{g_\lambda(\varphi)}
    \approx \lambda(\left\langle \varphi \right\rangle) \left[ 1 + g_\lambda(\varphi) \right],
\end{align}
the second equality at leading order in small $g_\lambda(\varphi)$.
Consequently, the scalar shifts $\lambda$ from its VEV by
\begin{align}
    \label{eqn:variation-lambda-ito-g_lambda}
    \frac{\Delta \lambda}{\lambda(0)}
    \equiv \frac{\lambda(\varphi)-\lambda(0)}{\lambda(0)}
    &\approx g_\lambda(\varphi)
\end{align}
to leading order, where we take $\left\langle \varphi \right\rangle = 0$ as for
\cref{eq:general-lagrangian}.
Since the value of the cosmological background field today, $\varphi_0 \equiv \varphi(t_0)$, is not
its VEV in general, $\lambda(0)$ may not coincide with the coupling
$\lambda_0 \equiv \lambda(\varphi_0)$ that would be measured by a contemporary experiment.
Tests of equivalence principle violation and other scalar phenomenology beyond the present-day value
of fundamental constants alone can, in principle, distinguish between $\lambda(0)$ and $\lambda_0$.

The coupling functions can be expanded as a power series around $\varphi = 0$,
\begin{align}
    g_\lambda(\varphi)
    &= \sum_{n=1} \frac{d_\lambda^{(n)}}{n!} \varphi^{n}.
    \label{eqn:coupling-function-series-expansion}
\end{align}
At small $\varphi$ and still assuming that $\varphi$ is slowly varying, such that its derivative interactions are relatively suppressed, the coupling functions are parametrically dominated by the lowest powers of the expansion, motivating the existing
studies of linear and quadratic couplings. Tests of the equivalence principle stringently limit a light scalar's linear couplings to the SM to far below gravitational strength,
independently of its cosmological abundance~\cite{Hees:2018fpg} (see
\cref{sec:UFF}).
Furthermore, in the perturbative regime $\vert g_\lambda \vert$ monotonically increases with
$\varphi$, which in turn generally decays in amplitude as the Universe expands.
We therefore expect cosmological signatures at high redshift to be larger in magnitude compared to
present-day ones, to an extent that is greater for higher-order interactions.

The set of specific ``constants'' $\lambda$ that one considers ``fundamental'' depends on the energy scales under
consideration.
In this work, we restrict our attention to the fine-structure constant $\alpha\equiv e^2/4\pi$ and the electron mass $m_e$ because
of their particular relevance to cosmology and recombination.
We thus replace $\alpha$ and $m_e$ with $\alpha(\varphi)$ and $m_e(\varphi)$ in the SM
Lagrangian:
\begin{align}
\label{eq:DD_lagrangian}
    \mathcal{L}_\mathrm{SM}[\varphi, \ldots]
    &= - \frac{1}{16 \pi \alpha(\varphi)} F^{\mu \nu} F_{\mu \nu}
        + \bar{e} \left[ i \slashed{D} - m_e(\varphi) \right] e
     + \cdots,
\end{align}
where the ellipsis designates the rest of the SM Lagrangian below the scale of electroweak symmetry
breaking.
Note that \cref{eq:DD_lagrangian} takes the ``Yang-Mills'' normalization with the (inverse) gauge
coupling multiplying the kinetic terms of gauge fields.

In the remainder of this work we aim to treat scalar couplings as generally as is practical; based on the above considerations, however, we typically anchor the discussion to quadratic
coupling functions with a vanishingly small linear term,
\begin{eq}
    \label{eq:quadratic_couplings}
    g_\lambda(\varphi)
    \approx d_\lambda^{(2)} \varphi^{2} / 2,
\end{eq}
and electromagnetically coupled [\cref{eq:DD_lagrangian}] scalars, as encoded by the Lagrangian\footnote{
    We replace the subscript $\alpha$'s with $e$'s for the photon couplings, e.g.,
  $d^{(n)}_e \equiv d^{(n)}_\alpha$, to conform with prior literature.
}
\begin{align}
    \label{eq:quadratic_lagrangian}
    \mathcal{L}_\mathrm{SM}[\varphi, \ldots]
    \approx
        \frac{\varphi^2}{2}
        \left(
            \frac{d^{(2)}_e}{16 \pi \alpha(0)} F^{\mu \nu} F_{\mu \nu}
            - m_e(0) \bar{e} e
        \right)
        + \mathcal{L}_\mathrm{SM}[0, \ldots],
\end{align}
where $\mathcal{L}_\mathrm{SM}[0, \ldots]$ denotes the usual SM Lagrangian.
The corresponding $\varphi$-dependent electron mass and fine-structure constant are
\begin{subequations}
\label{eq:quadratic_variations_in_constants}
\begin{align}
    m_e(\varphi)
    &\approx m_e(0) \left( 1 + d_{m_e}^{(2)} \varphi^2 / 2 \right),
    \\
    \alpha(\varphi)
    &\approx \alpha(0) \left( 1 + d_{e}^{(2)} \varphi^2 / 2 \right).
\end{align}
\end{subequations}
The assumed suppression of terms linear in $\varphi$ could be achieved with a $\mathbb{Z}^\varphi_2:
\varphi \mapsto -\varphi$ symmetry; we mention models that feature such a symmetry in the following.

\subsubsection*{Model-building considerations}\label{sec:model-building-considerations}

Before proceeding to the phenomenology of hyperlight scalars in the early Universe, we comment on several aspects of these models in the context of particle theory.
A more complete understanding the class of theories we consider calls for a UV completion,
which we do not specify here.
We focus only on a subset of possible couplings (to $\alpha$ and $m_e$); the others pertinent at
lower energies are the strong confinement scale $\Lambda_\text{QCD}$ and the up and down quark
masses, $m_u$ and $m_d$~\cite{Damour:2010rp, Damour:2010rm}.
These parameters would ultimately inherit $\varphi$ dependence from higher-scale (i.e.,
electroweak-scale) physics---the Higgs's mass, vacuum expectation value, and Yukawa couplings to
fermions along with the three $\mathrm{SU}(3) \times \mathrm{SU}(2) \times \mathrm{U}(1)$ gauge
couplings~\cite{Olive:2002tz, Green:2012pqa, Damour:1994ya, Uzan:2010pm, Uzan:2024ded}.
At yet higher energies, the scalar couplings in \cref{eq:DD_lagrangian} are characteristic of so-called ``dilaton'' and moduli fields found in string theory compactifications and supersymmetric models, although typically with an unsuppressed linear term~\cite{Damour:1994ya, Dimopoulos:1996kp,
Kaplan:2000hh, Damour:2010rm, Damour:2010rp}.

Many UV completions thus would, in addition to electron and photon couplings, introduce couplings to quarks and gluons.
In the Higgs portal model, in which the scalar interacts with the SM Higgs field $H$ via renormalizable operator
$\varphi^2 H^\dagger H$, the Yukawa interactions between the Higgs and SM fermions $f$ generate interactions of the form
$m_f(\varphi) \bar{f} f$ for all fundamental SM fermions, each with the same $\varphi$ dependence.
In particular, $g'_{m_e}(\varphi) = g'_v(\varphi)$ for the electron.
The QCD scale $\Lambda_\text{QCD}$ also depends on heavy quark masses via the renormalization of the strong gauge coupling $\alpha_s$ and introduces variations in nucleon masses at the same order.
Specifically, the dependence on the Higgs VEV is $g'_{m_q}(\varphi) = g'_v(\varphi)$ for quark masses and $g'_{\Lambda_\text{QCD}} = (2/9)g'_v$ for the QCD scale, yielding $g'_{m_N} \approx 0.79 g'_{\Lambda_\text{QCD}} + 0.24 g'_v = 0.42 g'_v$ for nucleon masses, with the second term dominated by the strange quark's contribution to nucleon masses~\cite{Coc:2006sx}.
The ``axi-Higgs'' model, the cosmology of which is studied in Refs.~\cite{Fung:2021wbz,
Fung:2021fcj, Luu:2021yhl}, is a recent example of this class of models.
References~\cite{Fung:2021wbz, Fung:2021fcj, Luu:2021yhl} claim that, because nucleon masses are
parametrically dominated by $\Lambda_\text{QCD}$, their relative
variations are suppressed as $g'_{m_N} \approx g_{m_{u, d}}' / \Lambda_\mathrm{QCD} \approx 10^{-3}
g'_v$; they therefore neglect variations in baryon masses when modeling the physics of recombination
and nucleosynthesis.
However, as outlined above, any model derived from
Higgs couplings is not well described by variations in the $m_e$ alone at low energies.
We confine ourselves to scalar field interactions with the electron Yukawa coupling and the
fine structure constant [\cref{eq:DD_lagrangian}] and leave the interesting question of the
effect of baryon mass variations on cosmology to future work.

Depending on the assumed symmetries, scalars may interact with fermions and gauge bosons via
operators other than those parametrized by \cref{eq:DD_lagrangian}.
For instance, a shift-symmetric pseudovector coupling is characteristic of (pseudo)Nambu-Goldstone
bosons such as axionlike particles~\cite{Kim:1979if,Shifman:1979if,Dine:1981rt,Zhitnitsky:1980tq}.
A pseudoscalar coupling is also allowed by symmetries and is equivalent to the pseudovector coupling
alone only in processes whose Feynman diagrams have a single $\varphi$ per
electron~\cite[Sec.~90]{ParticleDataGroup:2022pth}.
Pseudoscalar (PS) and pseudovector (PV) operators effectively contribute to those explicitly included in
\cref{eq:DD_lagrangian}.
Consider a background field $\varphi$ with general electron couplings of the form
\begin{align}
    \label{eq:general_electron_lagrangian}
    \mathcal{L}
    &\supset i\bar{e} \slashed{D} e
        - m_{e}(0) \left[ 1 + G_\text{S}(\varphi) \right] \bar{e} e
        + i m_{e}(0) G_\text{PS}(\varphi) \bar{e} \gamma^5 e
        + G_\text{PV}(\varphi)\partial_\mu \varphi \bar{e} \gamma^\mu \gamma^5 e
        + \cdots,
\end{align}
with the ellipsis designating Lorentz structures with higher derivatives of $\varphi$.
The scalar functions $G_\text{S}(\varphi)$, $G_\text{PS}(\varphi)$, and $G_\text{PV}(\varphi)$ are
not uniquely defined---they mix into one another under chiral rotations.
One can show that a chiral rotation $e^{i \gamma^5\theta(\varphi) / 2} = \cos[\theta(\varphi) / 2]
I_4 + i \gamma^5 \sin[\theta(\varphi) / 2]$ of the electron acts as a two-dimensional rotation
[SO(2)] on the two-component vector with entries $1 + G_\text{S}(\varphi)$ and
$G_\text{PS}(\varphi)$.
With an appropriate choice of $\theta(\varphi)$, one can therefore rotate away the pseudoscalar or
the scalar term (including the bare mass term) entirely, at the cost of introducing a pseudovector
coupling $i \theta'(\varphi) \partial_\mu \varphi \bar{e} \gamma^\mu \gamma^5 e / 2$.
The correction to the electron mass for slowly varying $\varphi(t,\three{x})$ is
\begin{align}
    g_{m_e}(\varphi)
    &= \sqrt{\left[ 1 + G_\text{S}(\varphi) \right]^2 + G_\text{PS}(\varphi)^2} - 1,
\end{align}
which is invariant under chiral rotations.
One may alternatively obtain this result, with additional corrections proportional to gradients of
the field from the pseudovector coupling, by computing the poles of the electronic two-point
function for \cref{eq:general_electron_lagrangian} with $\varphi$ as a background field.

A similar discussion holds for photon couplings; for example
\begin{eq}
\label{eq:general_photon_lagrangian}
    \mathcal{L}
    \supset -\frac{1 -G_{FF}(\varphi)}{16 \pi \alpha(0)} F_{\mu \nu} F^{\mu \nu}
        + G_{F \tilde{F}}(\varphi) \frac{F_{\mu \nu} \tilde{F}^{\mu \nu}}{16 \pi \alpha(0)}
        + \cdots,
\end{eq}
where again $G_{FF}$ and $G_{F \tilde{F}}$ are scalar functions of $\varphi$ and $\cdots$ designates
Lorentz structures with higher derivatives of $\varphi$.
The phenomenology of the latter coupling has been studied extensively and for
instance can source cosmic birefringence~\cite{Carroll:1998zi,Lue:1998mq,Arvanitaki:2009fg,Agrawal:2019lkr,Diego-Palazuelos:2022dsq,Yin:2024fez,Sullivan:2025btc}.
Models can also simultaneously feature both couplings, which transform differently under
parity~\cite{Flambaum:2009mz,Smith:2024ayu,Smith:2024ibv}.
For our purposes, the electromagnetic exchange potential (i.e., the amplitude for $ee \rightarrow
ee$) receives corrections from all interaction terms, resulting in a coupling function
$g_e(\varphi) \equiv g_\alpha(\varphi)$ that depends on $G_{F\tilde{F}}$ in addition to $G_{FF}$. However, $F_{\mu \nu} \tilde{F}^{\mu \nu}$ by itself is a total derivative term, which holds for $G_{F \tilde{F}}(\varphi) F_{\mu \nu} \tilde{F}^{\mu \nu}$ unless $\varphi$ varies in spacetime. For an Abelian gauge boson like the photon, any physical effect must therefore be proportional to derivatives of $\varphi$.

Thus, both the PV coupling in \cref{eq:general_electron_lagrangian} and the $F\tilde F$ coupling in \cref{eq:general_photon_lagrangian} are ``derivative'' couplings, the corrections from which are suppressed for the nonrelativistic,
ultralight fields we consider here and the non-derivative, scalar couplings fully captures the low-energy physics.

Finally, we note that a light mass for a scalar with the non--shift-symmetric couplings we consider is not protected from
radiative corrections.
For instance, if the theory is ultraviolet (UV) completed by new physics at some cutoff scale
$\Lambda$, the electron coupling contributes a one-loop effective potential
$m_e(\varphi)^2 \Lambda^2 / 16 \pi^2$ at leading order in large $\Lambda$.\footnote{
    The photon coupling does not modify the parameters of the free SM Lagrangian and therefore
    generates no effective potential at one loop~\cite{Cyncynates:2024yxm}; the two-loop
    contributions for such noncanonical theories are technically challenging to compute and we
    do not consider them here.
}
As discussed in the introduction, we focus on cosmologies featuring fundamental constants that vary only after recombination and thus
$m_\varphi$ near or below $10^{-28}~\eV$---not merely ultralight, but \emph{hyper}light.
Even for Planckian scalar misalignments, the scale of the bare potential is only of order
$\mathrm{eV}^4$, implying some drastic cancellation of quantum corrections
(which naively are at the very least $m_e^4$ in magnitude) to achieve effective masses so light.
For the quadratic couplings we consider, quantum corrections nominally shift the mass by
$\sim d_{m_e}^{(2)} m_e(0)^2 \Lambda^2 / 8\pi^2\Mpl^2$; an order-unity coupling renders masses
$m_\varphi \lesssim 10^{-11}~\eV$ unnatural for $\Lambda = 100~\GeV$~\cite{Arvanitaki:2014faa,
Brzeminski:2020uhm, Banerjee:2022sqg, Bouley:2022eer}.
Moreover, radiative corrections generate higher-order self interactions, since
$m_e(\varphi)^2 / m_e(0)^2
= e^{2 g_{m_e}(\varphi)}
\approx 1 + 2 g_{m_e}(\varphi) + 2 g_{m_e}(\varphi)^2$.
Corrections of yet higher order arise beyond one loop and at different orders in $\Lambda$ but are
in general suppressed by powers of $g_{m_e}(\varphi) < 1$.
A quadratic coupling, for instance, induces a quartic interaction
$\sim \left( d_{m_e}^{(2)} \right)^2 m_e(0)^2 \Lambda^2 \varphi^4/8\pi^2$ at one loop, which, while a
problem in its own right, requires less tuning [by a factor of $g_{m_e}(\varphi)$] relative to the
bare potential.

The cosmological probes we develop here thus test scalars that are nominally unnatural, as is the
case for many existing probes of interacting, ultralight scalars.
In principle, the phenomenology we consider only requires the effective theory to be valid at energy
scales below $100~\eV$ or so (compared to signatures from nucleosynthesis, for instance, which is
sensitive to temperatures up to $\sim \MeV$ scales~\cite{Giudice:2000ex, Kawasaki:2000en, Kawasaki:2004qu,
Hannestad:2004px, Ichikawa:2005vw, deSalas:2015glj, Hasegawa:2019jsa}).
Whether this translates into a less severely tuned theory (that also remains a suitable description
of known physics at moderate energy scales) ultimately can only be assessed in concrete UV models.
Apparently unnatural scalars could arise accidentally in models with anthropic solutions for
the cosmological constant~\cite{Arvanitaki:2016xds}, have naturally light masses from discrete
symmetry structures~\cite{Brzeminski:2020uhm} (see also Ref.~\cite{Banerjee:2022sqg} for technically
natural scalars with quadratic couplings), or simply be fine tuned.
Nevertheless, perhaps more important than specific tuning considerations is the fact that an ever-growing
set of precision cosmological data allows us to test the existence and unique signatures of
hyperlight fields, with abundances consistent with observational constraints~\cite{Hlozek:2014lca,
Hlozek:2016lzm, Hlozek:2017zzf, Lague:2021frh,Rogers:2023ezo} and couplings as weak as
gravitational strength~\cite{Baryakhtar:2024rky}.

\subsection{Effective field theory at low energies}
\label{sec:Effective Lagrangian for phi-matter coupling}

In this section, we discuss the effective field theory description of the interaction between
scalars and composite matter.
Around the time of primordial nucleosynthesis, the scalar interacts with nucleons and bound nuclei.
After recombination, some electrons and nuclei are free while others are bound as neutral atoms.
An effective description in terms of these composite states is therefore pertinent to study
the scalar's interactions with the early Universe plasma and present-day experimental apparatuses.
Here we build upon and expand the discussions in Refs.~\cite{Damour:2010rp,Damour:2010rm}.

The coupling of $\varphi$ to composite states is not readily apparent from a theory written in terms
of fundamental SM degrees of freedom, as in \cref{eq:DD_lagrangian}.
Consider the fundamental coupling of $\varphi$ to the electromagnetic tensor
$F_{\mu \nu} F^{\mu \nu} \propto \three{E}^2 - \three{B}^2$, for example.
As noted in Ref.~\cite{Bekenstein:1982eu}, freely propagating electromagnetic radiation neither sources nor interacts with $\varphi$ because $\langle \three{E}^2 - \three{B}^2 \rangle$ vanishes for a traveling plane wave.
Although $\langle \three{E}^2 - \three{B}^2 \rangle$ can be nonzero for, e.g., a singular standing wave, it vanishes for an \emph{incoherent} superposition of plane waves such as the cosmological blackbody radiation.
Conversely, $F_{\mu \nu} F^{\mu \nu}$ can be nonzero at order $\alpha$ in the presence of charged
matter, as discussed in \cref{sec:matter-potentials}.
Thus even if the ``bare'' particle constituents of matter do not directly couple to $\varphi$, composite states can nevertheless inherit a coupling to $\varphi$, e.g., through their electromagnetic binding energy~\cite{Bekenstein:1982eu,Olive:2001vz,Olive:2007aj}.
In light of this, we define a new set of $\varphi$-dependent parameters---namely, the masses $m_X$ of matter species $X$ and potential new couplings $c_i$ between those species:
\begin{eq}
    \mathcal{C}_+
    = \{m_e, m_p, m_n, m_\text{H}, m_\text{He}, m_{\text{He}^{2+}}, \dots\}
        \cup \{\alpha, c_1, c_2,\dots \}.
\end{eq}
A new coupling function is associated to each such parameter.

A proper treatment of atomic species as an effective field theory associates a unique field and mass parameter to every atomic excited state considered in the theory, labeled by its spectroscopic quantum numbers
$n^{2 S + 1} L_J$.
Since in many applications the distinction between atomic states makes a negligible difference to the interaction of the species with $\varphi$, we only consider ground-state atoms, assigning each a fermionic or bosonic field of the appropriate spin representation:
\begin{align}
\begin{split}
    \label{eq:effective_lagrangian}
    \mathcal{L}_\mathrm{SM}[\varphi, \ldots]
    &=
        - \frac{1}{16\pi\alpha(\varphi)^2} F^{\mu \nu} F_{\mu \nu}
        + \sum_{f}^{ \{e, p, n, \hefour^+, \prescript{3}{}{\mathrm{He}}, \ldots\} }
        \bar{f} \left[ i \slashed{D} - m_f(\varphi) \right] f
    \\ &\hphantom{ {}={} }
        + \sum_{b}^{ \{ \text{H}, \hefour, \hefour^{2+}, \ldots \} }
        \left[ \left\vert D_\mu b \right\vert^2 - m_b^2(\varphi) b^\ast b \right]
        + \cdots
        + \mathcal{L}_\text{int}[c_1(\varphi), c_2(\varphi), \ldots],
\end{split}
\end{align}
where the fields $f$ are spinors, $b$ complex scalars, and
the first ellipsis denotes the quadratic Lagrangian for fields of higher spin representation.
Some species remain charged under electromagnetism, while $\mathcal{L}_\text{int}$ allows for short-range interactions between species, such as collisions between neutral species and ionization and recombination processes.

The remaining task is to relate the scalar coupling functions for the set $\mathcal{C}_+$ to
the set of coupling functions to fundamental SM parameters $\mathcal{C}$.
These relations, the so-called ``dilatonic charges''~\cite{Damour:2010rp, Damour:2010rm},\footnote{
    Our definition differs slightly from that of Refs.~\cite{Damour:2010rp, Damour:2010rm}, which
    define the dilatonic charge only as a property of matter species rather than as a relation
    between Lagrangian parameters (which would include, e.g., $\alpha$).
    Because the concept developed here is so proximate, we retain this nomenclature.
}
are assembled into a charge matrix $Q$ with elements
\begin{eq}
    (Q_{\sigma})_\lambda
    = \frac{\lambda}{ \sigma}\frac{\partial \sigma}{\partial \lambda},
    \qquad \lambda \in \mathcal{C}, \sigma \in \mathcal{C}_+,
\end{eq}
which relates the new and old coupling functions via
\begin{eq}
    g'_\sigma(\varphi)
    = \sum_{\lambda \in \mathcal{C}}(Q_\sigma)_\lambda g'_\lambda(\varphi).
    \label{eq:dilatonic_relations}
\end{eq}
For a composite species $X$, the coupling functions $g'_{m_X}$ provide a convenient expression of the dependence of $m_X$ on $\varphi$ through \cref{eq:kappa_as_function_of_phi}.
They also determine the effective potential $V_\text{matter}(\varphi)$ which dictates the reaction of $\varphi$ to matter, whether it be a localized source mass (\cref{sec:UFF}) or the matter content of the Universe averaged over cosmological scales (\cref{sec:late_universe_potential,sec:early_universe_potential}).

As an example, consider an atomic species $X$ with atomic number $Z$ and mass number $A$.
Somewhat heuristically, its total mass comprises the masses of the constituent protons, neutrons,
and electrons as well as nuclear and atomic binding energies,
\begin{align}
\begin{split}
    m_\text{X}
    &\simeq Z m_p\left( \alpha \right) + (A - Z) m_n \left( \alpha \right)
        + Z m_e
        + E_{\text{nuclear}, X}\left( \alpha \right)
        + E_{\text{atomic}, X}\left( \alpha, m_e \right),
\end{split}
\end{align}
where $E_{\text{nuclear},X}$ is the nuclear binding energy and $E_{\text{atomic}, X}$ the atomic
binding energy.
Here we highlight the dependence on $\alpha$ and $m_e$ relevant to this work; in addition, the
proton and neutron masses and nuclear energies themselves depend on $\Lambda_\text{QCD}$ and
quark masses.
The dilatonic charge $(Q_{m_X})_\lambda$ is roughly the fraction of the mass of $X$ that originates
from the electron's mass or electromagnetic energy. The distinction between atomic excited states is often unnecessary: the
difference in their atomic binding energies $E_{\text{atomic}, X}(\alpha, m_e) \simeq \mathcal{O}(\alpha^2 m_e)$ is $E_{\text{atomic},X}/m_X \sim 10^{-8}$, yielding a
negligible change to the dilatonic charge of the species.
An exception is precision measurements of energy level differences, such as in atomic clock systems and astrophysical spectroscopy.

The value of the electron dilatonic charge at leading order in $\alpha$ is the ratio of masses,
\begin{equation}
    (Q_{m_X})_{m_e} \approx Z m_e / m_X.
\end{equation}
The form of the electromagnetic dilatonic charge $(Q_{m_X})_e$ is less certain due to nuclear effects. Historically, Ref.~\cite{Bekenstein:1982eu} is the first attempt to estimate $(Q_{m_X})_e$ in the context of time-varying fundamental constants, arriving at $4 (Q_{m_X})_e \approx 1.3\times 10^{-2}$ by using a classical charge distribution model of the proton [or equivalently by extrapolating the large $Z$ limit of \cref{eqn:dilatonic_charges} below to $Z=1$].
More modern methods yield $(Q_{m_X})_e\sim 10^{-4}$; see Ref.~\cite[Sec.~IIA]{Tohfa:2023zip} for a summary.
References~\cite{Damour:2010rp,Damour:2010rm} estimate $(Q_{m_X})_e$ using the semi-empirical mass formula for nuclei (SEMF or the Bethe-Weizs\"acker formula), as well as the value for the electromagnetic energies of the proton and neutron from~\cite{Gasser:1982ap}.
While improvements in nuclear modeling and measurements have further refined these
estimates, they nevertheless capture $\mathcal{O}(1)$ of the effects and are sufficient for the
present discussion. Therefore, we use Eqs.~(16-21) of Ref.~\cite{Damour:2010rm}, extended to
ionized species:
\begin{subequations}
\label{eqn:dilatonic_charges}
\begin{align}
    (Q_{m_X})_{m_e}
    &= F_A \left( 5.5 \frac{Z_e}{A} \right) \times 10^{-4}
    \\
    (Q_{m_X})_{e}\,
    &= F_A \left(
            -1.4
            + 8.2 \frac{Z_p}{A}
            + 7.7 \frac{Z_p (Z_p - 1)}{A^{4/3}}
        \right)
        \times 10^{-4},
\end{align}
\end{subequations}
where $Z_p$ is the number of protons, $Z_e$ the number of electrons, $A$ the total number of
nucleons, and $F_A \equiv A m_\text{amu} / m_X$ with $m_\text{amu} = 931~\MeV$.

Note that because the dilatonic charge $(Q_{m_X})_{e}$ is nonzero,
\cref{eq:kappa_as_function_of_phi} implies that a scalar coupled only to the photon still modulates
the masses of nucleons and atoms.
However, because $(Q_{m_p})_{e}$ and $(Q_{m_p})_{m_e}$ are both of order $10^{-4}$, the relative variations in $m_e$ and in
$\alpha$ are much larger than the relative variation in the nucleon and atomic masses for comparable Lagrangian couplings; we therefore
take the approximation $m_X(\varphi) = m_X(0) $ for any nucleon, nucleus, or atom $X$ in our treatment of the CMB anisotropies~\cite{Baryakhtar:2024rky}.
On the other hand, we take into account the $\alpha$ dependence of nucleon, nuclear, and atomic
masses on the potential for $\varphi$ sourced by matter (\cref{sec:matter-potentials}), primordial
nucleosynthesis (\cref{sec:BBN}), and $\varphi$-mediated violations of the equivalence principle
(\cref{sec:UFF,sec:clocks}), where it is a leading effect.
These considerations motivate a more complete study of cosmological dynamics with varying nucleon
masses, including $\varphi$ dependence in $m_u$, $m_d$, and $\Lambda_\text{QCD}$, which is beyond the
scope of the present work.
(See Ref.~\cite{Smith:2024ayu} for exploration of such scenarios in the context of axio-dilaton
scalar-tensor theories.)

\subsection{Matter potentials}\label{sec:matter-potentials}

Having established the effects a coupled scalar can have on the fundamental parameters of the SM, we
next consider the effects of SM matter on $\varphi$ itself.
The matter content generates an effective, in-medium potential
$\partial \mathcal{L}_\mathrm{SM} / \partial \varphi \equiv - V_\text{matter}'(\varphi)$ that encodes
how $\varphi$ is sourced or scattered by matter distributions.
The magnitude of $V_\text{matter}'(\varphi)$ compared to $m_\varphi^2 \Mpl^2 \varphi$ and
$3 H \dot{\varphi}$ determines relevance of matter effects to the dynamics of $\varphi$ at any
particular epoch.
Even though matter-driven dynamics for $\varphi$ and the associated variations in $\alpha$ or $m_e$ are
not inadmissible \textit{a priori}, our analysis of the CMB in Ref.~\cite{Baryakhtar:2024rky}
assumes that the scalar evolution is driven only by the bare mass term $m_\varphi^2 \varphi$ and
cosmological expansion.
We compute in some generality the effective matter potential at various cosmological epochs in order to
delineate the parameter space where it can or cannot be neglected.
In this section we establish the formalism and important results; we discuss the consequences
thereof---namely, for which couplings the scalar field, and therefore its effect on the
fundamental constants, remains effectively frozen until after recombination---in
\cref{sec:cosmological-dynamics}.

A convenient way to compute the scalar's effective potential due to a background of matter with
finite temperature and density is the ``method of background fields''~\cite{Dolan:1973qd,
Weinberg:1974hy, Schwartz:2014sze} applied in thermal field theory.
This approach evaluates the (thermal) path integral for all coupled degrees of freedom,
treating $\varphi$ as constant and restoring its spacetime dependence after the fact; this
approximation is appropriate if the scales most important to the calculation are sufficiently
separated from those over which $\varphi$ varies.
For the scalar couplings encoded by $\alpha(\varphi)$ and $m_e(\varphi)$ [\cref{eq:DD_lagrangian}], this amounts to computing the thermodynamic pressure $P(\alpha, \{m_X\}, \{ T_X \}, \{\mu_X\}, \ldots)$ for a collection of species $X$ with
nonzero temperatures $T_X$ and chemical potentials $\mu_X$ (or equivalently, nonzero temperatures $T_X$ and number densities $n_X$),
treating the fundamental parameters as constant and replacing each with its $\varphi$-dependent form
at the end.
Then $V_\text{matter}(\varphi)=-P(\varphi)$, i.e., the matter potential
encodes how thermodynamic properties of a medium---whether the primordial thermal bath or a
volume of bulk material---vary as $\varphi$ changes the fundamental properties of its constituents.
Scalar-SM interactions thus enter the equation of motion simply as
\begin{eq}
    -\pd{V_\text{matter}}{\varphi} = \pd{P}{\varphi}
    = \sum_\lambda \pd{P}{\ln \lambda} g_\lambda'(\varphi).
 \label{eqn:def-matter-potential-ito-free-energy}
\end{eq}
In practice, we evaluate \cref{eqn:def-matter-potential-ito-free-energy} at leading order in
$\varphi$---i.e., with $P(\{\lambda(\varphi)\}) \rightarrow P(\{\lambda(0)\})$ and, for quadratic
couplings, $g_\lambda'(\varphi) \rightarrow d_\lambda^{(2)} \varphi$.
Note that $\varphi$ enters in the effective potential only through the SM parameters it modulates.
The smallness of variations in SM fundamental constants therefore guarantees that higher order
interactions generated by the quadratic coupling (i.e., terms proportional to higher powers of
$d_\lambda^{(2)}$) may be self-consistently neglected.

In what follows, we evaluate the thermodynamic pressure of a multispecies system as a function of
$\varphi$.
In \cref{sec:matter_potentials_free_gas}, we approximate the cosmological bath as a gas of noninteracting matter and radiation, and use the well-known expressions for the total thermodynamic pressure $P_\text{non-int}$ to obtain the corresponding matter potential $V_\text{non-int}(\varphi)$.
This noninteracting approximation is sufficient in the late Universe, i.e., after electron-positron
annihilation; however, a consistent calculation at $\mathcal{O}(\alpha)$ at earlier times must
account for long-range electromagnetic interactions between charged species (i.e., ``plasma
effects'').
In \cref{sec:plasma} we therefore compute the explicit $\alpha$-dependent contributions to the
pressure to obtain the potential $V_\text{int}(\varphi)$ due to interactions.
We then apply our results to assess the dominant contributions to
$V_\text{matter}(\varphi) = V_\text{non-int}(\varphi) + V_\text{int}(\varphi)$ in the late and early
Universe in \cref{sec:late_universe_potential,sec:early_universe_potential}, respectively.
Note that our discussion of $V_\text{non-int}$ and its evolution through the course of cosmological
history is similar to that of Ref.~\cite{Damour:1994ya}.

\subsubsection{Noninteracting gas}\label{sec:matter_potentials_free_gas}

We first consider a noninteracting gas of fermions and bosons of the different species $X$ [i.e., neglecting all interaction terms in the effective Lagrangian \cref{eq:effective_lagrangian}].
Then~\cite{Dolan:1973qd, Weinberg:1974hy, Kapusta:2006pm, Batell:2021ofv}
\begin{align}
    P_\text{non-int}
    &= \sum_{X}^{ \{ \gamma, e, \bar{e}, p, \bar{p}, n, \bar{n}, \text{H}, \hefour, \cdots\} }
        P_{\text{non-int}, X}
\end{align}
with
\begin{align}
    P_{\text{non-int},X}(\varphi)
    &= \frac{2s_X+1}{6\pi^2} T_X^4 \int_0^\infty \ud x \,
        \frac{
            x^4 / \sqrt{x^2 + [m_{X}(\varphi) / T_X]^2}
        }{
            \exp\left( \sqrt{x^2 + [m_{X}(\varphi) / T_X]^2} - \mu_X / T_X \right) \pm 1
        }
    ,
    \label{eqn:free-energy-non-int}
\end{align}
where $s_X$ is the spin of species $X$, $\mu_X$ is its chemical
potential, and the plus and minus signs are taken for fermions and bosons, respectively.
Even though a free photon gas contributes to the total pressure of the Universe, it does not depend
on any scale other than $T_\gamma$ [i.e., since $m_\gamma(\varphi) = 0$]; the partial pressure
$P_{\text{non-int},\gamma}$ of a free photon gas therefore does not depend on $\varphi$ and does not
contribute to its dynamics.

Next, recall that the spatially averaged energy density of a species $X$ is
\begin{align}
    \bar{\rho}_{\text{non-int},X}(\varphi)
    &= \frac{2s_X+1}{2\pi^2} T_X^4 \int_0^\infty \ud x \,
        \frac{
            x^2 \sqrt{x^2+[m_{X}(\varphi)/T_X]^2}
        }{
            \exp\left( \sqrt{x^2 + [m_{X}(\varphi) / T_X]^2} - \mu_X / T_X \right) \pm 1
        }.
\end{align}
One can then check that the pressure differential yields
\begin{align}
    \label{eq:matter_potential}
    \left(\pd{P_{\text{non-int},X}}{\varphi}\right)_{\mu_X,T_X}
    = - g_{m_X}'(\varphi) \Theta_{\text{non-int},X}(m_X(\varphi))
\end{align}
where
\begin{align}
\label{eqn:rho-P-noninteracting-gas}
    \Theta_{\text{non-int},X}(m_X(\varphi))
    &= \bar{\rho}_{\text{non-int},X}(\varphi) - 3P_{\text{non-int},X}(\varphi)
    \\
    &= \frac{2s_X+1}{2\pi^2}m_X(\varphi)^2 T_X^2 \mathcal{J}_\pm(m_X(\varphi), T_X, \mu_X)
\end{align}
is the trace the (noninteracting) stress-energy tensor of the species, and we defined
\begin{align}
    \mathcal{J}_\pm(m, T, \mu)
    &\equiv \int_0^\infty \ud x
        \frac{
            x^2 / \sqrt{x^2 + (m / T)^2}
        }{
            \exp\left( \sqrt{x^2 + (m / T)^2} - \mu / T \right) \pm 1
        }.
    \label{eqn:J-integral}
\end{align}
As observed in Refs.~\cite{Damour:1994ya, Erickcek:2013dea, Sibiryakov:2020eir, Bouley:2022eer}, the
result that $\varphi$ effectively couples to $\sum_X \Theta_{\text{non-int},X}$ can also be obtained
by noting that in the interaction Lagrangian $\varphi$ couples to the bilinear
mass operators $m_f \bar{f} f$, $m_b^2 b^\ast b$, etc.
In the thermal state these operators correspond to the trace of the thermal energy-momentum tensor
of each free species at $\varphi = 0$.

In the relativistic limit $T_X \gg m_X,\mu_X$, the matter potential reduces to
\begin{eq}
\label{eq:relativistic_trace}
    \Theta_{\text{non-int},X}(\varphi)
    &\approx
        \frac{2s_X+1}{12}m_{X}(\varphi)^2T_X^2 \times\begin{cases}
            1, &\text{bosons,}\\
            \frac{1}{2}, &\text{fermions}.
        \end{cases}
\end{eq}
In the nonrelativistic and dilute (nondegenerate) limit, $T_X \ll m_X$ and $T_X\ll m_X - \mu_X$,
\begin{eq}
\label{eq:non-relativistic_trace}
    \Theta_{\text{non-int},X}(\varphi)
    &\approx
       m_{X}(\varphi) n_{X}.
\end{eq}
where the number densities $n_X$ are averaged over cosmological scales.
In summary, the effective source term $V_\text{matter}'(\varphi)$ depends on $\varphi$ simply via
the coupling functions $g'_{m_X}(\varphi)$, weighted by the trace of species $X$'s energy-momentum
tensor, \cref{eq:matter_potential}.
Each $g'_{m_X}(\varphi)$ is related to the fundamental couplings $g'_e(\varphi)$ and
$g'_{m_e}(\varphi)$ by the dilatonic charges of species $X$, which heuristically quantify how much
of the mass $m_X$ is of electronic or electromagnetic origin.

\subsubsection{Interacting plasma}\label{sec:plasma}

As discussed in \cref{sec:Effective Lagrangian for phi-matter coupling}, the electromagnetic energy
content in matter depends on $\alpha$, meaning the contributions to $g'_{m_X}(\varphi)$ in
\cref{eq:relativistic_trace,eq:non-relativistic_trace} from the photon coupling are suppressed by
powers of $\alpha$.
In other words, the dilatonic charges $(Q_{m_X})_e$ are really $\mathcal{O}(\alpha)$ quantities, the
same order as the leading contributions from electromagnetic interactions.
We now compute the contribution $P_\text{int}$ from interactions to the pressure by
treating the matter system as a quantum electrodynamics (QED) gas or plasma of photons and charged
matter.

The pressure of an interacting QED gas is known at $\mathcal{O}(e^2)$~\cite{Kapusta:2006pm} and beyond~\cite{Arnold:1994eb}.
Formally, the contribution to the pressure due to the electromagnetic interaction between particles
at $\mathcal{O}(\alpha)$ corresponds to the sum of ``sunset''-type bubble Feynman diagrams of
individual charged matter fields (i.e., single loops of a matter field of a given type with an
internal photon line inserted).
Because particle-antiparticle pairs are created and annihilated in a QED plasma, the full expression
contains products of the phase space distribution functions of particles and antiparticles of the same
type, with $\mu_X = -\mu_{\bar{X}}$.
In standard cosmological thermodynamics, however, SM species are either relativistic with a
negligible matter-antimatter asymmetry ($|\mu_X| \ll T_X$) or nonrelativistic and dilute but with
a large matter-antimatter asymmetry.
In the former, symmetric case, we denote the total interaction of particles and antiparticles of the same type with $P_{\text{int},X} + P_{\text{int},\bar{X}}$.
In the latter, nonrelativistic case, one can simply take $P_{\text{int},\bar{X}}\approx 0$ for antimatter and $P_{\text{int},X}$ the nonrelativistic limit of the term corresponding to matter-matter interactions.

Taking these assumptions, working at leading order in $\alpha$, and assuming all species share a
common plasma temperature $T_p$, $P_\text{int}$ may be written as a sum over the pressure of each matter species:
\begin{eq}
    P_\text{int}
    \approx \sum_{X}^{ \{ e, \bar{e}, p, \bar{p}, \text{H}^+, \hefour^{2+}, \cdots\} }
        P_{\text{int},X}.
\end{eq}
To evaluate the relativistic limit while still capturing the exponential suppression at
annihilation, we take Eq.~(5.58) from Ref.~\cite{Kapusta:2006pm} for the contribution of a
single type of particle/antiparticle sector to the interaction pressure and substitute $m_X = 0$
everywhere except inside distribution functions:
\begin{align}
    P_{\text{int}, X}
    + P_{\text{int}, \bar{X}}
    &\approx \frac{2 q_X^2 \alpha T_p^4}{3 \pi}
        \left[
            \mathcal{J}_+(m_X, T_p, 0)
            + \frac{3}{\pi^2} \mathcal{J}_+(m_X, T_p, 0)^2
        \right],
    \label{eqn:interacting-plasma-energy-rel}
\end{align}
where $q_X$ is the electromagnetic charge of $X$.

As before, the source term in the equation of motion of $\varphi$ is given by the pressure differential
$\sim \partial P_\text{int}/\partial \varphi$, replacing $\alpha$ and $m_X$ with their
$\varphi$-dependent values.
The contribution from $X$ particles together with $\bar{X}$ antiparticles is
\begin{subequations}
\begin{align}
\begin{split}
    \left( \pd{P_{\text{int},X}}{\varphi} + \pd{P_{\text{int},\bar{X}}}{\varphi} \right)_{T_p,\mu_X}
    &\approx -g'_e(\varphi)
       \frac{2 q_X^2 \alpha(\varphi)  T_p^4}{3 \pi}
       \left[
           \mathcal{J}_+(m_X, T_p, 0)
           + \frac{3}{\pi^2} \mathcal{J}_+(m_X, T_p, 0)^2
       \right].
    \label{eq:plasma_tot}
\end{split}
\end{align}
In the relativistic limit $T_p \gg m_X$, the interaction contribution to the matter potential is then
\begin{align}
    \left( \pd{P_{\text{int},X}}{\varphi} + \pd{P_{\text{int},\bar{X}}}{\varphi} \right)_{T_p,\mu_X}
    &\approx -g'_e(\varphi) \frac{5\pi}{72} q_X^2 \alpha(\varphi) T_p^4.
        \label{eq:plasma_rel}
\end{align}
The nonrelativistic ($T_p \ll m_X$) and dilute (nondegenerate) limit with a matter-antimatter asymmetry
($n_X \gg n_{\bar{X}}$)
is $T_p \ll m_X-\mu_X \ll m_{\bar{X}} - \mu_{\bar{X}}$.
The interaction contribution is given by~\cite[see Eq.~(5.61)]{Kapusta:2006pm},
\begin{align}
    \left( \pd{P_{\text{int},X}}{\varphi} \right)_{T_p,\mu_X}
    &\approx  - g'_e(\varphi)  m_X(\varphi) n_X
        \left[
            q_X^2 \alpha(\varphi)
            \left(
                \frac{4\pi}{12} \frac{T_p^2}{m_X(\varphi)^2}
                -\frac{n_X \lambda_{T_p,X}^3}{4 \sqrt{2\pi}} \sqrt{\frac{T_p}{m_X(\varphi)}}
            \right)
        \right],
    \label{eq:plasma_non_rel}
\end{align}
\end{subequations}
where $\lambda_{T_p,X} = \sqrt{2 \pi / m_X T_{p}}$ is the thermal de Broglie wavelength of species
$X$.
The first term in large brackets in \cref{eq:plasma_non_rel} is equal to
$\omega_{p,X}^2 T_{p}^2 / 12$, where $ \omega_{p,X}^2 = 4 \pi \alpha q_X^2 n_X / m_X$
is the classical plasma frequency of a plasma of the charged species $X$.
Heuristically, photons propagating in a plasma are ``dressed'' in the field of $X$ and acquire a
``plasma mass'' $\omega_{p, X}$.
As a result, the electromagnetic operator $F_{\mu \nu} F^{\mu \nu} = \three{E}^2 - \three{B}^2$ is
nonzero, meaning plasma photons can source $\varphi$.
The second term in large brackets in \cref{eq:plasma_non_rel} is quantum in origin as it involves the
thermal de Broglie wavelength.

\subsubsection{After electron-positron annihilation}
\label{sec:late_universe_potential}

After annihilation and nucleosynthesis, species are nonrelativistic and nondegenerate and antimatter is negligible, so the effective potential in the noninteracting limit is given by \cref{eq:non-relativistic_trace} summed over species:
\begin{eq}
    \label{eq:non_rel_potential}
    -\pd{V_\text{non-int}}{\varphi} = \pd{P_\text{non-int}}{\varphi}
    &\approx \sum_{X}^{ \{ e, \bar{e}, p, \bar{p}, \text{H}, \hefour, \cdots\} }
        -g'_{m_X}(\varphi) m_{X}(0) n_{X}.
\end{eq}
Turning to the potential from interactions, first note that for species with a common
temperature, $\partial P_\mathrm{int}/\partial \varphi$ [\cref{eq:plasma_non_rel}] is dominated by the
lightest species (namely, the electron)---an ubiquitous feature of nonrelativistic plasma physics.
Moreover, the explicit photon coupling term in \cref{eq:plasma_non_rel} is multiplied by $\alpha$,
and the implicit photon contribution to $g_{m_X}(\varphi)$ in \cref{eq:non_rel_potential} is
proportional to the dilatonic charge $(Q_{m_X})_e$, which is formally $\mathcal{O}(\alpha)$ as well.
Thus, the plasma contributions are indeed necessary for a consistent treatment of the effective
potential under the photon coupling.
However, these terms [in \cref{eq:plasma_non_rel}] are further suppressed by $(T_p/m_X)^2$ or
$n_X \lambda_{T_p,X}(0)^3 \sqrt{T_p / m_X(0)}$, which are both small at late times, and also by the hierarchy $\sim m_e / m_p$ between the electron and baryon energy densities.
The plasma contribution of electrons is therefore highly suppressed at late times relative to the nonplasma contributions of nuclei.
Thus, we conclude that plasma effects in the late Universe are subdominant and take
$-V_\text{matter}'(\varphi) \approx P_\text{non-int}'(\varphi)$ after electron-positron
annihilation.

In principle, the sum in \cref{eq:non_rel_potential} distinguishes between neutral atoms, free nuclei, free neutrons and free electrons.
After primordial nucleosynthesis, the Universe contains no free neutrons---the content of the
Universe is made up of a fraction $Y_\mathrm{He} \equiv \bar{\rho}_\mathrm{He} / (\bar{\rho}_\mathrm{H} + \bar{\rho}_\mathrm{He})$ of helium nuclei by mass, with the remainder
almost entirely in free protons.
Moreover, as explained in \cref{sec:Effective Lagrangian for phi-matter coupling}, atomic binding
energies make a negligible contribution to dilatonic charges.
For nonrelativistic species, the thermal potential then only depends on the \emph{number} of
particles of each species, as is evident in \cref{eq:non_rel_potential}.
Since the total numbers of electrons and nuclei are the same whether or not they are bound into
atoms (taking the Universe to be completely charge neutral), \cref{eq:non_rel_potential} may be
computed as if all nuclei and electrons were bound into atoms (again when neglecting contributions
from binding energy).
As a consequence, the effective potential in this limit is unchanged in form throughout
recombination and reionization.

An additional subtlety is that, after photon decoupling, the electrons and photons do not have a
common temperature as assumed in \cref{eq:plasma_non_rel}.
Namely, after recombination the temperature of matter redshifts as $1/a^2$ rather than $1/a$.
However, we only expect the appropriate extension of the bracketed term in \cref{eq:plasma_non_rel} to remain a suppression factor for any temperatures below the electron mass.
We therefore ignore this distinction; determining whether the scalar rolls substantially leading up
to recombination does not require treating this regime regardless.

\begin{figure}[t!]
\begin{centering}
    \includegraphics[width=\textwidth]{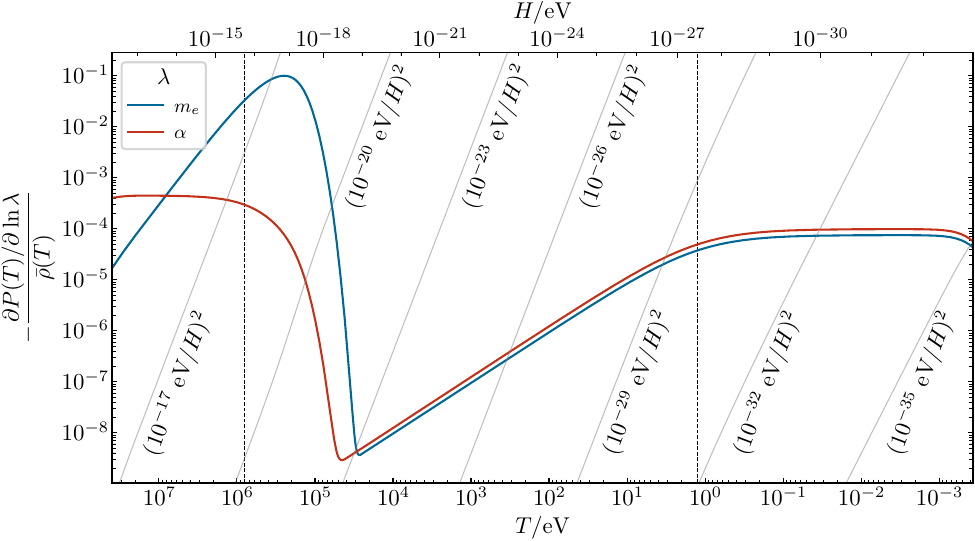}
    \caption{
        Contributions to the effective matter potential
        \cref{eqn:def-matter-potential-ito-free-energy} for a scalar coupled to the electron (blue)
        or photon (red) relative to the total energy density of the Universe.
        For a quadratically coupled scalar, the vertical axis is equal to
        $2 \delta m_T^2 / 3 H^2 d_{\lambda}^{(2)}$ where $\delta m_T$ is its in-medium or
        ``thermal'' mass.
        Grey lines depict various mass scales relative to the Hubble rate (both squared) as labeled
        on the figure; where they intersect the red and blue curves indicates when the in-medium
        mass takes the labeled value (up to a factor of $3 d_{\lambda}^{(2)} / 2$).
        Vertical lines indicate when the photon temperature $T = m_e$ and
        $T = T_\mathrm{eq} \approx 0.80~\eV$.
    }
    \label{fig:sm-free-energy-dlambda}
\end{centering}
\end{figure}

In summary, we may conveniently parametrize the thermal potential in terms of effective dilatonic
charges for the total baryonic content, i.e., as a weighted sum of those for hydrogen and helium
atoms:
\begin{align}
    \left( Q_b \right)_\lambda
    &\equiv \left( 1 - Y_\mathrm{He} \right) (Q_{m_\mathrm{H}})_{\lambda}
        + Y_\mathrm{He} (Q_{m_\mathrm{He}})_{\lambda}
    \approx 10^{-4}
        \cdot
        \begin{cases}
            4.8, & \lambda = m_e,
            \\
            6.3, & \lambda = e,
        \end{cases}
    \label{eqn:dilatonic-charge-baryonic-content}
\end{align}
for a fiducial helium mass fraction $Y_\mathrm{He} = 0.25$.
The thermal force term is then
\begin{align}
    \label{eqn:non_rel_potential-rho-b}
    \pd{V_\text{matter}}{\varphi}
    &= \left[ (Q_b)_{m_e} g'_{m_e}(\varphi) + (Q_b)_{e} g'_{e}(\varphi) \right]
        \bar{\rho}_b.
\end{align}
\Cref{fig:sm-free-energy-dlambda} depicts the temperature evolution of the potential
[specifically, the coefficients of the coupling functions in \cref{eqn:non_rel_potential-rho-b}].
In \cref{fig:sm-free-energy-dlambda}, we obtain the noninteracting electron  pressure differential (for the
electron coupling) by integrating \cref{eqn:rho-P-noninteracting-gas} directly with $\mu_e$ and
$\mu_{\bar{e}} = 0$, and likewise for the contribution to the pressure differential from plasma interactions
[\cref{eq:plasma_rel}]; see \cref{eqn:matter-potential-early} and the surrounding discussion.
The nonrelativistic limit [\cref{eqn:non_rel_potential-rho-b} divided by $g_\lambda'(\varphi)$] is
simply added to the relativistic result (since the detailed evolution in the brief interval where
they are comparable in size is unimportant).

Specializing to quadratic couplings, the gradient of the effective potential is
\begin{align}
    4\pi G \pd{V_\text{matter}}{\varphi}
    &=
        \frac{\bar{\rho}_b}{\bar{\rho}_\mathrm{crit}}
        \frac{3 H^2}{2}
        \left[
            \left( Q_b \right)_{m_e} d_{m_e}^{(2)}
            + \left( Q_b \right)_{e} d_{e}^{(2)}
        \right]
        \varphi
    \equiv \delta m_T^2 \varphi,
    \label{eqn:dVmatter-dphi}
\end{align}
where $\bar{\rho}_\mathrm{crit} = 3 H^2 \Mpl^2$ is the critical energy density of a flat Universe.
Evidently, a quadratic coupling sources an effective in-medium mass for the scalar, often referred
to as a thermal mass:
\begin{align}
    \delta m_T^2
    &= \sum_{\lambda \in \{ e, m_e \}}
        \left( 5 \times 10^{-36}~\eV \right)^2
        d_{\lambda}^{(2)}
        a^{-3}
        \frac{(Q_b)_{\lambda}}{5 \times 10^{-4}}
        \frac{\omega_b}{0.0224},
    \label{eqn:post-annhihilation-thermal-mass-parametrics}
\end{align}
where the present-day baryon density $\omega_b \equiv \bar{\rho}_{b, 0} / 3 H_{100}^2 \Mpl^2$.
Its parametrization in terms of the Hubble rate squared in \cref{eqn:dVmatter-dphi} facilitates
identifying under what conditions thermal effects are important to the scalar's dynamics.
Since $\bar{\rho}_b / \bar{\rho}_\mathrm{crit} \lesssim 1/6$ at all times~\cite{Planck:2018vyg}
(with the inequality saturated during the matter-dominated era), matter effects never dominate the
scalar's dynamics after electron-positron annihilation if both quadratic couplings satisfy
$d_\lambda^{(2)} \lesssim 1/(Q_b)_\lambda \sim 10^{4}$, as also evident in
\cref{fig:sm-free-energy-dlambda}.
In \cref{sec:cmb-matter-potential} we discuss more precise conditions under which even subdominant
effects can be neglected within the context of our analysis of variations of fundamental constants
as relevant to analyses of cosmological data.

\subsubsection{Before electron-positron annihilation}
\label{sec:early_universe_potential}

Before electrons and positrons annihilate efficiently [i.e., at temperatures $T \gg m_e(0)$], they are
relativistic and freely produced in pairs ($\mu_{e} \approx - \mu_{\bar{e}} \approx 0$), drastically
increasing their comoving number density relative to later times.
The electron-coupling contribution to the matter potential must be computed directly from
\cref{eq:matter_potential}, whose relativistic limit is $g'_{m_e}(\varphi) m_e(\varphi)^2 T^2 / 6$.
Of course, in numerical solutions one may compute the full matter potential (without a relativistic
approximation) by numerically integrating \cref{eqn:rho-P-noninteracting-gas}, as in
\cref{fig:sm-free-energy-dlambda}.

The potential from the plasma effects of baryons remains significantly suppressed relative to their
nonplasma one, but that of electrons and positrons is much enhanced.
Namely, electrons and positrons each contribute \cref{eq:plasma_rel}.
Since $\bar{\rho}_b / T^4 \approx 2 \eta \zeta(3) m_p / \pi^2 / T \approx (T / 0.14~\eV)^{-1}$
(where $\eta \approx 6 \times 10^{-10}$ is the baryon-to-photon number ratio), the baryonic
contributions are substantially smaller than those from the electron's plasma interactions at
$T > m_e(0)$.
On the other hand, the relativistic limit of $P_\mathrm{int}$
[\cref{eqn:interacting-plasma-energy-rel}] should have subleading corrections
$\sim \alpha m_e^2 T^2$, which would induce a forcing term proportional
to $\alpha(\varphi) g'_{m_e}(\varphi)$.
However, this contribution is smaller than the noninteracting result by a factor
$\alpha(\varphi)$ and may be neglected.

In principle, $\partial P_\text{non-int.}/\partial \ln \alpha$ of the nonrelativistic, noninteracting baryons evolves nontrivially as
neutrons decay and nucleosynthesis occurs.
However, because the baryon-to-photon ratio is so small, electron-positron annihilation does not
complete until temperatures $T \approx 20~\mathrm{keV}$~\cite{Grayson:2023flr, Thomas:2019ran},
after nucleosynthesis completes.
The matter potential from relativistic electrons (for either the electron or
photon coupling) therefore dominate over the baryonic contributions in this epoch in spite of the
exponential suppression that begins at $T \approx m_e$, rendering the details of nucleosynthesis
irrelevant.

In sum, the full effective potential up to $e^+$-$e^-$ annihilation is well approximated as that
due just to the electron:
\begin{align}
\begin{split}
    \pd{V_\text{matter}}{\varphi}
    = - \pd{P}{\varphi}
    &\approx
       \frac{2 m_e(\varphi)^2 T^2}{\pi^2} \mathcal{J}_+(m_e, T, 0) g'_{m_e}(\varphi)
    \\ &\hphantom{ {}={} }
        +
        \frac{2 \alpha(\varphi) T^4}{3 \pi}
        \left[
            \mathcal{J}_+(m_e, T, 0)
            + \frac{3}{\pi^2} \mathcal{J}_+(m_e, T, 0)^2
        \right]
        g_e'(\varphi)
\end{split}
    \label{eqn:matter-potential-early}
\end{align}
with $\mathcal{J}_+$ defined in \cref{eqn:J-integral}.
In the limit $T \gg m_e$, $\mathcal{J}_+(m_e, T, 0) \approx \pi^2 / 12$.
Substituting the radiation-era result $H^2 = \pi^2 g_\star(T) T^4 / 90 \Mpl^2$, where $g_\star$ is
the effective number of relativistic degrees of freedom in the Universe, and again specializing to
quadratic couplings, the in-medium mass is
\begin{align}
    \delta m_T^2
    &= \frac{H^2}{g_\star(T)}
        \left[
            \frac{15 d_{m_e}^{(2)}}{2 \pi^2} \left( \frac{T}{m_e(0)} \right)^{-2}
            + \frac{25 \alpha(0) d_{e}^{(2)}}{8 \pi}
        \right]
    \label{eq:thermal_mass_early}
\end{align}
in the relativistic limit.
Shortly before annihilation, $g_\star \approx 10.75$.
As illustrated by \cref{fig:sm-free-energy-dlambda}, $\delta m_T^2$ is largest relative to $H^2
\propto \bar{\rho}$ around the time of electron-positron pair annihilation (i.e., $T\approx m_e$).
Requiring the impact of matter to be irrelevant to the scalar's dynamics in the early Universe
imposes $d^{(2)}_{m_e} \lesssim 10$ and $d_e^{(2)}\lesssim 10^3$.
Again, we defer more precise discussion of subdominant matter effects within the context our
analysis of variations of fundamental constants to \cref{sec:cmb-matter-potential}.

\subsection{Standard Model sources of variations in fundamental constants}\label{sec:sm-sources-of-variation}

The plasma effects studied in \cref{sec:plasma} induce environment dependence of $\alpha$ and $m_e$
purely via SM processes.
As discussed earlier, in thermal field theory photons and charged particles are ``dressed'' in the
surrounding bath of other charged particles and photons.
The term $\partial P_{\text{int},X}/\partial\varphi$ due to interactions, \cref{eq:plasma_non_rel}, takes the form
of the number density times an effective mass.
Heuristically, we might interpret the coefficient of $n_X$ in \cref{eq:plasma_non_rel}
as an estimate of the mass correction for species $X$ due to plasma effects:
\begin{align}
    \frac{\delta m_X(T_p)}{m_X}
    &= q_X^2 \alpha \left(
            \frac{4\pi}{12} \frac{T_p^2}{m_X^2}
            - \frac{n_X \lambda_{T_p,X}^3}{4 \sqrt{2\pi}} \sqrt{\frac{T_p}{m_X}}
        \right)
    .
    \label{eq:thermal_mass}
\end{align}
For the electron, writing $n_e = 2 \eta \zeta(3) T_p^3 / \pi^2$,
\begin{align}
    \frac{\delta m_e(T_p)}{m_e}
    &= \alpha \left(
            \frac{4 \pi}{12}
            - \frac{\eta \zeta(3)}{\pi}
        \right)
        \frac{T_p^2}{m_e^2},
\end{align}
which is order $10^{-11}$ when $T_p \sim \alpha^2 m_e$ (i.e., the binding energy of hydrogen), and
even smaller by the time of photon decoupling, $T_p \approx 0.26~\eV$.

Similarly, electric charges are screened in plasmas.
In a nonrelativistic and dilute plasma, the electrostatic Coulomb potential between two charges,
which in vacuum is $\alpha / r$, is suppressed by a factor
$\exp(- r / \lambda_D) \approx 1 - r / \lambda_D$, where
\begin{eq}
    \lambda_D
    &= \sqrt{\frac{1}{4\pi \alpha}\frac{T_p}{\sum_X q_X^2 n_X}}
\end{eq}
is the Debye length of the plasma~\cite{thorne2017modern}.
This screening effect may be interpreted as a shift
$\vert \delta \alpha(T_p)/\alpha \vert = r/\lambda_D$ of the fine-structure constant for processes taking place over a characteristic length scale $r$.
In other words, plasma charges appear as neutral on scales greater than the Debye length.

There are at least two relevant length scales around recombination.
The first is the Bohr radius $r_B \sim 1 / \alpha m_e$: $\alpha(1-r_B/\lambda_D)$ is essentially the
fine structure constant ``seen'' by atomic charges.
For example, in the extreme case $\lambda_D \ll r_B$, proton charges would be screened on scales smaller than atomic sizes and atoms could not form.
Parametrically, $r_B/\lambda_D\sim \alpha^{-1/2}(n_p/ T_p^3)^{1/2}(T_p/m_e)\sim \alpha^{-1/2}\sqrt{\eta}(T_p/m_e) \sim 10^{-8}$ at the time of recombination.
The second important length scale is the wavelength of Thomson-scattered photons.
Electrons appear neutral to photons with wavelengths longer than $\lambda_D$, leading to a suppression of scattering.
In principle the bath contains photons of all wavelengths, most conveniently parametrized in units of the plasma temperature $T_p$ for which $1/\lambda_D T_p\sim \alpha^{1/2} (n_e / T^3)^{1/2} \sim \alpha^{1/2} \sqrt{\eta} \sim 10^{-6}$.
Photons with longer wavelengths experience a larger suppression but represent an exponentially suppressed fraction of the bath.
SM plasma effects are thus subdominant to the level of precision with which variations in $m_e$ and
$\alpha$ can be tested at present.

\subsection{Cosmological dynamics}\label{sec:cosmological-dynamics}

We conclude our overview of the theory of models of varying fundamental constants by applying the
preceding results to the cosmological dynamics of new, coupled scalars.
We consider scenarios in which $\varphi$ has a nonzero, spatially homogeneous, ``misaligned'' initial
condition for which an expansion into a homogeneous component and a small spatial perturbation,
$\varphi(t,\three{x}) = \bar{\varphi}(t) + \delta \varphi(t,\three{x})$, is appropriate.
For the purposes of this work (except those results relying on Ref.~\cite{Baryakhtar:2024rky}), only
the homogeneous component has important cosmological implications.
Per \cref{eq:scalar-eom}, $\bar{\varphi}(t)$ evolves according to the homogeneous
Klein-Gordon equation,
\begin{subequations}\label{eqn:general-klein-gordon}
\begin{align}
    0
    &= \ddot{\bar{\varphi}} + 3 H \dot{\bar{\varphi}} + \meff(a)^2 \bar{\varphi}
\intertext{with}
    \meff(a)^2
    &\equiv m_\varphi^2 + \delta m_T(a)^2,
\end{align}
\end{subequations}
where the ``thermal mass'' $\delta m_T$ is given by \cref{eqn:dVmatter-dphi,eq:thermal_mass_early}.
Note that $\delta m_T(a)^2$ is proportional to the quadratic couplings $d_\lambda^{(2)}$ and
therefore is not necessarily positive.
The field $\varphi$ itself contributes to the Friedmann equation for the Hubble parameter $H$ but is
subdominant so long as $\meff^2 \varphi^2 \ll H^2$, which we assume to be the case.
To first approximation, expansion thus proceeds as in standard $\Lambda$ cold dark matter (\LCDM{}) cosmology.

\subsubsection{Free evolution}

In the bare-mass--dominated limit ($m_\varphi \gg \vert \delta m_T \vert$),
\cref{eqn:general-klein-gordon} reduces to the standard equation of motion for a massive scalar, a
regime that has been extensively studied in the context where $\varphi$ is, e.g., a pre-inflationary
axion, some other light scalar dark matter candidate, or a dynamical subcomponent of dark energy.
In such a context, solutions to \cref{eqn:general-klein-gordon} are effectively frozen at the
initial condition $\bar{\varphi}_i$ until $H$ drops below $m_\varphi$, at which point damped, harmonic
oscillations begin.
If the field is sufficiently light at recombination
($m_\varphi < H_\star \approx 3 \times 10^{-29}~\eV$) but is also heavy today ($m_\varphi > H_0$)
oscillations begin between recombination and the present day.
The fundamental constants are then spacetime independent over the Universe's history up to (and
including) recombination but different from their current values.
Hence, the free evolution of a hyperlight ($10^{-33}~\eV < m_\varphi < 10^{-28}~\eV$) field realizes
the scenario considered in previous phenomenological studies of shifted fundamental constants at big bang nucleosynthesis (BBN)
and recombination.
Here and in Ref.~\cite{Baryakhtar:2024rky} we focus on this ``hyperlight'' regime.

Since recombination occurs after matter-radiation equality and
matter--dark-energy equality was relatively recent, the scalar begins oscillating in the
matter-dominated era in the parameter space we consider.
In a matter-dominated Universe (for which $H = 2/3t$), the solution to
\cref{eqn:general-klein-gordon} with $\delta m_T = 0$ and initial conditions
$\varphi \rightarrow \bar{\varphi}_i$ and $\dot{\varphi} \rightarrow 0$ as $m_\varphi t \rightarrow 0$ is
$\bar{\varphi}(t) = \bar{\varphi}_i \sin(m_\varphi t) / m_\varphi t$.
As anticipated, $\varphi(t)$ is nearly constant until $t \sim 1/m_\varphi$ and oscillates with
frequency $m_\varphi$ afterwards.
We therefore define the time $t_\osc$ when oscillations begin by
$H(t_\osc) = 2 / 3 t_\osc = 2 m_\varphi / 3$.
The associated scale factor $a_\osc$ and redshift $z_\osc \equiv 1 / a_\osc - 1$ are
determined via the first Friedmann equation by
\begin{align}
\label{eq:def_a_osc}
    m_\varphi^2
    = \frac{9 H(t_\osc)^2}{4}
    \approx \frac{3}{4}
        \frac{\bar{\rho}_c(t_\osc) + \bar{\rho}_b(t_\osc)}{\Mpl^2}
    \approx \frac{3}{4}
        \frac{\bar{\rho}_{c,0} + \bar{\rho}_{b,0}}{a_\osc^3 \Mpl^2}.
\end{align}
Solutions in the oscillatory regime may be extended until after matter-$\Lambda$ equality via
\begin{align}
    \bar{\varphi}(t)
    &=\varphiamp(t) \sin\left( m_\varphi t \right),
\end{align}
where the slowly varying oscillation amplitude is
\begin{align}
    \varphiamp(t)
    &\approx \bar{\varphi}_i \left(\frac{a(t)}{a_\osc}\right)^{-3/2}
\label{eq:evolution_free_amplitude}
\end{align}
at times $t > t_\osc$.

\subsubsection{Impact of matter potentials}
\label{sec:matter_corrections_to_evolution}

We now study modifications to the scalar dynamics incurred at larger couplings for which the
effective in-medium potential is nonnegligible, specializing to interactions quadratic in $\varphi$.
\Cref{sec:matter-potentials} shows that electron-positron annihilation separates two qualitatively
distinct regimes.
All probes we consider only depend on dynamics after electron-positron annihilation, except for BBN;
the impact of matter potentials before annihilation on BBN was studied in detail in
Refs.~\cite{Sibiryakov:2020eir, Bouley:2022eer}.
The dynamics before annihilation do impact the relationship between the post-annihilation initial
condition $\bar{\varphi}_i$ and the initial conditions at much earlier times.
We discuss analytic solutions for the pre-annihilation dynamics and their implications in
\cref{app:before-annihilation} and here consider only dynamics after annihilation, for which the
thermal mass is given by \cref{eqn:dVmatter-dphi,eqn:post-annhihilation-thermal-mass-parametrics}.

Since the thermal mass evolves with the energy density of SM matter, $\dot{m}_\mathrm{eff}$ is
generically proportional to $H \meff$ (except at electron-positron annihilation); when $\meff
\gtrsim H$, \cref{eqn:general-klein-gordon} is then amenable to a Wentzel-Kramers-Brillouin (WKB)
approximation~\cite{Bouley:2022eer, Sibiryakov:2020eir, Olive:2001vz}.
The solutions are therefore well described by an oscillatory function times a slowly varying
amplitude,
\begin{subequations}
\label{eq:WKB}
\begin{align}
    \bar{\varphi}(t)
    &= \varphiamp(t)
        \sin\left( \int_{t_\text{ref}}^t \meff(t') \ud t' + \vartheta(t_\text{ref}) \right),
\end{align}
where $t_\text{ref}$ is some reference time for which $\meff^2 > H^2$, $\vartheta(t_\text{ref})$
the phase at that reference time, and
\begin{align}
    \varphiamp(t)
    &\propto \frac{1}{a(t)^{3/2}}\frac{1}{\meff(t)^{1/2}}.
\end{align}
\end{subequations}
In the limit that $m_\varphi \gg \vert \delta m_T \vert$ and $m_\varphi \gg H$, we recover $\varphiamp
\propto a^{-3/2}$ as for free evolution [\cref{eq:evolution_free_amplitude}].
On the other hand, if $\delta m_T$ is much larger than both $H$ and $m_\varphi$, the envelope evolves
as $\varphiamp \propto a^{-3/2} \bar{\rho}_b^{-1/4} \propto a^{-3/4}$ (after electron-positron
annihilation).\footnote{
    We do not treat the regime of negative $\delta m_T$ much larger in magnitude than $m_\varphi$;
    the resulting tachyonic instability makes an analysis unwieldy and the phenomenology sensitive
    to initial conditions and dynamics at earlier times.
    An understanding thereof is also not required to describe the parameter regimes our analysis
    ultimately constrains.
}

When $\meff(a)^2 \lesssim H^2$, the WKB method does not apply.
Because $\delta m_T$ redshifts, generally $m_\varphi \ll \vert \delta m_T \vert$ during some period of
time before $t_\osc$ unless couplings are very small or the bare mass is relatively large.
Because \cref{eqn:general-klein-gordon} then describes an oscillator whose mass and damping
coefficient both evolve with time, the scalar is not necessarily frozen prior to $t_\osc$.
In \cref{app:thermal_field_evolution}, we obtain analytic solutions to the equation of motion at
times $t \lesssim t_\osc$ when the bare mass $m_\varphi$ is irrelevant to the dynamics.
These solutions are uniquely parametrized by the combination (see \cref{app:after-annihilation})
\begin{align}
    \mathcal{D}
    &\equiv \frac{3}{2} \frac{(Q_b)_\lambda d_\lambda^{(2)}}{1 + \omega_c / \omega_b}
    ,
    \label{eq:coupling-combo-late}
\end{align}
where the dilatonic charges of baryonic matter $(Q_b)_\lambda$ are given by
\cref{eqn:dilatonic-charge-baryonic-content} and are of order $10^{-4}$.
Here $\omega_X = \bar{\rho}_{X, 0} / 3 H_{100}^2 \Mpl^2$ denotes the rescaled, present-day energy
density of species $X$; in \LCDM{} cosmologies,
$\omega_b / \omega_c = 0.1855 \pm 0.0028$~\cite{Planck:2018vyg}.

Note that $\vert \delta m_T^2 \vert / H^2$ increases during radiation domination and is constant and
equal to $\vert \mathcal{D} \vert$ during matter domination, behavior reflected in
\cref{fig:sm-free-energy-dlambda}.
\Cref{app:after-annihilation} shows that the field begins evolving before matter-radiation equality
if $\vert \mathcal{D} \vert \gtrsim 1$, eventually reaching the WKB regime at larger couplings.
On the other hand, when $\vert \mathcal{D} \vert \lesssim 1$, $\vert \delta m_T^2 \vert$ is always
smaller than $H^2$ and the thermal mass never causes the field to oscillate.
Instead, the solutions presented in \cref{app:after-annihilation} are well approximated by
\begin{align}
    \bar{\varphi}(a)
    &\simeq \bar{\varphi}_i \left( 1 + 2 a / 3 a_\text{eq} \right)^{-2 \mathcal{D} / 3}
    \label{eqn:apx-soln-matter-effects-late}
\end{align}
(determined empirically; see \cref{app:after-annihilation} for genuine but more cumbersome analytic
solutions).
Moreover, in this regime the thermal mass drops below the bare mass in magnitude before
$t = t_\osc = m_\varphi^{-1}$, so the field only ever oscillates due to its bare mass alone.

\subsubsection{Evolution of energy density and fundamental constants}

We conclude our treatment of the theoretical description of coupled scalar models by delineating
useful parametrizations of their phenomenology---both their impact on fundamental constants and
gravitation.
The energy density in the homogeneous field $\bar{\varphi}$ with effective mass $\meff$ is
$\bar{\rho}_\varphi = \Mpl^2 \dot{\bar{\varphi}}^2 + \meff^2 \Mpl^2 \bar{\varphi}^2$.
We quantify the scalar's present-day energy density relative to that in cold dark matter via
\begin{align}
    \fphi
    &\equiv \frac{\bar{\rho}_{\varphi, 0}}{\bar{\rho}_{c,0}},
\end{align}
with
$\bar{\rho}_{c, 0} \equiv 3 H_{100}^2 \Mpl^2 \omega_c \approx (1.76~\meV)^4$~\cite{Planck:2018vyg}.
When matter effects are negligible, the scalar and CDM redshift in tandem once oscillations begin
and
\begin{align}
    \fphi
    = \frac{\bar{\rho}_{\varphi}(a)}{\bar{\rho}_{c}(a)}
    &= \frac{m_\varphi^2 \Mpl^2 \bar{\varphi}_i^2 / (a / a_\osc)^3}{\bar{\rho}_c(a_0) / a^3}
    = \frac{3 \left( 1 + \omega_b / \omega_c \right) \bar{\varphi}_i^2}{4}
    ,
    \label{eqn:Xi-ito-varphi}
\end{align}
which notably is independent of $m_\varphi$.
More generally, the late-time evolution of the scalar is well described by the WKB approximation in
\cref{eq:WKB} unless the field is extremely light ($m_\varphi < H_0$) or has a negative thermal mass
squared that is larger in magnitude than $m_\varphi$.
The scalar's energy density is then $\bar{\rho}_{\varphi} \approx 2 \meff^2 \Mpl^2 \varphiamp^2$ and
evolves according to
\begin{align}
    \frac{\bar{\rho}_\varphi(a)}{\bar{\rho}_c(a)}
    &= \fphi \frac{\meff(a)}{\meff(a_0)}
\end{align}
in the WKB regime.
Extracting the amplitude from the energy density,
\begin{align}
    \varphiamp(a)
    \approx \sqrt{ \frac{\rho_{\varphi}(a)}{\meff(a)^2 \Mpl^2} }
    = 1.28 \times 10^{-3} \left( \frac{\meff(a)}{10^{-30}~\eV} \right)^{-1}
        \left(\frac{\rho_\varphi(a)}{\bar{\rho}_{c,0}}\right)^{1/2}.
    \label{eqn:varphi-ito-rho-phi}
\end{align}
At the present, the scalar's dynamics may therefore be specified in terms of its frequency
$\meff(a_0)$, its abundance relative to cold dark matter $\fphi$, and a phase $\vartheta_0$.
For quadratic couplings [\cref{eq:quadratic_couplings}], the shift in the cosmological value of
fundamental constants with time is
\begin{align}
    \frac{\lambda(t) - \lambda_0}{\lambda_0}
    &\approx \frac{
            \left( d_\lambda^{(2)}/2 \right)
            \left[
                \varphiamp(t)^2 \sin^2 \left( \int_{t_0}^t \meff(t') \ud t' + \vartheta_0 \right)
                - \varphiamp_0^2 \sin^2(\vartheta_0)
            \right]
        }{
            1 + d_\lambda^{(2)} \varphiamp_0^2 \sin^2(\vartheta_0) / 2
        },
\end{align}
where $\lambda_0$ is the cosmological value of the fundamental constant today.

How present-day conditions map into the initial condition $\bar{\varphi}_i$ (and therefore the
early-time fundamental constants $\lambda_i$) depends on whether $\vert \mathcal{D} \vert$ is large
or small compared to unity.
In the latter case, \cref{eqn:Xi-ito-varphi} holds and
\begin{align}
    \bar{\varphi}_i
    &\approx \sqrt{\frac{4}{3}\frac{\fphi}{1+\omega_b/\omega_c}}
    \label{eqn:varphi-i-ito-Xi}
    \\
    \frac{\lambda_i-\lambda_0}{\lambda_0}
    &\approx \frac{d_\lambda^{(2)} \bar{\varphi}_i^2}{2}
    \approx \frac{2}{3} \frac{d_\lambda^{(2)} \fphi}{1 + \omega_b / \omega_c},
    \label{eqn:dlambda-i-ito-Xi}
\end{align}
taking $\bar{\varphi}_i \gg \varphiamp_0$, i.e., $a_\osc \ll 1$.
The small corrections due to matter couplings with $\vert \mathcal{D} \vert \lesssim 1$ may be
accounted for via \cref{eq:WKB,eqn:apx-soln-matter-effects-late}.
Perhaps surprisingly, a scalar with Planck-suppressed couplings (i.e., $d_\lambda^{(2)} \sim 1$) can
shift the early-time fundamental constants substantially---that is, to the same extent as the scalar
contributes to the total dark matter density.

On the other hand, when $\vert \mathcal{D} \vert \gtrsim 1$, the thermal mass induces substantial
evolution before $t_\osc$, as described in \cref{sec:matter_corrections_to_evolution}.
In particular, $\bar{\rho}_\varphi(a) / \bar{\rho}_c(a) \propto a^{-3/2}$ until at least
$a_\mathrm{eq}$, the scale factor of matter-radiation equality.
Thus, the abundance of the scalar relative to cold dark matter can be much larger than $\fphi$ at
earlier times.
Requiring $\bar{\rho}_\varphi(a) < \bar{\rho}_c(a)$ yields
\begin{align}
    (Q_b)_{\lambda} d_{\lambda}^{(2)}
    &\leq
        \frac{m_\varphi^2}{3 \omega_b H_{100}^2 / 2}
        \begin{dcases}
            \frac{\fphi^2 + 1}{\fphi^2 / a^3 - 1}
            & \fphi > a^{3/2}
            \\
            \infty
            & \fphi < a^{3/2}
        \end{dcases}
    .
    \label{eqn:rho-phi-below-rho-c}
\end{align}
No bound applies when the scalar cannot reach $\fphi(a) = 1$ even if its density redshifts like
$a^{-3/2}$ (due to its in-medium mass) from $a$ all the way until the present.
Evaluated at matter-radiation equality and for $a_\mathrm{eq}^3 \ll \fphi^2 \ll 1$,
\cref{eqn:rho-phi-below-rho-c} restricts
\begin{align}
    (Q_b)_{\lambda} d_{\lambda}^{(2)}
    &\lesssim \max\left[
        \left( \frac{m_\varphi / \fphi}{7.8 \times 10^{-29}~\eV} \right)^2,
        \frac{2}{3} \left( 1 + \omega_c / \omega_b \right)
        \right]
    ,
    \label{eqn:coupling-bound-fphi-equality}
\end{align}
including the truncation of the bound when $\mathcal{D}$ drops below order unity (i.e., when the
scalar only ever redshifts under the influence of its bare mass).

The impact of in-medium masses on dynamics affects the interpretation of a certain density in $\varphi$ today in terms of its initial conditions and energy density at earlier times.
Conversely, a limit (i.e., from CMB data) on the abundance of an interacting scalar around matter-radiation equality translates to a more stringent limit on its present-day abundance relative to CDM for larger couplings. The hyperlight scalar parameter space, including the constraint from \cref{eqn:coupling-bound-fphi-equality}, is displayed in
\cref{fig:quadratic-coupling-non-cmb-constraints-Xi-1e-2}.
\Cref{fig:quadratic-coupling-non-cmb-constraints-Xi-1e-2} also includes a conservative limit that
requires perturbativity of the coupling expansion at the present day, i.e.,
$g_\lambda(\varphi_0) < 1$ with $\varphi_0$ estimated by \cref{eqn:varphi-ito-rho-phi}.

\section{Late-time probes}
\label{sec:late-probes}

\begin{figure}[t!]
\begin{centering}
    \includegraphics[width=\textwidth]{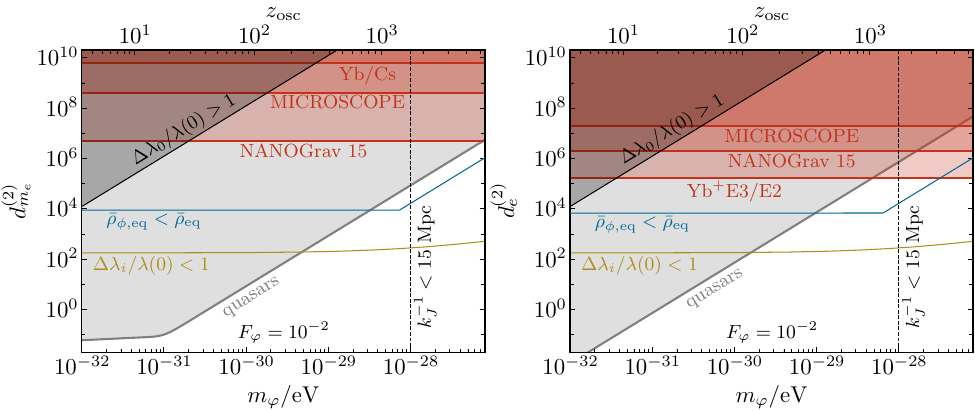}
    \caption{
        Constraints on a hyperlight scalar's quadratic couplings to the electron (left) and photon
        (right) from late-time astrophysical (gray) and laboratory (red) measurements.
        Bounds assume the scalar comprises a present-day cosmological energy density
        $\fphi \equiv \bar{\rho}_{\varphi, 0} / \bar{\rho}_{c, 0} = 10^{-2}$ relative to that of cold
        dark matter.
        Also shown are theoretical consistency curves delineating the minimum couplings for which the perturbative expansion would break down at the present time (black) and where the scalar energy density extrapolated from the present $\fphi$ back to matter-radiation equality would dominate the CDM [\cref{eqn:coupling-bound-fphi-equality}, blue].
        Finally, the gold line displays the coupling values for which the scalar coupling would
        violate perturbativity at its initial misalignment (which is inversely proportional to
        $\fphi$).
        The limits from atomic clocks, quasars, and perturbativity of the coupling expansion scale inversely with $\fphi$, while
        bounds from tests of the universality of free fall scale with $\fphi^{-1/2}$.
        The matter-radiation equality constraint is proportional to $\fphi^{-2}$ above the break in scaling.
        The dependence of these bounds on $\fphi$ is further detailed in \cref{sec:late-probes}.
        For simplicity we only report bounds on positive couplings; bounds from quasars on negative couplings differ by an order-unity factor.
        The vertical dashed lines indicate the heaviest mass for which the scalar's present-day
        Jeans length is larger than the size of the local supercluster; at higher masses, the scalar
        would cluster with the rest of the dark matter to some degree, making the reported bounds conservative in this regime.
        \Cref{sec:early-probes} shows that early-time data from the CMB and BBN supersede these late-time constraints for all but the lightest masses.
    }
    \label{fig:quadratic-coupling-non-cmb-constraints-Xi-1e-2}
\end{centering}
\end{figure}

In this section, we assess the constraining power of numerous late-time ($z \lesssim 10$)
probes of variations in fundamental constants as applied to hyperlight dark matter. We discuss the
impact of a hyperlight scalar's slow oscillations in \cref{sec:phase-marginalization} and the
spatial profile of a scalar sourced by a central body in
\cref{sec:spatial-variations}.
We then evaluate the constraining power of quasar absorption spectra (\cref{sec:quasars}), tests of
the universality of free fall (\cref{sec:UFF}), the Oklo phenomenon (\cref{sec:oklo}), atomic clocks
and pulsar timing arrays (\cref{sec:clocks}), and stellar emission (\cref{sec:stars}).
\Cref{fig:quadratic-coupling-non-cmb-constraints-Xi-1e-2} summarizes of the results of this section.

\subsection{Coherent oscillations}\label{sec:phase-marginalization}

Local signatures of scalar dark matter depend on its amplitude, and therefore its energy density, near Earth.
The density in \emph{ultralight} scalars is often measured relative to the energy density of
virialized galactic dark matter at the Earth,
$\rho_{c, \oplus} \approx 0.4~\GeV/\cm^3 \approx (41.9~\meV)^4$~\cite{McMillan:2011wd}, which is
about $\sim 10^6$ times the cosmological background density.
A hyperlight scalar, however, is inherently cosmological: it does not cluster on galactic scales and
its local amplitude is specified strictly as a fraction of the average cosmological density.
Specifically, density perturbations of a scalar do not grow below the Jeans length
$1/k_J \gtrsim 15~\mathrm{Mpc} \sqrt{\meff / 10^{-28}~\eV}$~\cite{Marsh:2015xka}.

Interpolating between the regimes in which the scalar does and does not track the local dark matter
density would require a more quantitative understanding of the dynamics of hyperlight subcomponents
of dark matter.
Here, we restrict our results to the mass range $m_{\varphi} \ll 10^{-28}~\eV$ and furthermore to couplings small enough that $\vert \delta m_T^2 \vert \lesssim m_\varphi^2$ holds before $z = 10$.
The scalar's Jeans length at and after the time of galaxy formation is then larger than the local
supercluster; this happens to be the same mass range relevant to the cosmological analysis of
Ref.~\cite{Baryakhtar:2024rky}.
In this regime, the field is entirely specified by the dimensionless amplitude of the cosmologically
homogeneous field, \cref{eqn:varphi-ito-rho-phi}, which we rewrite as
\begin{align}
    \varphiamp
    = 0.720 \left( \frac{\meff}{10^{-30}~\eV} \right)^{-1}
        \left( \frac{\rho_\varphi}{\rho_{c, \oplus}} \right)^{1/2},
    \label{eqn:varphiamp-for-conversion}
\end{align}
which, for ease of comparison with probes at heavier masses, we've normalized to the local CDM
energy density.
The scalar oscillates over a period
$T_\varphi = 2 \pi / m_\varphi \approx 1.3~\Myr \left( 10^{-28}~\eV / m_\varphi \right)$, which is much
longer than the duration of any experiment.

Most experimental measurements also take place over timescales very short compared to the coherence
time of ultralight, galactic dark matter candidates.
The local amplitude of virialized bosonic dark matter is then effectively stochastic---i.e., there
is some probability that the Earth resides in a region that is underdense compared to the
expectation for the Milky Way~\cite{McMillan:2011wd}.
Current bounds on ultralight scalar dark matter account for the distribution of possible scalar
amplitudes by reducing the expectation from \cref{eqn:varphi-ito-rho-phi} by a factor
$f_\mathrm{stoch} \approx 3$~\cite{Centers:2019dyn}.
Because hyperlight scalars do not cluster on galactic scales, their local amplitude is instead set
deterministically by \cref{eqn:varphiamp-for-conversion}.

Consider a measurement taken over a period of time $T$ at some cosmological time $t$.
When $t \gg T_\varphi \gg T$, it is intractable---if not impossible---to know where the scalar is in
its oscillation at $t$, due to the limited precision with which we know both cosmological parameters
and the scalar's dynamics.
For example, the scalar's present-day phase scales as $m_\varphi t \propto 2 m_\varphi / 3 h H_{100}$, with
$t$ the age of the Universe and $H_{100} \approx 2.133 \times 10^{-33}~\eV$.
Thus, even fixing all other cosmological parameters, percent-level variations in $h$ result in
$\sim 2 \pi (m_\varphi / 10^{-30}~\eV)$ uncertainties in the scalar's phase.
Since observables depend on, e.g.,
$d_\lambda^{(n)} \varphi(t)^n \approx d_\lambda^{(n)} \varphiamp(t)^n \sin(m_\varphi t)^n$,
direct constraints on the combination $d_\lambda^{(n)} \varphiamp(t)^n$ are intrinsically limited
by the unknown phase---namely, one cannot exclude the possibility that the signal vanishes during
the experiment due to the scalar being presently near an antinode.

To concretely reason about the effect of phase marginalization, we ground our analysis with a
likelihood.
To capture the variety of signals we consider, we begin with a general parametrization of a
Gaussian likelihood with measurement mean $\hat{x}$ and variance $\sigma_x$.
We decompose the model prediction at parameters $\bm{\theta}$ into amplitude and phase factors
$A(t; \bm{\theta}) P(t; \bm{\theta})$, where $P$ is a periodic function bounded by
unity in magnitude and $A$ encodes the signal magnitude if $P(t; \bm{\theta})$ were at an antinode.
In the relevant limit where the phase $\vartheta$ of $P$ is effectively unknown and treated as a
uniform variate, the phase factor simply reduces to $P(t; \bm{\theta}) \to P(\vartheta)$.
The phase-marginalized posterior is then
\begin{align}
    p\left(
        \bm{\theta}
        \middle\vert \hat{x}, \sigma_x
    \right)
    &= \pi(\bm{\theta})
        \int_{0}^{2 \pi} \frac{\ud \vartheta}{2 \pi} \,
            \frac{1}{\sqrt{2 \pi \sigma_x^2}}
            \exp \left[
                - \frac{1}{2} \left(
                \frac{
                    A(t; \bm{\theta}) P(\vartheta)
                    - \hat{x}
                }{
                    \sigma_x
                }
            \right)^2
        \right]
    \label{eqn:toy-posterior-for-measurement}
    ,
\end{align}
with $\pi(\bm{\theta})$ the prior over all other parameters.
The $95\%$ highest density interval, used to place upper limits, depends on the behavior of the
phase-marginalized posterior in its tails, i.e., where
$\vert A(t; \bm{\theta}) \vert \gg \vert \hat{x} \pm \sigma_x \vert$.
In this regime, the integrand of the marginal posterior is dominated where
$\vert P(\vartheta) \vert \ll 1$.
The only important property of $P(\vartheta)$ is therefore how quickly it approaches zero at its
steepest node.
If $P(\vartheta) \propto \vartheta^m$ (taking the relevant node to be at $\vartheta = 0$ for
simplicity), the integral's support is then dominated where
$\vert \vartheta \vert^m < \vert \hat{x} \pm \sigma_x \vert / A(t; \bm{\theta})$.
Approximating the Gaussian as constant over this interval reveals that the marginal posterior
scales with $1 / \sqrt[m]{A(t; \bm{\theta})}$ when $A(t; \bm{\theta})$ is large, a finding
reproduced by numerical quadrature.
Given that $A(t; \bm{\theta})$ is often linear in an unknown coupling $d_\lambda^{(n)}$, the
marginal posterior over the coupling is improper (since the integral over $d_\lambda^{(n)}$ does
not converge at infinity).

Meaningful posterior constraints thus require regularization by some means.
If one simply takes $P(\vartheta) > P_\mathrm{min}$ \textit{a priori} (i.e., excluding the fraction
of the scalar's period where its oscillation amplitude is below some threshold), then the integrand
of \cref{eqn:toy-posterior-for-measurement} falls off exponentially when
$\vert A(t; \bm{\theta}) \vert \gtrsim \vert \hat{x} \pm \sigma_x \vert / P_\mathrm{min}$.
Such a prior, however, is entirely arbitrary without concrete justification.
On the other hand, experimental measurements with duration $T$ are guaranteed to observe the scalar
when it is at least $T / 2 T_\varphi$ away from a node in phase for some portion of the experiment's
duration, effectively imposing a prior that $P(\vartheta) \gtrsim (T / 2 T_\varphi)^{m}$ for some
interval.
Therefore, marginalizing over the unknown phase of a scalar with period $T_\varphi \gg T$ penalizes
bounds by a factor $\propto m_\varphi^{-m}$.
Another simple solution is to combine more than $m$ independent measurements with spacing in time
greater than $T_\varphi$, each marginalized over their own independent phase; the posterior's tails
then fall off faster than $1 / d_\lambda^{(n)}$, making it proper.
Given that in our parameter space $T_\varphi$ is much longer than the history of human civilization,
this option is only practical for distant astrophysical signals such as quasar absorption
(\cref{sec:quasars}).

Finally, we reiterate that phase marginalization does not so severely penalize the actual discovery
potential of a given search: while a null signal can be consistent with a large coupling if the
scalar were near a node, a measurement of a nonzero signal cannot be explained by an arbitrarily
small coupling, no matter the scalar's phase.
The above arguments concern measures of the tail of posterior distributions and still apply in that
regime, but the posterior would still localize about finite couplings if the model could match the
detected signal.
When observing only a small fraction of an oscillation, however, it may be challenging to
disentangle the phase factor $P(\vartheta)$ from the actual amplitude $A(t; \bm{\theta})$ and
measure the latter at better than the order-unity level.

\subsection{Spatial variations about a central body}\label{sec:spatial-variations}

Just as a cosmological abundance of matter sources an effective potential for the scalar
(\cref{sec:matter-potentials}), localized bulk matter distributions---such as the Earth, or
any central mass---source a potential that affects the scalar's spatial distribution.
The field profile generated by a central mass depends on the specific form of the $\varphi$-matter
interaction and is obtained by solving the classical equation of motion for $\varphi$
[\cref{eq:scalar-eom}] with the effective potential $V_\text{matter}(\varphi)$ generated by the localized matter distribution.
A central mass thus generally introduces spatial variations in fundamental constants which, among
other phenomenology, leads to an apparent ``fifth force'' towards the central mass (\cref{sec:UFF}).

For couplings that are linear in $\varphi$, the space-dependent profile sourced by a localized matter volume simply adds to
any ambient field constituting dark matter.
For a spherical, time-independent, uniform distribution of nonrelativistic matter~\cite{Adelberger:2003zx}:
\begin{align}
    \varphi(t,\three{x})
    = \bar{\varphi}(t) - d_{M_C}^{(1)}\frac{GM_C}{\vert \three{x} \vert}
    e^{-m_\varphi \vert \three{x} \vert},
    \label{eq:field_from_sphere_linear}
\end{align}
for a central body $C$ with total mass $M_C$ and radius $R_C$, for $\vert \three{x} \vert>R_C$ and in the limit $m_\varphi R_C \ll 1$.
Here the effective dilatonic coupling of $C$ is defined as
$d_{M_C}^{(n)} = \sum_\lambda \left( Q_{M_C} \right)_\lambda d_{\lambda}^{(n)}$, where $\lambda$ is
summed over SM parameters.
The boundary condition at $\vert \three{x} \vert\rightarrow \infty$ corresponds to the unvirialized cosmological abundance of $\varphi$, as is appropriate for a hyperlight scalar.
Notably, the sourced contribution to \cref{eq:field_from_sphere_linear} is independent of the
background field, yielding signatures regardless of whether the scalar makes up any dark matter at all;
searches for equivalence principle violation stringently constrain linear couplings for this reason.

For the quadratic couplings we focus on, however, the matter distribution scatters the incident
ambient dark matter; experiments then probe the combination of the incident and scattered profiles.
For $\vert \three{x} \vert > R_C$~\cite{Hees:2018fpg},
\begin{align}
    \varphi(t,\three{x})
    &\approx \bar{\varphi}(t)
        \left[
            1 - d_{M_C}^{(2)}
            \frac{GM_C}{\vert \three{x} \vert}
            s_{C}^{(2)}\bigg( d_{M_C}^{(2)} \frac{G M_C}{R_C}\bigg)
        \right],
    \label{eq:field_from_sphere_quadratic}
\end{align}
which satisfies the same boundary condition as \cref{eq:field_from_sphere_linear}.
The screening factor $s^{(2)}_{C}(y)$ is a function of the quadratic coupling times the
compactness (or surface gravity) $GM_C/R_C$ of the source mass, which can have an important effect
for, e.g., terrestrial probes of variations in fundamental constants.
Because $s^{(2)}_{C}(y \gg 1) \approx y^{-1}$, strongly coupled compact objects yield a field profile
\begin{align}
\varphi(t,\three{x}) \approx \bar{\varphi}(t) \left[1-\frac{R_C}{\vert \three{x} \vert}\right]
\end{align}
for $\vert \three{x} \vert \geq R_C$ and assuming $d_{M_C}^{(2)} > 0$.
In this limit, the amplitude of temporal variations in fundamental constants is strongly suppressed near the surface of the central mass and independent of the coupling strength of the central mass at any $\vert \three{x} \vert$, while the size of spatial gradients $\bm{\nabla} \varphi$ is also independent of the coupling.
However, if the matter couplings are small compared to the inverse compactness,
\begin{align}
    d^{(2)}_{M_C}
    &\ll R_C/G_NM_C,
    \label{eq:screening_condition}
\end{align}
then the relevant limit is $s_{C}(y \rightarrow 0) \approx 1$.

Recall from \cref{eq:dilatonic_relations} that the couplings of composite bulk matter to $\varphi$ are
suppressed relative to the fundamental Lagrangian couplings by the dilatonic charges
\cref{eqn:dilatonic_charges}, reflecting that only a fraction of the total energy couples to $\varphi$.
The dilatonic charge of an atomic mixture is the average of the dilatonic charges of the atomic
constituents, weighted by mass.
Earth is approximately $32\%$ iron and $67\%$ silicate $\text{SiO}_2$ by
mass~\cite{Su:1994gu,Damour:1995gi,morgan1980chemical}.
Using \cref{eqn:dilatonic_charges} and accounting for isotopic abundances~\cite{Hees:2018fpg}, one has
$(Q_{m_\text{Fe}})_{m_e} \approx 2.5 \times 10^{-4}$ and
$(Q_{m_{\text{SiO}_2}})_{m_e} \approx 2.7 \times 10^{-4}$, as well as
$(Q_{m_\text{Fe}})_{e} \approx 2.6 \times 10^{-3}$ and
$(Q_{m_{\text{SiO}_2}})_{e} \approx 1.6 \times 10^{-3}$.
All in all, $(Q_{M_\oplus})_{m_e} \approx 2.6 \times 10^{-4}$ and
$(Q_{M_\oplus})_{e} \approx 1.9 \times 10^{-3}$.
For the Earth, the screening condition \cref{eq:screening_condition} translates to
$d^{(2)}_{M_\oplus} \ll 10^{9}$; because the dilatonic charges of Earth are small, screening is
only relevant at fundamental couplings much larger than those considered here.
In conclusion, the scalar amplitude and time dependence are well approximated by the background
density in measurements in the vicinity of the Earth, as relevant for terrestrial probes
(\cref{sec:clocks,sec:oklo}), while the spatial gradients remain proportional to
$d_{M_\oplus}^{(2)}$.

Finally, very light scalars can be thermally produced in stars and compact objects (\cref{sec:stars}).
Possible processes includes pair production of scalars from Standard Model particles, or single scalar production in the presence of the cosmological background field. In the latter case, central-body effects can suppress (or enhance) the background field inside the star. We discuss this further in \cref{sec:stars}.

\subsection{Quasar absorption lines}\label{sec:quasars}

The precision measurement of quasar absorption spectra has played a key role in the search for
variations of fundamental constants due to their cosmologically nontrivial
distances~\cite{Ubachs:2015fro, Uzan:2024ded, savedoff1956physical, petitjean2009constraining,
srianand2009probing, Kanekar:2009kaa}.
Comparisons of atomic transition lines allow for a search for variations of $\alpha$ and the
rotational and vibrational transitions of molecules for variations of the proton-to-electron mass
ratio $\mu \equiv m_p/m_e$ by comparing the line frequencies to those measured today in the
laboratory.
The redshift $z_\mathrm{abs}$ at which distant quasar light passes through the absorptive medium of
interest (like a molecular cloud) is obtained by measuring multiple lines or even a broad spectrum,
which enables distinguishing underlying differences in fundamental constants from the determination
of the redshift and additional systematics such as Doppler shifts and instrumental effects~\cite{Uzan:2024ded}.

\begin{table}[t!]
\begin{center}
    \caption{
        Quasars absorption measurements of variations in the proton-to-electron mass ratio $\mu$
        (left) and the fine structure constant $\alpha$ (right), tabulated from references listed in
        each table.
        The reported uncertainties add systematic and statistical errors in quadrature where needed.
    }
\begin{tabular}[t]{l r r r r r}
    \toprule
    Quasar & $z_\mathrm{abs}$ & $10^6 \Delta \mu(z_\mathrm{abs}) / \mu$ \\
    \midrule
    HE0027-1836~\cite{Ubachs:2015fro} & $2.4$ & $-7.6 \pm 10.2$ \\
    Q0347-383~\cite{Ubachs:2015fro} & $3.02$ & $5.1 \pm 4.5$ \\
    Q0405-443~\cite{Ubachs:2015fro} & $2.59$ & $7.5 \pm 5.3$ \\
    Q0528-250~\cite{Ubachs:2015fro} & $2.81$ & $-0.5 \pm 2.7$ \\
    B0642-5038~\cite{Ubachs:2015fro} & $2.66$ & $10.3 \pm 4.6$ \\
    Q1232+082~\cite{Ubachs:2015fro} & $2.34$ & $140 \pm 60$ \\
    J1237+064~\cite{Ubachs:2015fro} & $2.69$ & $-5.4 \pm 7.2$ \\
    J1443+2724~\cite{Ubachs:2015fro} & $4.22$ & $-9.5 \pm 7.5$ \\
    J2123-005~\cite{Ubachs:2015fro} & $2.05$ & $7.6 \pm 3.5$ \\
    Q2348-011~\cite{Ubachs:2015fro} & $2.43$ & $-6.8 \pm 27.8$ \\
    B0218+357~\cite{Murphy:2008yy} & $0.685$ & $0 \pm 0.9$ \\
    PKS1830-211~\cite{Bagdonaite:2013sia} & $0.886$ & $-0.1 \pm 0.13$ \\
    PMNJ0134-0931~\cite{Kanekar:2012fy} & $0.765$ & $-3.31 \pm 2.74$ \\
    PKS1413+135~\cite{Kanekar:2018mxs} & $0.247$ & $-1 \pm 1.3$ \\
    \bottomrule
\end{tabular}
\quad
\begin{tabular}[t]{l r r r r r}
    \toprule
    Quasar & $z_\mathrm{abs}$ & $10^6 \Delta \alpha(z_\mathrm{abs}) / \alpha$ \\
    \midrule
    J0058$+$0041~\cite{Murphy:2016yqp} & $1.072$ & $-1.35 \pm 7.16$ \\
    PHL957~\cite{Murphy:2016yqp} & $2.309$ & $-0.2 \pm 12.9$ \\
    J0108$-$0037~\cite{Murphy:2016yqp} & $1.371$ & $-8.45 \pm 7.34$ \\
    J0226$-$2857~\cite{Murphy:2016yqp} & $1.023$ & $3.54 \pm 8.87$ \\
    J0841$+$0312~\cite{Murphy:2016yqp} & $1.342$ & $5.67 \pm 4.71$ \\
    J1029$+$1039~\cite{Murphy:2016yqp} & $1.622$ & $-1.7 \pm 10.1$ \\
    J1237$+$0106~\cite{Murphy:2016yqp} & $1.305$ & $-4.54 \pm 8.66$ \\
    Q1755$+$57~\cite{Murphy:2016yqp} & $1.971$ & $4.72 \pm 4.71$ \\
    Q2206$-$1958~\cite{Murphy:2016yqp} & $1.921$ & $-4.65 \pm 6.41$ \\
    J0120$+$2133~\cite{Murphy:2017xaz} & $0.576$ & $-9.12 \pm 40$ \\
    J0120$+$2133~\cite{Murphy:2017xaz} & $0.729$ & $0.73 \pm 6.42$ \\
    J0120$+$2133~\cite{Murphy:2017xaz} & $1.048$ & $5.47 \pm 18.7$ \\
    J0120$+$2133~\cite{Murphy:2017xaz} & $1.325$ & $2.6 \pm 4.19$ \\
    J0120$+$2133~\cite{Murphy:2017xaz} & $1.343$ & $8.36 \pm 12.2$ \\
    J1944$+$7705~\cite{Murphy:2017xaz} & $1.738$ & $12.7 \pm 16.2$ \\
    HS1549$+$1919~\cite{Evans:2014yva} & $1.143$ & $-7.49 \pm 5.53$ \\
    HS1549$+$1919~\cite{Evans:2014yva} & $1.342$ & $-0.7 \pm 6.61$ \\
    HS1549$+$1919~\cite{Evans:2014yva} & $1.802$ & $-6.42 \pm 7.25$ \\
    J1120$+$0641~\cite{Wilczynska:2020rxx} & $7.05852$ & $128 \pm 684$ \\
    J1120$+$0641~\cite{Wilczynska:2020rxx} & $6.17097$ & $-102 \pm 152$ \\
    J1120$+$0641~\cite{Wilczynska:2020rxx} & $5.95074$ & $-228 \pm 174$ \\
    J1120$+$0641~\cite{Wilczynska:2020rxx} & $5.50726$ & $74.2 \pm 111$ \\
    \bottomrule
\end{tabular}
    \label{tab:quasars}
\end{center}
\end{table}

In \cref{tab:quasars} we gather measurements of $\mu(z_\mathrm{abs}) / \mu_0 - 1$ and
$\alpha(z_\mathrm{abs}) / \alpha_0 - 1$ from an ensemble of quasar results that mitigate
the systematic distortions of wavelength calibration~\cite{Whitmore:2014ina}.
As described in \cref{sec:phase-marginalization}, the uncertainty in the scalar's oscillation phase
at $z_\mathrm{abs}$ and today should be taken into account.
A complication beyond the discussion of \cref{sec:phase-marginalization} is that the
comparison of values of $\lambda$ means each quasar measurement is also sensitive to the
scalar's unknown \emph{present-day} phase:
\begin{align}
    \frac{\lambda[\varphi(t)]}{\lambda[\varphi(t_0)]} - 1
    \approx g_\lambda[\varphi(t)] - g_\lambda[\varphi(t_0)]
    &= \frac{d_{\lambda}^{(n)}}{n!}
        \left[
            \varphiamp(t)^n \sin^n \vartheta(t)
            - \varphiamp(t_0)^n \sin^n \vartheta(t_0)
        \right].
    \label{eqn:constant-shift-ito-amplitude-phase}
\end{align}
Given that we consider quadratic couplings and that most of the absorbers in \cref{tab:quasars}
reside at redshifts $z_\mathrm{abs} > 1$, usually
$\varphiamp(t_\mathrm{abs})^n > 10 \varphiamp(t_0)^n$, suppressing the impact of marginalization
over $\vartheta(t_0)$.
Moreover, combining an ensemble of measurements with independent phases makes the posterior steeper
about its peak and less sensitive to such effects deeper in the tails.
Ignoring the present-day contribution greatly simplifies the analysis, as one may marginalize
the phase for $N$ measurements independently---otherwise, marginalization requires an
$N + 1$ dimensional integral that does not factorize.

\begin{figure}[t!]
\begin{centering}
    \includegraphics[width=\textwidth]{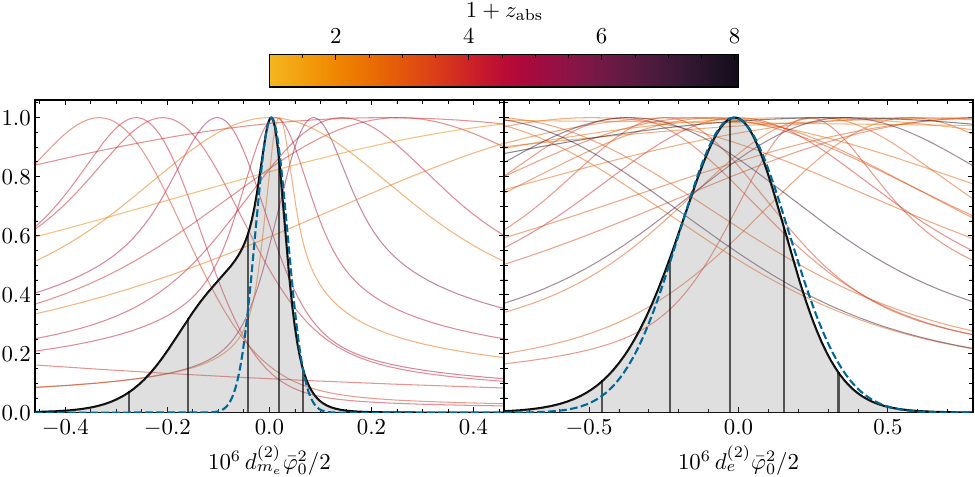}
    \caption{
        Individual and combined posteriors from various observations of quasar absorption lines
        for a scalar coupled quadratically to the electron (left) or photon (right).
        Posteriors are plotted relative to their maximum value and against the present-day shift in
        a parameter relative to its vacuum value (evaluated at the antinode),
        $\Delta \lambda_0 / \lambda(0) = d_{\lambda}^{(2)} \varphiamp_0^2 / 2$; the shift at the
        absorption redshift $z_\mathrm{abs}$ is given by
        $(1 + z_\mathrm{abs})^3 \Delta \lambda_0 \sin^2 \theta(z_\mathrm{abs})$.
        The transparent curves depict posteriors from individual observations (despite the fact that
        they are improper), colored by $1 + z_\mathrm{abs}$ per the colorbar.
        The black curves depict the joint posterior for the quasar observations in \cref{tab:quasars}, independently
        marginalized over each unknown phase $\theta(z_\mathrm{abs})$.
        Vertical lines denote the $\pm 1 \sigma$ and $\pm 2 \sigma$ quantiles and the median
        for the joint posterior.
        The dashed blue curves depict the would-be joint posteriors were one to instead assume
        $\sin^2 \theta(z_\mathrm{abs}) = 1/2$ at each $z_\mathrm{abs}$.
    }
    \label{fig:marginalized-quasar-posteriors-quadratic}
\end{centering}
\end{figure}

For a collection of observations at multiple redshifts, their spacing in time is also important,
since the scalar's phases for two nearly coincident (relative to $T_\varphi$) observations are
deterministically related.
To accurately account for these correlations, and to provide realistic bounds for scalars so light
that their phases at the relevant times are not effectively stochastic, an analysis must fully model
the scalar's dynamics and marginalize over the background cosmology.
Here we perform a simpler analysis where we treat the scalar's phase at each $z_\mathrm{abs}$ in the
sample as independent and stochastic and perform a few cross checks on subsets of these samples.
\Cref{fig:marginalized-quasar-posteriors-quadratic} demonstrates the impact of phase marginalization
on combined quasar constraints.
We take each measurement to specify a simple Gaussian posterior over $\Delta \lambda_0 / \lambda_0$
and independently marginalize over the phase of the scalar's oscillation for each measurement [as in
\cref{eqn:toy-posterior-for-measurement}].
\Cref{fig:marginalized-quasar-posteriors-quadratic} assumes quadratic couplings to the electron
or photon, insofar as the measured bounds on $\Delta \lambda(z_\mathrm{abs}) / \lambda_0$
are converted to present-day shifts via
$\Delta \lambda(z_\mathrm{abs}) = (1 + z_\mathrm{abs})^3 \Delta \lambda_0$
and the phase factor is taken to be $\sin^2 \theta(z_\mathrm{abs})$.

\Cref{fig:marginalized-quasar-posteriors-quadratic} includes curves that combine all measurements by
either marginalizing over each phase $\theta(z_\mathrm{abs})$ independently or by fixing
$\sin^2 \theta(z_\mathrm{abs}) = 1/2$ for each absorption redshift $z_\mathrm{abs}$.
For the varying-$\alpha$ results, the two results are quite similar near the peak, reflecting that
the larger collection of $\Delta \alpha$ constraints are effectively randomly scattered about zero.
(As explained in \cref{sec:phase-marginalization}, the tails of the distribution do fall off as
a power law in the phase-marginalized case rather than exponentially.)
The difference is much more striking for constraints on the electron mass: while the
phase-marginalized result resembles the phase-fixed one near the peaks of the distributions, it
exhibits greatly enhanced support at $\Delta m_{e, 0} / m_{e, 0} \sim - 10^{-7}$.
This skew arises due to a cluster of higher-redshift observations that themselves skew to negative
$\Delta m_{e, 0} / m_{e, 0}$, which are evident in
\cref{fig:marginalized-quasar-posteriors-quadratic}.
When marginalizing over each phase, other measurements centered near zero or positive values of
$\Delta m_{e, 0} / m_{e, 0}$ each only contribute a square-root suppression at negative
$\Delta m_{e, 0}$ as opposed to an exponential one (when the phases are instead fixed).

\Cref{fig:marginalized-quasar-posteriors-quadratic} displays $1 \sigma$ intervals of order
$10^{-7}$ for $d_\lambda^{(2)} \varphiamp_0^2 / 2$.
As a cross check, we compare these results to those with subsets of the quasar samples partitioned
by redshift.
For both couplings, selecting only quasars with $z_\mathrm{abs} > 1.5$ hardly affects the width of the
posteriors.
However, for this subsample the electron coupling constraints shift further toward negative values
(the median lying $1.4 \sigma$ rather than $0.5 \sigma$ below zero); these observations source the
skew evident in \cref{fig:marginalized-quasar-posteriors-quadratic}.
(All other tested subsample exhibit an expected degree of scatter in central values.)
Only four of the 14 quasar measurements of $\Delta \mu / \mu$ reside at $z_\mathrm{abs} < 1.5$, and
the bound deriving from them alone degrades by a factor of $\sim 10$.
Twelve of the 22 $\Delta \alpha / \alpha$ measurements reside at $z_\mathrm{abs} < 1.5$, broadening
the posterior by a factor of $\sim 2$.
Many of the $\Delta \alpha / \alpha$ measurements have $z_\mathrm{abs}$ near $1$, with numerous
pairs at relatively similar redshifts.
Taking subsets that are separated in scale factor by at least $1-10\%$ again only degrades
constraints by a factor $\lesssim 2$.
Moreover, unlike for the electron coupling, the photon coupling bounds are hardly affected by phase
marginalization for any subsample.

We reiterate that marginalizing over each phase independently is conservative for deriving upper
limits---in the absence of a systematic or physical effect, combining multiple measurements at
nearby times with random scatter about zero would improve constraining power.
In principle, however, doing so could obfuscate correlations that in fact favor nonzero $\Delta
\lambda_0$ (if multiple measurements at nearby redshifts favor similar $\Delta \lambda$).
Given that the joint, phase-marginalized posterior for $\Delta m_{e, 0}$ in
\cref{fig:marginalized-quasar-posteriors-quadratic} displays nonnegligible skew, this possibility
merits a more dedicated analysis which we do not pursue here.

In conclusion, we take the result from \cref{fig:marginalized-quasar-posteriors-quadratic},
$- 0.3 \lesssim 10^6 d_{m_e}^{(2)} \varphiamp_0^2 / 2 \lesssim 0.07$, as our most
conservative estimate of the $95\%$ credible interval.
Because this bound is strongly influenced by higher-$z_\mathrm{abs}$ observations, it is
inappropriate for $m_\varphi \lesssim 10^{-31}~\eV$; we therefore take the $z_\mathrm{abs} < 1$ result,
$10^6 \vert d_{m_e}^{(2)} \vert \varphiamp_0^2 / 2 \lesssim 5$, to indicate the degree to which
bounds degrade at lower masses.
(Scalars with masses below $10^{-32}~\eV$ are not captured by our treatment here, as they would not
have redshifted sufficiently by the present day for their effect on the present-day values of the
constants to be negligible.)
Using \cref{eqn:varphi-ito-rho-phi}, the bounds on the quadratic electron coupling are
\begin{align}
    - 0.36
    \lesssim d_{m_e}^{(2)}
        \left( \frac{m_\varphi}{10^{-31}~\eV} \right)^{-2}
        \frac{\fphi}{10^{-2}}
        \frac{\bar{\rho}_{c, 0}}{(1.76~\meV)^4}
    &\lesssim 8.6 \times 10^{-2}
    \label{eqn:quasar-dme2-bound}
\end{align}
at $m_\varphi \gtrsim 10^{-31}~\eV$, with the bounds for negative and positive couplings respectively
degrading by relative factors of $\sim 70$ and $17$ at $m_\varphi \approx 10^{-32}~\eV$.
For the photon coupling we take
$10^6 \vert d_{e}^{(2)} \vert \varphiamp_0^2 / 2 \lesssim 0.6$ as a representative bound.
Given that this result is relatively insensitive to the choice of redshift cut (or even whether
phases are marginalized over), we take it to apply for all $m_\varphi \gtrsim 10^{-32}~\eV$.
The resulting bound on the quadratic photon coupling is
\begin{align}
    \left\vert d_{e}^{(2)} \right\vert
    &\lesssim 0.73
        \left( \frac{m_\varphi}{10^{-31}~\eV} \right)^2
        \left( \frac{\fphi}{10^{-2}} \right)^{-1}
        \left( \frac{\bar{\rho}_{c, 0}}{(1.76~\meV)^4} \right)^{-1}.
    \label{eqn:quasar-de2-bound}
\end{align}
The bounds from quasars established in \cref{eqn:quasar-dme2-bound,eqn:quasar-de2-bound} provide the most stringent constraints on the lightest end of hyperlight scalar parameter space, as seen in \cref{fig:quadratic-coupling-non-cmb-constraints-Xi-1e-2} and \cref{fig:quadratic-coupling-constraints-Xi-1e-2} in \cref{sec:discussion}.

\subsection{Universality of free fall}
\label{sec:UFF}

The effects considered so far derive from temporal variations in fundamental constants.
\emph{Spatial} variations in the value of fundamental constants, such as those induced by localized
matter distributions, exert forces on matter.
Because the effects of coupled scalars on matter can be composition dependent, they source apparent
violations of the universality of free fall (UFF) towards a central mass~\cite{Damour:1994ya,
Damour:2010rp, Damour:2010rm, Uzan:2010pm, Hees:2018fpg, Uzan:2024ded}.

A species $X$ whose effective mass [i.e., $m_X(\varphi)$ from \cref{sec:late_universe_potential}]
depends on space and time through $\varphi$ experiences an acceleration
\begin{align}
    \three{a}_X(t, \three{x})
    \approx -\frac{\bm{\nabla} m_X(\varphi)}{m_X(\varphi)}
    = -g_X'(\varphi(t, \three{x})) \bm{\nabla} \varphi(t, \three{x}),
\end{align}
with $g_X'$ defined by \cref{eq:coupling-function-def}. Additional terms proportional to temporal gradients or suppressed by $\dot{\three{x}}$ are omitted.
For the power-series expansion of $g_{m_X}(\varphi)$
[\cref{eqn:coupling-function-series-expansion}], the scalar sources an acceleration
\begin{align}
    \three{a}_X(t, \three{x})
    &\approx - \bm{\nabla} \varphi(t, \three{x})
        \sum_{n=1} \frac{d_{m_X}^{(n)}}{(n - 1)!} \varphi(t, \three{x})^{n-1}
    = - \bm{\nabla} \varphi(t, \three{x})
        \left[ d_{m_X}^{(1)} + d_{m_X}^{(2)} \varphi(t, \three{x}) + \cdots \right],
    \label{eq:acceleration-series}
\end{align}
as one might have inferred directly from the Lagrangian.
In the vicinity of a localized matter distribution (\cref{sec:spatial-variations}) the scalar's gradient $\bm{\nabla} \varphi$ generally has a component toward the central mass that may be
interpreted as a new, long-range ``fifth'' force.
Species with different compositions (different dilatonic charges) at the same location experience different accelerations towards the central mass.

The degree of UFF violation for test masses $A$ and $B$ attracted toward a central mass $C$ (often
taken to the be Earth) is quantified by the E\"otv\"os parameter
$\eta \equiv \vert \three{a}_A - \three{a}_B \vert / \vert \three{g} \vert$, where
$\three{g} = - G M_C \three{x} / r^3$ is the acceleration due to the gravitational attraction between
the test and central masses.
Linearly coupled scalars are directly sourced by a central body independent of their cosmological or
galactic abundance (if any), with local profile given by \cref{eq:field_from_sphere_linear} when the
scalar only otherwise has a homogeneous, cosmological component (and when $m_\varphi R_C \ll 1$).
The resulting E\"otv\"os parameter is~\cite{Wagner:2012ui, Hees:2018fpg}
\begin{align}
    \eta^{(1)}
    &\approx \left\vert d^{(1)}_{M_C} \left( d_{m_A}^{(1)} - d_{m_B}^{(1)} \right) \right\vert
        e^{- m_\varphi r}
        \left( 1 + m_\varphi r \right).
    \label{eqn:eotvos-linear}
\end{align}
On the other hand, the UFF violation due to a quadratically coupled scalar is suppressed by its
local field value (which, for hyperlight scalars, corresponds to its cosmological abundance).
From \cref{eq:field_from_sphere_quadratic}~\cite{Hees:2018fpg, Oswald:2021vtc},
\begin{align}
    \eta^{(2)}
    &\approx \left\vert d^{(2)}_{M_C} \left( d_{m_A}^{(2)} - d_{m_B}^{(2)} \right) \right\vert
        \bar{\varphi}(t)^2.
    \label{eqn:eotvos}
\end{align}

The preceding calculations are classical, arising at tree level in field theory.
The value of $\eta$ is independent of the location $\three{x}$ because the tree-level force
\cref{eq:acceleration-series} with the boundary condition \cref{eq:field_from_sphere_quadratic} goes
as $1/\vert \three{x} \vert^2$, like gravity.
Contributions to the force from loops are negligible as long as the cosmological abundance satisfies
$\varphi(t)^2\gg G/\vert \three{x} \vert^2$, or equivalently as long as the momentum exchange $\sim
\vert \three{x} \vert^{-1}$ is much less than $\varphi(t)$ \cite{Ferrer:1998ue,Ferrer:2000hm,Bauer:2023czj}.

Terrestrial~\cite{Schlamminger:2007ht} and space-based~\cite{Touboul:2022yrw} experiments have
obtained stringent constraints on $\eta$, requiring models with new scalars to suppress their
violation of the UFF by some means.
The MICROSCOPE experiment in particular measured $10^{15} \eta = - 1.5 \pm 2.7$ between test masses
made of a platinum-rhodium alloy and of a titanium-aluminium-vanadium alloy, in a satellite orbiting
the Earth at an altitude of $710~\mathrm{km}$~\cite{Touboul:2022yrw}.
Since the dilatonic charges of these alloys and of Earth are of order $10^{-3}$ and $10^{-5}$ (for
photon and electron couplings, respectively)~\cite{Hees:2018fpg}, MICROSCOPE limits the linear
couplings of light scalars (with $m_\varphi r \ll 1$) as
$d_{e}^{(1)} \lesssim 3 \times 10^{-5}$ and $d_{m_e}^{(1)} \lesssim 6 \times 10^{-4}$.
The dependence of UFF violation [\cref{eqn:eotvos}] for quadratically coupled scalars on their
background values weakens bounds on the couplings because the scalar is sub-Planckian
($\varphi \ll 1$) and, for hyperlight scalars, due to the penalty from marginalizing over the
scalar's unknown phase.
MICROSCOPE took data over a two year period; per \cref{sec:phase-marginalization}, the E\"otv\"os
parameter predicted at fixed couplings $d_\lambda^{(2)}$ and abundance $\fphi$ is suppressed by a
factor of $(m_\varphi / 6.6 \times 10^{-23}~\eV)^2$ (for $m_\varphi < 6.6 \times 10^{-23}~\eV$).
Simply placing an upper limit of $\vert \eta \vert \lesssim 2.7 \times 10^{-15}$ yields constraints
on quadratic couplings of
\begin{subequations}
\begin{align}
    \left\vert d_{m_e}^{(2)} \right\vert
    &\lesssim 4.1 \times 10^{8}
        \left( \frac{\fphi}{10^{-2}} \right)^{-1/2}
        \left( \frac{\bar{\rho}_{c, 0}}{(1.76~\meV)^4}\right)^{-1/2}
\intertext{and}
    \left\vert d_{e}^{(2)} \right\vert
    &\lesssim 1.9 \times 10^{7}
        \left( \frac{\fphi}{10^{-2}} \right)^{-1/2}
        \left( \frac{\bar{\rho}_{c, 0}}{(1.76~\meV)^4}\right)^{-1/2}
    .
\end{align}
\end{subequations}
As evident in \cref{fig:quadratic-coupling-non-cmb-constraints-Xi-1e-2}, these bounds are extremely
weak in comparison to, e.g., those from quasar absorption spectra
[\cref{eqn:quasar-dme2-bound,eqn:quasar-de2-bound}].
They are, however, mass independent in the hyperlight regime.

\subsection{The Oklo phenomenon}\label{sec:oklo}

The Oklo phenomenon refers to the only known, natural occurrence of a sustained nuclear fission
reaction, which occurred $1.8~\Gyr$ ago and lasted $\mathcal{O}(10^5)~\yr$ in what
is now an open-pit uranium mine in Oklo, Gabon,
Africa~\cite{naudet1991oklo, shlyakhter1976direct, Uzan:2010pm, Uzan:2024ded}.
Constraints on the value of $\alpha$ at redshift $z \simeq 0.14$ may therefore be obtained from the
relative abundance of nuclear isotopes in the Oklo mines with an analysis somewhat similar to that
for BBN (\cref{sec:BBN}).
The key process is neutron capture on certain isotopes under exposure to a flux of neutrons from the sustained nuclear fission process.
The conversion process is described by a coupled system of nuclear equations and depends on the cross section $\sigma_n$ for neutron capture.
The cross section for such a ``heavy nucleus plus slow neutron''~\cite{Shlyakhter1976} process exhibits a resonant structure $\sigma_n \propto (E - E_r)^{-1}$, where $E$ is the energy of the incoming neutron and $E_r$ the resonant energy.
Constraints on $E_r(z\simeq 0.14)-E_{r,0}$ are thus obtained by measuring relative isotopic abundances in the Oklo mines; by modeling the dependence of $E_r$ on fundamental constants, one can infer constraints thereof.
Because no leptons are involved, the Oklo phenomenon is insensitive to variations in $m_e$.

As emphasized by Ref.~\cite{Uzan:2010pm}, constraints on $\alpha$ from the Oklo rocks involve a number of assumptions, not least of which is the dependence of $E_r$ on fundamental constants.
Other sources of uncertainties are the temperature of the reaction, variability between the rock samples used, the specific shape of the neutron spectrum assumed for the analysis, and the geometry of the natural reactor. Bounds on $\alpha$ vary by $\mathcal{O}(1)$~\cite[and references therein]{Uzan:2010pm}, but
\begin{align}
    \left\vert \frac{\Delta \alpha(z\simeq 0.14)}{\alpha_0} \right\vert
    &\lesssim 10^{-8}
    \label{eqn:oklo-bound-alpha}
\end{align}
is a conservative estimate.
As discussed in \cref{sec:quasars}, converting this bound into constraints on $d_{e}^{(2)}$ requires
comparing the range of possible values taken on by the scalar at $z \simeq 0.14$ to those today
[via \cref{eqn:constant-shift-ito-amplitude-phase}].
Even over the $\sim 10^5~\yr$ that the reaction lasted, any scalar with mass
$m_\varphi \ll 10^{-27}~\eV$ would have been effectively constant.
On the other hand, the phase shift between $z = 0.14$ and today is only $\lesssim 2 \pi$ for
$m_\varphi < 10^{-31}~\eV$; otherwise, the scalar's phases at both epochs cannot be precisely related
to one another.
This interval in mass is precisely that which we seek to constrain.
Given that the squared amplitude of the scalar is only reduced by a factor of $1.14^3 \approx 1.5$
in this interval, for much of parameter space the shift in $\alpha$ could be negligible sheerly by a
coincidental cancellation in \cref{eqn:constant-shift-ito-amplitude-phase}.
Marginalizing over such possibilities (as described in \cref{sec:phase-marginalization} but for both
phases) impedes drawing any robust constraint from \cref{eqn:oklo-bound-alpha}.

\subsection{Atomic clocks and pulsar timing arrays}
\label{sec:clocks}

A dynamical scalar field coupled to the SM imparts spacetime dependence to atomic energy levels that
can be probed by laboratory experiments.
Searches for this effect compare two levels with different dilatonic charges or ``sensitivities'' in
order to identify a relative difference that can be attributed to a scalar field.
Atomic clocks are currently the most precise laboratory systems, and comparisons of different
transitions within the same system or different systems yield stringent constraints on oscillatory
signals of scalar dark matter~\cite{VanTilburg:2015oza, Hees:2016gop, Kennedy:2020bac,
BACON:2020ubh, Kozyryev:2018pcp, Kobayashi:2022vsf, Filzinger:2023zrs} and also linearly drifting
fundamental constants more generally~\cite{Fortier:2007jf, Rosenband:2008qgq, Guena:2012zz,
Leefer:2013waa, Tobar:2013pwa, Huntemann:2014dya, Godun:2014naa} (see Ref.~\cite{Safronova:2017xyt}
for a review).
In addition to atomic transitions, nature provides another precise clock: pulsars with a steady rotation rate.
The residuals in pulse arrival times at Earth may be compared to a local laboratory clock, providing another means to search for scalar dark matter~\cite{Graham:2015ifn,Kaplan:2022lmz,NANOGrav:2023hvm}.
Due to the effects discussed in \cref{sec:phase-marginalization}, both yield weaker
constraints than cosmological probes in the hyperlight regime we consider; as such, we only briefly
summarize the constraints on quadratically coupled, hyperlight scalars as inferred from bounds on
linearly coupled scalars that do follow the galactic dark matter profile.

A signal that is quadratic in an oscillating scalar has time dependence $\sim \sin 2 m_\varphi t$;
a bound on the linear coupling of a scalar with mass $m_\varphi$ therefore translates to one at
$m_\varphi / 2$ for a quadratic coupling.
Following the considerations of \cref{sec:phase-marginalization}, we may recast limits on a locally
stochastic, linearly coupled scalar with local density $\rho_{\varphi, \oplus}$ and mass $2 m_\varphi$ into
those for a cosmologically homogeneous, quadratically coupled scalar with mass $m_\varphi$ that
comprises a fraction $\fphi$ of the dark matter density by equating their
values for $\Delta \lambda_0 / \lambda(0) \approx g_\lambda(\varphi)$.
Namely, \cref{eqn:varphi-ito-rho-phi} sets
\begin{align}
    \frac{ {d_\lambda^{(2)}}_\text{cosmo.} }{ {d_\lambda^{(1)}}_\text{galactic} }
    &\approx \frac{4.45 \times 10^{13}}{f_\mathrm{stoch}}
        \sqrt{\frac{\rho_{\varphi, \oplus}}{0.4~\GeV/\cm^3}}
        \frac{m_\varphi}{10^{-22}~\eV}
        \frac{(1.76~\meV)^4}{\fphi \bar{\rho}_{c, 0}}
        \label{eqn:cosmo-galactic}
\end{align}
where $f_\mathrm{stoch}$ is the stochasticity factor discussed below
\cref{eqn:varphiamp-for-conversion}.

In the regime of hyperlight masses, atomic clocks effectively probe the rate of change of
fundamental constants---e.g., a linear drift rate, $\ud \ln \lambda / \ud t$.
Marginalization over the scalar's unknown phase still penalizes bounds in this case, but with
different parametric dependence on $m_\varphi$.
In general, the theoretical drift rate is
$\ud g_\lambda(\varphi) / \ud t = m_\varphi d_\lambda^{(n)} \sin^{n-1} \vartheta \cos \vartheta / (n-1)!$;
per \cref{sec:phase-marginalization}, the presence of nodes in phase that scale with
$\vartheta^{n-1}$ penalizes bounds by a factor of $m_\varphi^{-1}$ for $n = 1$ and $2$ and by
$m_\varphi^{-(n-1)}$ for larger $n$.
As such, converting bounds from linear couplings to quadratic couplings requires no additional
suppression with $m_\varphi$ in the regime where the experimental runtime is much shorter than the
scalar's oscillation period.

The strongest constraint on the photon coupling at low masses to date is that from an optical clock
frequency comparison of the electric octupole (E3) and electric quadrupole (E2) transitions of the
$^{171}\mathrm{Yb}^+$ ion with observations spanning over $\sim 6$ years~\cite{Filzinger:2023zrs},
giving
\begin{align}
    \label{eq:Ybion_de2}
    \left\vert d_{e}^{(2)} \right\vert
    &\lesssim 1.8 \times 10^{5}
        \left( \frac{\fphi}{10^{-2}} \right)^{-1}
        \left( \frac{\bar{\rho}_{c, 0}}{(1.76~\meV)^4} \right)^{-1}
\end{align}
The strongest clock constraint on the electron coupling at low masses is placed by a comparison
between a ${}^{171}\mathrm{Yb}$ optical lattice clock and a ${}^{133}\mathrm{Cs}$ fountain microwave
clock which yields, over a $\sim 298$-day duration~\cite{Kobayashi:2022vsf},
\begin{align}
    \label{eq:YbCs_dme2}
    \left\vert d_{m_e}^{(2)} \right\vert
    &\lesssim 7.7 \times 10^{9}
        \left( \frac{\fphi}{10^{-2}} \right)^{-1}
        \left( \frac{\bar{\rho}_{c, 0}}{(1.76~\meV)^4} \right)^{-1}.
\end{align}
Electron coupling limits are generally weaker as transition energies typically vary much less in
their sensitivity to the electron mass than to the fine-structure constant; that is, the dilatonic
charges of the transition energies are small.

Bounds from pulsar timing are driven by the $\varphi$ dependence of the cesium clocks used to
measure the pulsar residuals, which may therefore be interpreted as an ``Earth term'' in the
analysis~\cite{Kaplan:2022lmz,NANOGrav:2023hvm}.
The pulsar rotation rate itself does depend on fundamental parameters due to angular momentum
conservation; however, its sensitivity to the electron mass is five orders of magnitude weaker than
that of the clock transition because the fraction of electrons in pulsars by mass is small.
The dependence of the pulsar rotation rate on $\alpha$ is assumed to be negligible.
As such, the pulsars effectively provide a steady reference clock that is independent of
$\varphi$~\cite{Kaplan:2022lmz, NANOGrav:2023hvm}.
The recent NANOGrav 15-year results constrain the linear couplings of hyperlight
scalars~\cite{NANOGrav:2023hvm} which translate to
\begin{subequations}\label{eqn:nanograv-quadratic-bounds}
\begin{align}
    \label{eq:nanograv_de2}
    \left\vert d_{e}^{(2)} \right\vert
    &\lesssim 2.0 \times 10^{6}
        \left( \frac{\fphi}{10^{-2}} \right)^{-1}
        \left( \frac{\bar{\rho}_{c, 0}}{(1.76~\meV)^4} \right)^{-1}
    \\
    \label{eq:nanograv_dme2}
    \left\vert d_{m_e}^{(2)} \right\vert
    &\lesssim 4.9 \times 10^{6}
        \left( \frac{\fphi}{10^{-2}} \right)^{-1}
        \left( \frac{\bar{\rho}_{c, 0}}{(1.76~\meV)^4} \right)^{-1}
    .
\end{align}
\end{subequations}

The bounds listed in this section all only apply for nonclustering scalars with periods well below
the total time span covered by the experimental measurements.
As discussed in \cref{sec:phase-marginalization},
both conditions are satisfied below masses $m_\varphi \sim 10^{-28}~\eV$.
The ytterbium ion bound is the leading clock-based bound on the quadratic photon coupling, and the NANOGRAV bound is that for the quadratic electron coupling.

\subsection{Stars}
\label{sec:stars}

New light particle degrees of freedom weakly coupled to the SM can be produced efficiently in hot stellar plasma and their subsequent escape provides additional cooling pathways.
We find that stellar evolution is a relatively negligible constraint on hyperlight, quadratically coupled scalars  considered here, but we note interesting features that arise due to the effect of the cosmological scalar background.

Reference~\cite{Olive:2007aj} considers the production of two scalar quanta in stellar plasmas.
Higher dimensional operators like the ones considered here---$\varphi^2 F_{\mu \nu} F^{\mu \nu}$, $\varphi^2 \bar{e} e$, and $\varphi^2 \bar{N} N$ (for $N$ some nucleon)---are suppressed by higher powers of the Planck mass.
Thermal production rates therefore depend steeply on temperature, making core-collapse supernovae ($T \sim 30~\MeV$) the most constraining systems.
Reference~\cite{Olive:2007aj} limits $d_e^{(2)}\lesssim 10^{15}$ and $d^{(2)}_{m_N} \lesssim 10^{14}$ from photon-pair annihilations ($\gamma+\gamma \rightarrow\varphi+\varphi$) and nucleon-nucleon bremsstrahlung ($N+N \rightarrow N+N+\varphi+\varphi$).
Limits on $d_{m_e}^{(2)}$ from electron-pair annihilations ($e^+ + e^- \rightarrow \varphi + \varphi$) and Compton-like scattering ($e + \gamma \rightarrow e + \varphi + \varphi$) are further suppressed by $\sim (m_e/T)^2$ and $\sim \alpha (m_e/T)^2$, respectively, making the constraints irrelevant both compared to cosmological and laboratory bounds (\cref{fig:quadratic-coupling-non-cmb-constraints-Xi-1e-2}).

In the presence of a cosmological background $\bar{\varphi}(t)$, one of the ``legs'' of the emission process can be replaced with the background field.
The emission process in the stellar environment is then proportional to that of a single energetic $\varphi$ particle, which can give stronger constraints as
single particle production arises from relatively lower-dimension operators and suffers less phase space suppression. For example, energy loss through the Primakoff process ($\gamma + Ze \rightarrow \varphi + Ze$) in globular clusters constrain $d_e^{(1)}\lesssim 10^8$~\cite{Dolan:2022kul}.
Similarly, single-particle emission via electron-ion bremsstrahlung
($e^- + Z e \rightarrow e^- + Z e + \varphi$) in red giants, horizontal branch stars, and galactic white dwarves constrain both
$d_{m_e}^{(1)}$ and $d_{m_N}^{(1)} \lesssim 10^6$~\cite{Hardy:2016kme, Bottaro:2023gep}.
Given that the dilatonic charges of nuclei for the photon coupling are order $10^{-3}$, the latter translates to $d_{e}^{(1)} \lesssim 10^{6} / \left( Q_{m_N} \right)_{e} \sim 10^{9}$.

Rates for background-assisted single $\varphi$ emission may be obtained by substituting a slowly varying coupling $d^{(1)}(t) \rightarrow d^{(2)} \varphi_\text{core}(t)$, where $\varphi_\text{core}(t)$ is the scalar's value inside the core of the star.
For sufficiently small couplings to the star $d^{(2)}_{M_\star}$, the scalar at the core is well approximated by its cosmological value, $\varphi_\text{core}(t) \approx \bar{\varphi}(t)$.
However, above the critical coupling \cref{eq:screening_condition}, the field inside the star is exponentially suppressed~\cite{Hees:2018fpg}:
\begin{align}
    \varphi_\text{core}(t)
    \approx \frac{\bar{\varphi}(t)}{2} \exp\left(-\sqrt{d_{M_\star}^{(2)}\frac{GM_\star}{R_\star}}\right),
\end{align}
for $d_{M_\star}^{(2)}>0$ and a nonrelativistic, nondegenerate star; we expect a suppression to also occur in relativistic and/or degenerate systems.
Consequently, background-assisted emission in stellar cores drops off sharply above some (positive) critical coupling, in contrast to signals at the surface of the Earth which still monotonically increase with coupling even beyond the critical value. For a star like the Sun, the condition that the scalar amplitude is well approximated by the cosmological
background, \cref{eq:screening_condition}, translates to $d^{(2)}_{m_\odot}\ll 5\times 10^{5}$.
For a neutron star (with radius $\sim 10~\mathrm{km}$ and $1.4$ solar masses),
$d^{(2)}_{m_\mathrm{NS}} \ll 5$; while this can serve as a first estimate, \cref{eq:screening_condition} cannot be applied rigorously to a very degenerate system like a neutron star.
For $d_{M_\star}^{(2)}<0$, the solution of Ref.~\cite{Hees:2018fpg} in principle predicts that the scalar diverges at the core for a discrete set of couplings:
$\varphi_\text{core}(t) \approx \bar{\varphi}(t) \sec \sqrt{\vert d_{M_\star}^{(2)} \vert G M_\star / R_\star}$, but the theory becomes nonperturbative in this regime.

The nonmonotonic dependence on coupling of background-assisted emission in stars raises the possibility of a gap in what couplings are constrained, with a given star constraining a finite range of couplings, extending to larger couplings for less compact stars.
A rigorous bound would also depend on the integrated effect of $\bar{\varphi}(t)$ as it redshifts and oscillates over stellar lifetimes (including cosmological in-medium effects on its evolution, relevant at large couplings).
Moreover, even without taking screening into account, the above bounds on linear couplings and the fact that $\bar{\varphi}(t) \ll 1$ at all
times translate to constraints on $d_{e}^{(2)}$ and $d_{m_e}^{(2)}$ much above $10^{9}$ and $10^6$, respectively, significantly weaker than many depicted in \cref{fig:quadratic-coupling-non-cmb-constraints-Xi-1e-2}.

\section{Early-time probes}
\label{sec:early-probes}

We proceed by considering cosmological probes of the fundamental constants in the early Universe.
In \cref{sec:BBN} we review the dependence of nucleosynthesis on fundamental constants and derive
constraints on the couplings of hyperlight scalar fields.
\Cref{sec:cmb} then places bounds from the impact of varying fundamental constants on CMB
anisotropies, building off the results of Ref.~\cite{Baryakhtar:2024rky}.

\subsection{Big bang nucleosynthesis}\label{sec:BBN}

A deviation in the value of fundamental constants from their present-day values during BBN would change the predicted yields of primordial elements, which are currently
in excellent agreement with observations~\cite{Coc:2006sx, Dent:2007zu, Clara:2020efx,
Sibiryakov:2020eir, Bouley:2022eer}.
The vast majority of nuclei of primordial origin are either hydrogen or helium-4. Their partitioning is typically quantified by the helium yield by nucleon number,
$Y_p \equiv 4 n_\mathrm{He} / (n_\mathrm{H} + 4 n_\mathrm{He})$, which differs slightly from the
yield by mass $Y_\mathrm{He}$ used in \cref{sec:matter-potentials} and CMB literature.
We now review the basic elements of the nucleosynthesis calculations that predict $Y_p$ in order to
establish its expected scaling with various fundamental parameters,
focusing on $\alpha$ and $m_e$; we then derive approximate bounds on their early-time values from
astrophysical measurements of $Y_p$.
We assume the parameters are simply fixed to some time-independent values $\alpha_i$ and $m_{e, i}$
that differ from their present-day values, $\alpha_0$ and $m_{e, 0}$, as would be the case for
the models we consider in \cref{sec:models} when the scalar field is frozen at early times.
Our discussion follows that of Refs.~\cite{Mukhanov:2003xs, Bouley:2022eer}.

The primordial yield of $\hefour$ is largely determined by the abundance of free neutrons when
fusion to helium becomes efficient, as nearly all free neutrons become bound in primordial $\hefour$
nuclei at that time.
The neutron abundance at helium fusion is in turn determined in two stages~\cite{Mukhanov:2003xs}.
First, neutron-proton conversions $n + e^+ \leftrightarrow p + \bar{\nu}_e$ and
$n + \nu_e \leftrightarrow p + e^-$ become inefficient when the weak interactions decouple
(at a temperature $T_W \approx \MeV$), at which point their relative abundance freezes out.
Solving the Boltzmann equation for evolution of the relative neutron abundance sets
\begin{align}
    \label{eq:electroweak_freezeout}
    X_{n, W}
    \equiv \frac{\bar{n}_{n}(T_W)}{\bar{n}_{n}(T_W)+\bar{n}_{p}(T_W)}
    \approx \int_0^\infty \frac{\ud T}{T}
        \frac{m_{np} e^{-I(T)}}{2 T \left[ 1 + \cosh(m_{np}/T) \right]}
\end{align}
to a good approximation.
Here $m_{n p} = m_n - m_p$ is the neutron-proton mass difference and
\begin{align}
    I(T)
    = \int_0^T \frac{\ud T'}{T'}
        \frac{1 + 3 g_A^2}{\pi^3} \frac{G_F^2 {T'}^5}{H}
        J\left( m_{np} / T' \right) \left( 1 + e^{- m_{np} / T'} \right),
\end{align}
where the weak axial-vector coupling constant of the nucleon $g_A \approx
1.2755$~\cite{Czarnecki:2018okw}, $G_F \approx 1.166379 \times 10^{-5}~\GeV^{-2}$ is the Fermi
constant, and
\begin{align}
    J(x)
    &= \frac{45 \zeta(5)}{2}
        + \frac{21 \zeta(4)}{2} x
        + \frac{3\zeta(3)}{2} \left( 1 - \frac{m_e^2}{2 m_{np}^2} \right) x^2
\end{align}
is the phase space form factor associated with neutron-to-proton conversion.
During this epoch one may approximate $H^2 = \pi^2 g_\star(T) T^4 / 90 \Mpl^2$ using $g_\star
\approx 10.75$.\footnote{In trading integrals over time for ones over temperature, we approximate $a T$ and therefore also
$g_\star$ as constant, neglecting their actual time dependence near, e.g., electron-positron
annihilation.}

After weak decoupling, protons no longer convert efficiently to neutrons but the neutron abundance
decreases steadily via free neutron decay, $n \rightarrow p + e^-+ \bar{\nu}_e$.
Neutron decay continues until the so-called deuterium bottleneck is cleared, at which point
efficient fusion begins (at a time $t_\text{BBN}$) and
\begin{eq}
    \label{eq:neutron-abundance-bbn}
    X_{n,\text{BBN}}
    \approx X_{n,W} e^{-t_\text{BBN} / \tau_n},
\end{eq}
where $\tau_n$ is the lifetime of the neutron.
Virtually all neutrons present at $t_\text{BBN}$ then ultimately end up in primordial $\hefour$
nuclei, so
\begin{eq}
    Y_p^\text{th} \approx 2 X_{n, \text{BBN}}.
\end{eq}
In the Standard Model, the neutron lifetime is given by~\cite{Czarnecki:2018okw,Czarnecki:2004cw}
\begin{align}
    \frac{1}{\tau_n}
    &= \frac{1 + 3 g_A^2}{2 \pi^3} \left\vert V_\mathrm{ud} \right\vert^2 G_F^2 m_e^5
        \left( 1 + \mathrm{RC} \right) P(m_{np} / m_e),
    \label{eqn:neutron-lifetime}
\end{align}
where the quark mixing matrix element $V_\mathrm{ud} \approx 0.9742$ and the contribution from
electroweak radiative corrections
$\mathrm{RC} \approx 0.03886$~\cite{Czarnecki:2018okw,Czarnecki:2004cw}.
The phase space factor $P(m_{np} / m_e)$ is approximately given by~\cite{Mukhanov:2003xs}
\begin{align}
    P(x)
    &= \frac{1}{60} \left[
            \sqrt{x^2 - 1} \left( 2 x^4 - 9 x^2 - 8 \right)
            + 15 x \ln \left( x + \sqrt{x^2 - 1} \right)
        \right],
\end{align}
but more detailed calculations yield
$P(m_{np} / m_e) \approx 1.6887$~\cite{Czarnecki:2018okw,Czarnecki:2004cw,Wilkinson:1998hx}.
Experiments measure the present-day value of the neutron lifetime to be
$\tau_{n, 0} = 878.6 \pm 0.5~\s$~\cite{ParticleDataGroup:2022pth}.

\Cref{eq:electroweak_freezeout,eq:neutron-abundance-bbn,eqn:neutron-lifetime} show that the
electron mass affects the helium yield by setting the phase space available to all weak reactions.
Namely, heavier decay products reduce the phase space available to free neutron decay, extending the
lifetime of the neutron and increasing $Y_p$. (For example, in the extreme case that $m_e > m_{np}$, decay via weak interactions is forbidden.)
One can see that, evaluated near $(m_e/m_{np})^2\simeq 0.15$, a change in the electron mass introduces only a fractional difference to the form factor $J(x)$, at least when the argument $x$ is small ($\partial\ln P/\partial\ln m_e \ll 1$). Heuristically, this is because $J(x)$ comes about as an integral over the broad distribution of relativistic thermal leptons. In contrast, there are no temperature scales involved in the decay of free neutrons. This fact alone would not be sufficient to make the neutron lifetime sensitive to the electron mass scale were it not also the case that $m_{e}\sim 0.4 m_{np}$, such that electron products are only semirelativistic. As a result, $\tau_n^{-1} \sim m_e^5(m_{mn}/m_e-1)^{7/2}+\dots$, resulting in $\partial \ln \tau_n^{-1}/\partial \ln m_e \sim \mathcal{O}(1)$.

The fine-structure constant affects \cref{eq:electroweak_freezeout,eq:neutron-abundance-bbn} via the
electrodynamic contribution to the neutron-proton mass splitting,
\begin{eq}
    m_{np}
    \approx m_{np}^\text{(QED)} + m_{np}^\text{(QCD)}
    \approx b \, \alpha \,\Lambda_\text{QCD}+ (m_d-m_u),
\end{eq}
where $m_u$ and $m_d$ are the masses of the up and down quarks and $b$ is a proportionality constant.
Lattice calculations of the QED contribution to $m_{np}$ yield
$m_{np}^\text{(QED)} \approx - 1~\MeV$~\cite{BMW:2014pzb,Thomas:2014dxa},
with order $10\%$ precision.\footnote{
    Calculations by alternative means yield a value
    $m_{np}^\text{(QED)} \approx - 0.76~\MeV$~\cite{Gasser:1974wd} that has often been used in past
    literature; other recent approaches yield
    $m_{np}^\text{(QED)} \approx -0.58~\MeV$~\cite{Gasser:2020hzn,Gasser:2020mzy}.
    See, e.g., Ref.~\cite{Walker-Loud:2019qhh} for comparative discussion.
}
Assuming no variation in either $b$ or $\Lambda_\text{QCD}$, the neutron-proton mass splitting
then varies with $\alpha$ via
\begin{eq}
    \frac{\Delta m_{np}}{m_{np}}
    = \frac{m_{np}^\text{(QED)}}{m_{np}} \frac{\Delta\alpha}{\alpha}
    \approx - \frac{1~\MeV}{m_{np}} \frac{\Delta \alpha}{\alpha}.
\end{eq}
Like $m_e$, $m_{np}$ affects the phase space available to all weak reactions, but $m_{np}$
also controls the relative abundance of protons and neutrons in thermal equilibrium.
Consequently, electroweak freeze out is more sensitive to variations in $m_{np}$ then to variations in $m_e$.
An increase in $\alpha$ decreases the magnitude of the neutron-proton mass splitting, therefore increasing $Y_p$.
(In the extreme case that $m_{np} \rightarrow 0$, their abundances are equal and
$Y_p \rightarrow 1$.)
The neutron lifetime is also slightly more sensitive to variations in $m_{np}$ than $m_e$ because $m_{np}$ is the dominant dimensionful scale that enters relative to the Fermi coupling scale $G_F^{-1/2}$, as evidenced by considering neutron decay to massless leptons.

The net amount of neutron decay depends on the age of the Universe when the deuterium bottleneck
clears, which itself depends upon the expansion history up until that point.
It so happens that the bottleneck clears during electron-positron annihilation (which completes at
temperatures around $20~\mathrm{keV}$ in the SM~\cite{Grayson:2023flr, Thomas:2019ran}); the
resulting entropy transfer causes the SM temperature to redshift more slowly than $a^{-1}$ and the
Universe to expand slightly more slowly.
A heavier electron annihilates earlier, so the Universe is older when nucleosynthesis occurs and
neutrons are less abundant.
This effect slightly offsets the increase in neutron lifetime for a heavier electron, reducing the
sensitivity of the helium yield to variations in $m_e$.
While there may be other such effects on the dynamics of the plasma at play in an analysis more
detailed than ours, we account for this change in the thermal history since it only makes the bounds
more conservative.
We employ the fitting functions from Ref.~\cite{Saikawa:2018rcs} to calculate the effective degrees
of freedom $g_\star(T)$.

To assess the dependence of $Y_p$ on the fundamental constants $\lambda$, we numerically evaluate
\cref{eq:electroweak_freezeout,eq:neutron-abundance-bbn} at various values thereof.
Though the level of calculation described here in insufficient to accurately predict the actual
value of $Y_p$, we take it to provide a reasonable estimate of its scaling with the early-time
values $\lambda_i$.
We compute
\begin{eq}
    \frac{\Delta Y_p}{Y_p}
    &\approx 0.34 \frac{\Delta m_{e, i}}{m_{e, 0}}
        + 2.6 \frac{\Delta \alpha_i}{\alpha_0}.
    \label{eqn:Yp-dependence-on-constants}
\end{eq}
The coefficient of $\Delta m_{e, i}$ increases to $0.42$ if the aforementioned effects of shifting
$e^+$-$e^-$ annihilation are neglected, which is consistent with the findings of, e.g.,
Ref.~\cite{Dent:2007zu}.
For all fundamental constants at their present-day value during BBN, detailed numerical calculations
marginalized over \Planck{}'s determination of the baryon abundance~\cite{Planck:2018vyg} yield
\begin{eq}
    Y^\text{th}_p = 0.24672 \pm 0.00061,
\end{eq}
while the measured value is currently~\cite{ParticleDataGroup:2022pth}
\begin{eq}
    Y^\text{obs}_p
    &= 0.245 \pm 0.003.
\end{eq}
Converting the $\pm 2 \sigma$ interval into constraints on $\alpha$ and $m_e$ varied independently
yields
\begin{subequations}\label{eqn:bbn-bounds-alpha-me}
\begin{align}
    -1.2 \times 10^{-2}
    &\lesssim \frac{\Delta \alpha_i}{\alpha_0}
    \lesssim 6.9 \times 10^{-3},
    \\
    -9.2 \times 10^{-2}
    &\lesssim \frac{\Delta m_{e, i}}{m_{e, 0}}
    \lesssim 5.1 \times 10^{-2}
    .
\end{align}
\end{subequations}
Observe that the intervals are slightly skewed toward early-time decreases in $\alpha$ or $m_e$
because the standard theoretical prediction is slightly larger than the median of the observational results.

To translate \cref{eqn:bbn-bounds-alpha-me} into constraints on quadratic scalar couplings,
note first that the above analysis only applies to time-independent shifts in fundamental constants
and therefore to scalars that are frozen during BBN.
The pertinent mass range extends far beyond that for which the scalar remains frozen during
recombination as well, which is our main focus.
Constraints in this hyperlight regime ($m_\varphi \lesssim 10^{-28}~\eV$) are mass independent at fixed
$\fphi$ (see \cref{sec:cosmological-dynamics}) and are well described by
\cref{eqn:dlambda-i-ito-Xi}, which yields
\begin{subequations}\label{eqn:bbn-bounds-quad-couplings}
\begin{align}
    -2.1
    &\lesssim d_e^{(2)} \frac{\fphi}{10^{-2}}
    \lesssim 1.2
    \\
    -16
    &\lesssim d_{m_e}^{(2)} \frac{\fphi}{10^{-2}}
    \lesssim 9.1
    .
\end{align}
\end{subequations}
These bounds mostly exclude the regime in which in-medium contributions to the scalar's potential
are important (see \cref{app:before-annihilation}), but upper limits for scalar abundances $\fphi$
smaller than the fiducial one chosen in \cref{eqn:bbn-bounds-quad-couplings} would require a dedicated
analysis including such effects.

The abundance of deuterium is also sensitive to $\alpha$ and $m_e$~\cite{Dent:2007zu} and measured
at the percent level or better~\cite{Cooke:2017cwo, Kislitsyn:2024jvk}; uncertainties in nuclear
reaction rates important to theoretical predictions of the deuterium abundance have recently been
improved substantially~\cite{Mossa:2020gjc, Yeh:2020mgl}.
The deuterium abundance, unlike that of helium, is also sensitive to the baryon abundance (via the
baryon-to-photon number ratio $\eta$) and drives BBN-derived constraints on $\omega_b$.
In principle, the correlation of $\omega_b$ with $m_e$ imposed by CMB constraints would allow
deuterium measurements to constrain the early-time electron mass, but obtaining precise bounds is
nontrivial due to the intrinsic dependence on $\alpha$ and $m_e$ (via their effect on the neutron
lifetime and, for $\alpha$, on $m_{np}$ and nuclear binding energies)~\cite{Dent:2007zu}.
We defer such analyses to future work.

\subsection{Cosmic microwave background anisotropies}\label{sec:cmb}

The CMB anisotropies precisely measure the spatial inhomogeneities of the SM plasma at the time of
hydrogen recombination as viewed on the sky today.
The rates governing recombination dynamics and the diffusion of acoustic waves in the plasma depend
on the fine-structure constant and electron mass, providing a probe of their values at redshifts
$\gtrsim 1000$.
The hyperlight scalar responsible for early-time shifts in $\alpha$ or $m_e$ can also reduce the
distance to photon last scattering and the rate of structure growth; the former impacts the angular
extent of the acoustic peaks on the sky and the latter suppresses the extent to which they are
smeared by gravitational lensing.
We study these dynamics in detail and perform a quantitative analysis of current cosmological
data---measurements of the CMB anisotropies from \Planck{} and low-redshift distances via baryon
acoustic oscillations (BAO) and supernova (SN) brightnesses---in Ref.~\cite{Baryakhtar:2024rky}.

The CMB alone strongly constrains the early-time value of the fine-structure constant via its
modulation of the diffusion rate, which alters the degree to which small-scale anisotropies are
suppressed.
High-resolution observations like \Planck{} measure the so-called damping tail with sufficient
precision to constrain $\alpha_i$ at the permille level.
The fine-structure constant also affects the rate at which photons and free protons recombine into
hydrogen, which impacts the generation of CMB polarization in a manner that scales the overall
height of the polarization power spectrum.
\Planck{} also measured the $E$-mode component of the polarization spectrum (and its correlation
with temperature anisotropies), though less precisely than it did the temperature power spectrum.
On its own, \Planck{} yields a $68\%$ posterior interval
$\alpha_i / \alpha_0 - 1 \approx \left( 0.7 \pm 2.4 \right) \times 10^{-3}$, which tightens to
$\left( 1.3 \pm 2.1 \right) \times 10^{-3}$ when combined with baryon acoustic oscillation data~\cite{Baryakhtar:2024rky}.

Cosmological constraints on electron-mass variations are more complex because of the degeneracy with
\LCDM{} parameters that arise in cosmological observables.
The electron mass determines the temperature (and so redshift) of recombination by scaling the
hydrogen energy levels but, unlike the fine-structure constant, does not alter the relevant rate
of photon diffusion.
In fact, all of the physical effects apparent in the primary CMB anisotropies that are modulated by
the redshift of recombination may be compensated for by adjusting the energy densities in baryonic
matter, dark matter, and dark energy---all of which are free parameters in the Bayesian analysis.
These degeneracies are of contemporary interest because early recombination shrinks the sound
horizon of the photon-baryon plasma, for which CMB data then infer a larger Hubble constant; the
model is thus a modification of early-time physics that could explain the present tension of
Hubble constant measurements derived from the CMB and the distance ladder~\cite{Sekiguchi:2020teg,
Hart:2019dxi}.
We discuss this possibility (with updated constraints using the most recent BAO and SNe datasets) in detail in
Ref.~\cite{Baryakhtar:2024rky}.

The CMB by itself provides weak bounds on electron mass variations (at the $10\%$ level), but BAO
and SNe data break degeneracies when combined with CMB data by constraining the late-time expansion
history (irrespective of the Hubble tension, i.e., not including the SNe calibrations that lead to
the tension).
In particular, these low-redshift datasets measure the redshift at which dark energy overtook matter
in the Universe's energy budget; along the CMB's degeneracy between the electron mass and \LCDM{}
parameters, the redshift of matter--dark-energy equality changes substantially.
However, measurements of this redshift (or, equivalently, the fraction of Universe's energy in
matter) are mildly discrepant between different probes---in particular, BAO data prefer earlier
transitions to dark-energy domination (lower present-day matter fractions) than recent SNe
datasets~\cite{DESI:2024mwx, Baryakhtar:2024rky, Loverde:2024nfi}.
This discrepancy is puzzling even for standard \LCDM{} cosmology, given that both probes measure
cosmological distances over a common interval of the Universe's history.

Reference~\cite{Baryakhtar:2024rky} examines the manner in which these differences in low-redshift
datasets propagate to constraints on the early-time electron mass (and other \LCDM{} parameters,
including the Hubble constant); at the extreme ends, \Planck{} data combined with either SNe data
from the Dark Energy Survey~\cite{DES:2024jxu} or BAO data from the Dark Energy Spectroscopic
Instrument~\cite{DESI:2024mwx} yield
$10^2 \left( m_{e, i} / m_{e, 0} - 1 \right) \approx -1.6 \pm 1$ and $1.7 \pm 0.7$, respectively.
Using BAO data from the completed Sloan Digital Sky Survey (SDSS)~\cite{eBOSS:2020lta,
eBOSS:2020hur} and Six-degree Field Galaxy Survey (6dFGS)~\cite{Beutler:2011hx} instead yields
$0.8 \pm 0.7$, while using alternative SNe samples from Pantheon+~\cite{Brout:2022vxf,
Scolnic:2021amr} and Union3~\cite{Rubin:2023ovl} yield $-0.6 \pm 1.1$ and $-2 \pm 1.4$,
respectively.\footnote{
    Curiously, the measurement using an older version of the Pantheon
    sample~\cite{Pan-STARRS1:2017jku} is $1^{+1.5}_{-1.4}$, a result more consistent with
    current BAO than current SNe data.
}
This lack of concordance impedes a consensus cosmological constraint, as all dataset combinations
yield percent-level measurements that are offset by $1$ to $3 \sigma$ at worst.
Rather than arbitrarily selecting one measurement to highlight, we present results deriving from
each dataset combination in order to be forthright about the discordance.

The preceding discussion made no reference to the gravitational impact of a hyperlight scalar field
because current CMB data turn out to disfavor its suppression of structure.
While Ref.~\cite{Baryakhtar:2024rky} shows that the late-time increase to the matter density from
such a scalar by itself opens up a new parameter degeneracy (that extends not just to the primary
CMB anisotropies but BAO and SNe data simultaneously), \Planck{} data do not tolerate the degree of
structure suppression induced by a scalar field whose abundance can substantially realize this
degeneracy.
While this predicted parameter relationship is mildly evident in posterior constraints, its impact
on marginal measurements of $m_{e, i}$ or $\alpha_i$ is only moderate.
Similar conclusions apply to SM neutrinos~\cite{Baryakhtar:2024rky, Loverde:2024nfi}, whose masses
are truly free parameters and which have similar phenomenology to hyperlight scalars.
The constraints on the abundance of coupled hyperlight scalars from Ref.~\cite{Baryakhtar:2024rky}
thus do not differ substantially from prior bounds on uncoupled scalars (at the same
mass)~\cite{Rogers:2023ezo}: over the mass range
$10^{-31}~\eV \lesssim m_\varphi \lesssim 10^{-29}~\eV$, the $95$th posterior percentile is
$\fphi \approx 1\%$ to $2\%$.
On the other hand, Ref.~\cite{Loverde:2024nfi} showed that cosmological neutrino mass measurements
also vary substantially among more recent low-redshift datasets; constraints on the abundance of
uncoupled, hyperlight scalars are likely also subject to a similar degree of variability.

While the detailed relationships between the fundamental constants, standard \LCDM{} parameters, and
current datasets are interesting in their own right, for the purposes of this work we need only
understand how to extract bounds on the scalar's fundamental couplings $d_{\lambda}^{(n)}$ and
abundance $\fphi$.
Doing so requires understanding the regime in which the analysis of Ref.~\cite{Baryakhtar:2024rky}
is valid (\cref{sec:cmb-matter-potential}) and the relationship between constraints on the
fundamental constants and the fundamental couplings (\cref{sec:cosmo-constraints-couplings-and-Xi}).
In \cref{app:analysis-variants} we discuss the impact of analysis choices in cosmological parameter
inference, including the impact of priors (\cref{sec:priors}) and of consistently accounting for the
fundamental-constant dependence of the helium yield $Y_p$ per \cref{sec:BBN}
(\cref{sec:bbn-consistency}).

\subsubsection{Effects of the matter potential}\label{sec:cmb-matter-potential}

In Ref.~\cite{Baryakhtar:2024rky} we assume that the scalar remains, to a good approximation, constant in time at least until last scattering.
However, as discussed in \cref{sec:matter-potentials}, the effective in-medium mass is not
necessarily negligible compared to $H$ and $m_\varphi$
[see \cref{eqn:dVmatter-dphi,eq:thermal_mass_early}].
\Cref{fig:sm-free-energy-dlambda} depicts the time evolution of the squared in-medium mass relative to
$H^2 \propto \bar{\rho}$, up to an overall factor of the dimensionless coupling $d_\lambda^{(2)}$,
and shows that electron-positron annihilation separates two qualitatively distinct regimes.
While early-time dynamics inform the conditions of the field after inflation, our analysis of the
CMB is only strictly affected by dynamics after annihilation; here we focus on the latter regime,
deferring a detailed study of both to \cref{app:thermal_field_evolution}.
In particular, we quantitatively assess the impact of the matter potential as a function of coupling
on the simplifying assumptions made in the analysis of Ref.~\cite{Baryakhtar:2024rky}.

In the late Universe, matter-induced dynamics are dictated by the in-medium potential \cref{eqn:dVmatter-dphi} and
parametrized by the dimensionless combination $\mathcal{D}$ from \cref{eq:coupling-combo-late}.
\Cref{app:after-annihilation} shows that the field begins evolving before matter-radiation equality
if $\vert \mathcal{D} \vert \gtrsim 1$---that is, if the in-medium mass is comparable in size to
$H$.
Because the CMB visibility function peaks around $a_\star / a_\mathrm{eq} \approx 3$ (where
$a_\mathrm{eq}$ is the scale factor of matter-radiation equality), the fundamental constants would
evolve nonnegligibly during recombination for such values of the couplings.
These dynamics would be inconsistent with the treatment of Ref.~\cite{Baryakhtar:2024rky} (and prior
studies of the CMB with varying constants); they are not precluded \textit{a priori} but
are simply beyond the scope of the present work.
In the regime $\vert \mathcal{D} \vert \lesssim 1$, the approximate solution
\cref{eqn:apx-soln-matter-effects-late} sets an estimated shift in fundamental constants $\lambda$
between early times (post annihilation) and recombination of
\begin{align}
    \left\vert \lambda_\star / \Delta \lambda_i - 1 \right\vert
    = \left\vert \bar{\varphi}(a_\star)^2 / \bar{\varphi}_i^2 - 1 \right\vert
    &\approx \left\vert (Q_b)_{\lambda} d_\lambda^{(2)} \right\vert / 3.
    \label{eq:variation_in_lambda_from_matter}
\end{align}
Requiring the early-time shift in $\alpha$ or $m_e$ to remain constant at the $10\%$ level
then requires $\vert d_\lambda^{(2)} \vert \lesssim 500$.

As discussed in \cref{sec:cosmological-dynamics}, matter-induced dynamics can also modify the
relationship between the scalar's early-time value $\bar{\varphi}_i$ and its present-day abundance
$\fphi$, which we take to be given exactly by the free-evolution result, \cref{eqn:Xi-ito-varphi}, in this section.
The total deviation is the sum of that accumulated through \cref{eqn:apx-soln-matter-effects-late}
evaluated at $a(t_\osc)$ and the small deviation from an $a^{-3/2}$ oscillation envelope at
$t > t_\osc$ (see \cref{app:after-annihilation}).
The former effect is more pronounced for lighter fields that begin oscillating later.
Requiring the matter-free result \cref{eqn:Xi-ito-varphi} to be accurate at the $10\%$ level
constricts the regime of validity by a further factor of two to five for scalars with masses
$10^{-29}~\eV$ to $10^{-31}~\eV$.

\subsubsection{Cosmological constraints}\label{sec:cosmo-constraints-couplings-and-Xi}

The late-Universe probes discussed in \cref{sec:late-probes} each only constrain a particular combination of model
parameters---that is, they technically cannot disentangle the size of a dimensionless coupling
$d_\lambda^{(n)}$ from the present-day scalar amplitude $\bar{\varphi}_0$ on their own.
While $\bar{\varphi}_0$ is related to the scalar's mass $m_\varphi$ and abundance $\fphi$ relative to
CDM, in the hyperlight regime the probes of \cref{sec:late-probes} have no information on the scalar's
oscillation frequency.
Cosmological probes are unique in their sensitivity not just to the scalar's effect on the SM but
also to its gravitational effects.
The CMB, for instance, measures scale-dependent information through the integrated Sachs-Wolfe
effect that depends on both the scalar's abundance and its mass (see Ref.~\cite{Baryakhtar:2024rky}).
A hypothetical cosmological detection of a coupled hyperlight scalar could therefore measure not
just the magnitude of the scalar's effect on SM parameters but also its mass and abundance, allowing
for a direct constraint on its dimensionless coupling $d_\lambda^{(n)}$.

\begin{figure}[t!]
\begin{centering}
    \includegraphics[width=.495\textwidth]{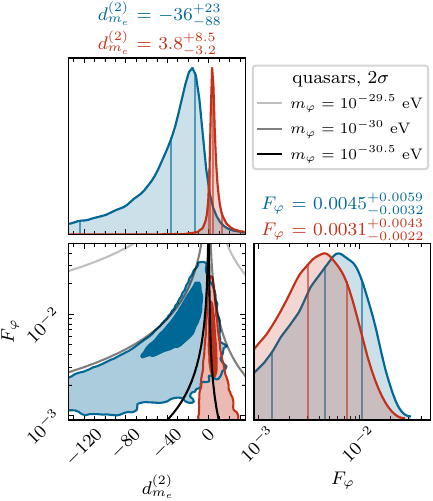}
    \includegraphics[width=.495\textwidth]{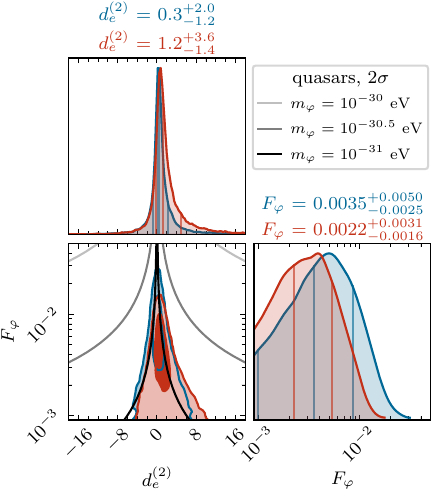}
    \caption{
        Posterior distributions over a hyperlight scalar field's present energy density relative to
        that of cold dark matter, $\fphi = \omega_\varphi / \omega_c$, and its quadratic couplings
        to the electron (left) and the photon (right).
        The diagonal panels depict one-dimensional marginal distributions, and the lower left panel
        depicts $1$ and $2 \sigma$ mass levels of the joint, two-dimensional posterior from
        parameter inference using \Planck{} 2018 data alone (blue) and combined with BAO data (red).
        Grey lines depict $2 \sigma$ limits from quasar absorption lines
        (derived in \cref{sec:quasars}) for a set of scalar masses $m_\varphi$ as labeled.
        For the early-time constraints, the scalar mass is fixed to $m_\varphi = 10^{-30}~\eV$, but
        these results do not change qualitatively within the range
        $10^{-32} \lesssim m_\varphi / \eV \lesssim 10^{-29}$.
        Note that parameter inference for the early-time constraints place flat priors on
        $\fphi$ and the early-time parameter value $\lambda_i$; the increase in posterior density with
        $\fphi$ at low values is the result of plotting the distribution over $\log_{10} \fphi$.
    }
    \label{fig:constraints-planck-bao-quasars}
\end{centering}
\end{figure}

The results of Ref.~\cite{Baryakhtar:2024rky} do not in fact demonstrate a preference for a nonzero
abundance of a new hyperlight scalar, thereby limiting the data's potential to independently probe
the couplings $d_\lambda^{(2)}$.
Nevertheless, to better assess any information on $d_\lambda^{(2)}$ contained in the cosmological
datasets, we display in \cref{fig:constraints-planck-bao-quasars} two-dimensional joint posteriors
on the abundance $\fphi$ and the coupling strengths $d_\lambda^{(2)}$, as well as the corresponding
marginalized one-dimensional posteriors.
The most salient feature in the two-dimensional posteriors is that, at lower $\fphi$ and away from
$d_{\lambda}^{(2)} \approx 0$, the mass levels of the joint posteriors follow lines of constant
$d_\lambda^{(2)} \fphi$, i.e., constant early-time shift in $\alpha$ or $m_e$ as per
\cref{eqn:dlambda-i-ito-Xi}.
The joint constraints in this regime are simply driven by the posteriors over $\alpha_i$ and
$m_{e, i}$ themselves.
As such, the constraints on the electron coupling are substantially narrower when BAO data is
combined with \Planck{}, reflecting the strong degeneracy of the early-time electron mass with
\LCDM{} parameters~\cite{Sekiguchi:2020teg, Baryakhtar:2024rky}.
Since \Planck{} data on their own generally prefer the electron to be lighter at recombination than
today, the posterior over $d_{m_e}^{(2)}$ skew toward negative values, but only at the $1 \sigma$
level; the posterior shifts to positive values when combined with BAO, but again only by
$\sim 1 \sigma$.

Since the CMB-derived bounds on coupled scalars are relatively insensitive to the scalar's
mass~\cite{Baryakhtar:2024rky}, \cref{fig:constraints-planck-bao-quasars} takes
$m_\varphi = 10^{-30}~\eV$ as a benchmark.
However, the bounds from quasars do depend on mass; we therefore overlay the $2 \sigma$ upper limits
from the analysis of \cref{sec:quasars} for a variety of masses.
Comparing these lines to the posteriors from \Planck{} and BAO data indicates the mass at which the
two probes have comparable constraining power.

Because \cref{fig:constraints-planck-bao-quasars} takes uniform priors over $\fphi$, the marginalized
posteriors over $\log_{10} \fphi$ grow exponentially at small $\fphi$; they are then sharply cut off at
values near half a percent, a feature shared with results for scalars with purely
gravitational interactions~\cite{Rogers:2023ezo} that derives solely from the disfavored effects of
hyperlight scalar fields on structure growth.
Namely, the small-$\fphi$ tail of the marginalized posterior is driven mostly by the prior, while the
sharp decline of posterior probability beyond $\fphi = \mathcal{O}(10^{-2})$ is driven by the data.
Further, because uniform priors disfavor extremely small values for $\fphi$, the bulk of each
posterior always lies within the regime of couplings where matter effects on the scalar's dynamics
(\cref{sec:cmb-matter-potential}) are negligible.
This holds even for the broad, \Planck{}-only constraints on the electron coupling, for which $95\%$
of posterior samples lie at values of $d_{m_e}^{(2)}$ above $-500$.
We discuss the impact of alternative priors in \cref{sec:priors}.

\subsection{Joint constraints}
\label{sec:discussion}

\begin{figure}[t!]
\begin{centering}
    \includegraphics[width=\textwidth]{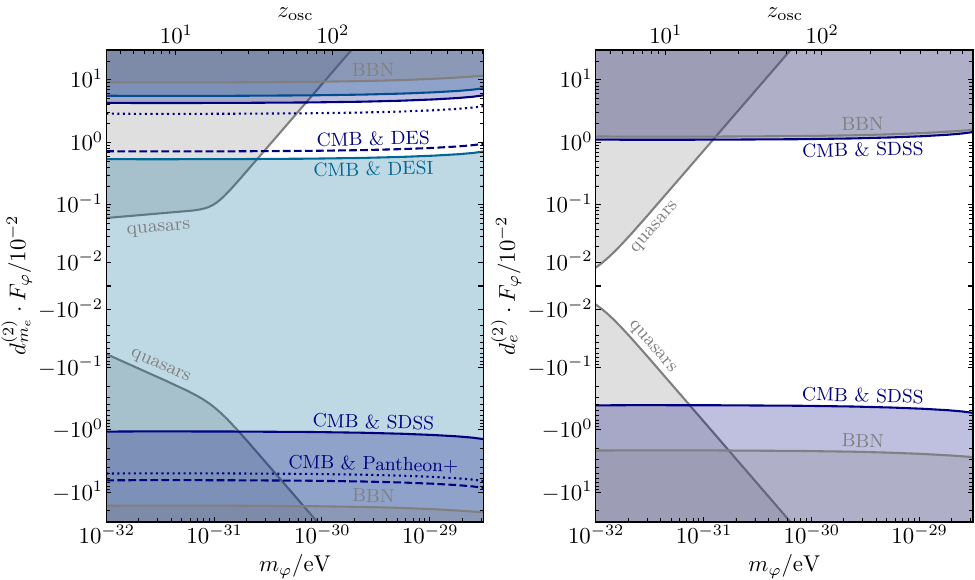}
    \caption{
        Exclusion regions for the quadratic couplings of a new scalar to the electron (left column)
        and photon (right column) that are positive (top row) or negative (bottom row).
        Results are shown for a fiducial abundance of the scalar (relative to that of CDM) of $\fphi
        \equiv \bar{\rho}_{\varphi, 0} / \bar{\rho}_{c, 0} = 10^{-2}$, which is roughly the largest
        allowed in our analysis for the mass range depicted here~\cite{Baryakhtar:2024rky}.
        The bounds on quadratic couplings, however, are mostly driven by the effect on the
        early-time values of $\alpha$ and $m_e$, as are the constraints from quasars
        (\cref{sec:quasars}) and BBN (\cref{sec:BBN}).
        The CMB data are from the 2018 \Planck{} release, which are presented in combination with
        BAO data from SDSS~\cite{eBOSS:2020lta, eBOSS:2020hur} and 6dFGS~\cite{Beutler:2011hx} (dark
        blue), BAO data from DESI~\cite{DESI:2024mwx} (light blue), and SNe data from either
        DES~\cite{DES:2024jxu} (dashed) or Pantheon+~\cite{Brout:2022vxf, Scolnic:2021amr} (dotted).
        Observe that the combination with DESI yields an apparent preference for nonzero, positive
        electron-mass couplings at $2 \sigma$ that is not particularly consistent with the other
        dataset combinations, all of which are consistent with zero; see \cref{sec:cmb} and
        Ref.~\cite{Baryakhtar:2024rky} for exposition on the cosmological sensitivity of these
        measurements.
        At masses higher than those depicted, the scalar deviates from its initial condition by more
        than $10\%$ at recombination, a regime which requires a dedicated analysis.
        Results approximately account for the impact on the relationship between mass and initial
        misalignment for scalars that begin oscillating around or before matter-radiation equality.
        Results are plotted on a symmetric logarithmic scale that interpolates linearly about zero
        between values of $\pm 10^{-2}$ via a transformation using an inverse hyperbolic function
        ($\mathrm{arcsinh}$).
    }
    \label{fig:quadratic-coupling-constraints-Xi-1e-2}
\end{centering}
\end{figure}

We now explore the joint constraining power of the early- and late-time datasets discussed in
\cref{sec:early-probes,sec:late-probes}.
The constraints from nucleosynthesis, the cosmic microwave background, quasar absorption spectra, and
laboratory experiments span a broad range of cosmic history and are therefore sensitive to distinct
regions of parameter space.
As explained in \cref{sec:late-probes}, present-day probes (whether from equivalence principle
violation or atomic clocks) are ultimately not competitive in their sensitivity to hyperlight
scalars that are too light to make up all of dark matter.
\Cref{fig:quadratic-coupling-constraints-Xi-1e-2} displays the remaining constraints from quasars,
BBN, and cosmological observations.
The impressive precision and redshift coverage of quasar absorption spectra enables strong constraints at the lightest
masses.
Sensitivity falls off with increasing mass simply because heavier scalars have a smaller amplitude
at fixed present-day abundance $\fphi$.
In the mass range for which our treatment is valid, constraints from \Planck{} (combined with BAO
observations to break degeneracies with \LCDM{} parameters, as discussed in \cref{sec:cmb}) improve
upon those from BBN by upwards of an order of magnitude, depending on the coupling and its sign, and
dominate that from quasars at masses $m_\varphi \gtrsim 10^{-31}~\eV$.

Measurements of the helium yield constrain the scalar's value at very high redshifts
$z \gtrsim 10^9$, at which point it is frozen unless $m_\varphi \gtrsim 10^{-17}~\eV$ or the quadratic
couplings are large enough that the effects of matter on the scalar's dynamics are important
(\cref{sec:cmb-matter-potential}; see also Ref.~\cite{Bouley:2022eer}).
Our analysis of \Planck{} data combined with BAO or SNe distance measurements (to break degeneracies
with \LCDM{} parameters) in Ref.~\cite{Baryakhtar:2024rky} also applies only when the scalar is
frozen through the end of recombination, which is the case for the entire mass range depicted in
\cref{fig:quadratic-coupling-constraints-Xi-1e-2}.
At a fiducial $\fphi = 10^{-2}$, both the CMB and BBN constraints on $d_e^{(2)}$ and $d_{m_e}^{(2)}$
are strong enough that the effective potentials due to matter are always irrelevant.
Larger $\fphi$ are ruled out by a scalar's independent gravitational effects on the CMB.
The constraints on couplings weaken at smaller $\fphi$; for $\fphi$ smaller than $10^{-3}$ or so,
the bounds on couplings are large enough that the matter effects discussed in
\cref{sec:cmb-matter-potential} can be relevant around recombination.
However, a hyperlight scalar with so small an abundance is gravitationally irrelevant.
Our analysis is therefore only inapplicable for $\fphi$ small enough such that (at fixed $\alpha_i$ or
$m_{e, i}$) the scalar's quadratic couplings are large enough that the fundamental constants are
dynamical during recombination.
Per \cref{sec:cmb-matter-potential}, the limits reported in
\cref{fig:quadratic-coupling-constraints-Xi-1e-2} only enter this regime for $\fphi \lesssim 10^{-4}$
or $10^{-5}$, depending on the coupling under consideration (and its sign).

\section{Discussion and conclusions}
\label{sec:conclusions}

New scalar fields can couple to the Standard Model as effective spacetime modulations of its
fundamental parameters.
In this work, we investigate scenarios in which a new field---light enough to be cosmologically
frozen in amplitude until after recombination, but heavy enough to constitute a subcomponent of the
dark matter today---modulates the mass of the electron and the electromagnetic fine-structure
constant on cosmological timescales.
We demonstrate that a scalar in this hyperlight mass range, even when
comprising a fraction of the energy density of dark matter and with couplings to the SM as weak as gravity, may still have an observable impact on the values of fundamental
constants throughout cosmological history.
Furthermore, our companion work shows that such scalars, which inevitably contribute to expansion
and the growth of structure, open up new degeneracies in cosmological parameters by adding to the
matter density late in cosmological history~\cite{Baryakhtar:2024rky}.
We now summarize the theoretical considerations and phenomenological constraints presented in this work, and discuss a number of avenues for future study they motivate.
The leading constraints on the hyperlight parameter space are presented in \cref{fig:quadratic-coupling-constraints-Xi-1e-2}.

\subsection{Models of cosmologically varying constants}

A substantial body of work has studied the cosmological implications of varying constants on a
purely phenomenological basis; our work grounds this literature in concrete microphysical
models and enables an assessment of their consistency.
\Cref{sec:models} reviews the field-theoretic framework of dynamical variations of
fundamental constants induced by a new, light scalar field.
We first outline a general description that promotes SM parameters to spacetime-dependent quantities via a scalar in \cref{sec: Varying constants with new scalars}.
In \cref{sec:Effective Lagrangian for phi-matter coupling} we then review the low-energy effective
theory of the scalar's interactions with composite matter, providing a mapping from microphysical
couplings to phenomenological consequences for both cosmological dynamics of SM matter and
signatures in laboratory experiments~\cite{Damour:2010rp, Damour:2010rm}.

Any evolution in fundamental constants is determined by the product of the cosmologically evolving amplitude of a scalar field (to some power $n$) and its interaction strength with the relevant SM operator, $d_\lambda^{(n)}$.
The interactions responsible for the variation of fundamental constants, in turn, inevitably modify the scalar's own dynamics.
Using background-field methods in thermal field theory, \cref{sec:matter-potentials} derives the
effective scalar potential sourced by the cosmological abundance of SM matter in the relativistic
and nonrelativistic regimes, presenting results relevant before and after electron-positron
annihilation (see \cref{fig:sm-free-energy-dlambda}).
We then apply these results in \cref{sec:cosmological-dynamics} to delineate the regimes
of the scalar's cosmological dynamics, where we show that the
approximations of prior work---time-independent shifts in fundamental constants leading up to
recombination---are valid only in the small-coupling (and thus nonnegligibly gravitating) regime.
After electron-positron annihilation and for quadratic couplings to both photon and electron, the magnitude of the effective potential is suppressed relative to the Hubble scale by
the small fractional contribution to the net mass in SM matter from electrons and QED effects; specifically, matter effects may be
neglected for dimensionless couplings $d_\lambda^{(2)} \ll 10^4$.
We study the dynamics of the regime of stronger coupling in more detail in
\cref{app:thermal_field_evolution}.
\Cref{sec:sm-sources-of-variation} also comments on the (currently undetectably small) extent to which in-medium effects within the SM itself effectively modify the fundamental constants.

\subsection{Local and late-time searches}

Numerous astrophysical and local probes---including quasar absorption spectra, atomic clocks, the
Oklo phenomenon, stars, and tests of the universality of free fall---constrain variations of
fundamental constants at times well after recombination.
The constraining power of these probes relative to those from early-Universe cosmology
depends on the form of the scalar's potential and interactions, insofar as they determine both its
cosmological redshifting and its response to matter.
In particular, strong bounds from equivalence principle tests (\cref{sec:UFF}) on light, linearly
coupled scalars are unique in being independent of the scalar's local or cosmological abundance,
leaving no parameter space with observable cosmological signatures.
The signatures of quadratically coupled scalars, on the other hand, do depend on the scalar's
abundance, leaving open the possibility of unconstrained parameter space with observable effects in
the early Universe.

In \cref{sec:late-probes} we extend the analysis of these late-time probes to handle several
qualitatively distinct features of hyperlight scalars, with results summarized in
\cref{fig:quadratic-coupling-non-cmb-constraints-Xi-1e-2}.
Measurements long after a hyperlight scalar begins oscillating are plagued by its extremely
long oscillation period: the statistical possibility that the measurement
occurs near a node of the signal steeply penalizes the reach of single, finite-duration probes.
As a result, atomic clocks and tests of the equivalence principle are limited in constraining power.\footnote{
    The penalty does not apply to the discovery \textit{reach} of such searches; see
    \cref{sec:phase-marginalization} for further discussion.
}
\Cref{sec:phase-marginalization} argues that the incurred penalty depends on the
signal's scaling with phase about its nodes.
Equivalence principle tests (\cref{sec:UFF}) for quadratic couplings are therefore penalized by two powers of mass times inverse experimental duration,
while atomic clock and pulsar timing arrays (\cref{sec:clocks}), which essentially probe the drift rate of fundamental constants, incur a penalty of one power for both linear and quadratic couplings.
Combinations of measurements can mitigate this penalty only if spaced in time by a substantial
fraction of the scalar's oscillation period because of the greatly suppressed chance that the
scalar's amplitudes sampled at random points in its oscillation all occur at nodes.
Constraints on varying constants from quasar absorption lines (\cref{sec:quasars}) span redshifts as high as $7$ and
therefore remain competitive when analyzed as an ensemble
(\cref{fig:marginalized-quasar-posteriors-quadratic}), even if during epochs when the field
oscillates, setting the leading constraints at the lowest masses.

\subsection{Cosmological signatures and constraints}

Searches for hyperlight, coupled scalars in the early Universe have the crucial advantage of
observing a hyperlight scalar at earlier times when it is not only more abundant but also frozen
rather than oscillating.
Namely, the amplitude of misaligned scalars (and therefore also that of their signatures) redshifts with expansion and is therefore greater at earlier times.
Moreover, regimes for which the scalar is frozen during the pertinent epoch, like nucleosynthesis
(for $m_\varphi \lesssim 10^{-17}~\eV$) or recombination ($m_\varphi \lesssim 10^{-28}~\eV$), are
not plagued by any uncertainty in the scalar's instantaneous amplitude as in the oscillatory regime.

\Cref{sec:early-probes} shows that the quadratic couplings of hyperlight scalars
are best constrained by probes at higher redshift, including measurements of the
primordial abundances of light elements (\cref{sec:BBN}) and the cosmic microwave background
anisotropies (\cref{sec:cmb}).
An ensemble of quasar absorption line measurements from various redshifts exceeds cosmological
probes in reach for masses $m_\varphi \lesssim 10^{-31}~\eV$ (see
\cref{fig:constraints-planck-bao-quasars,fig:quadratic-coupling-constraints-Xi-1e-2}) since such
scalars remain frozen longer and therefore redshift less by the time of the absorption that produces
the lines.
We studied the complementary power of these probes in \cref{sec:discussion}, showing that results
deriving from CMB data in Ref.~\cite{Baryakhtar:2024rky} provide leading bounds over nearly two
decades in mass.
The photon coupling is robustly constrained via its effects on the polarization and small-scale
anisotropies of the CMB, limited to $-0.4 < d_e^{(2)} < 1.1$ (as quantified relative to gravity) for scalars that make up a percent of
the CDM density (roughly the largest allowed fraction~\cite{Baryakhtar:2024rky}).
The electron coupling, on the other hand, is subject to strong degeneracies with cosmological
parameters~\cite{Baryakhtar:2024rky}; low-redshift data from baryon acoustic oscillations or supernovae can effectively break
them, but current datasets do so inconsistently.
Combined with CMB data, DESI has a preference for a positive, nonzero quadratic electron coupling whereas DES, SDSS, and Pantheon datasets rule out $\vert d_{m_e}^{(2)} \vert$ smaller than a few.
Despite the lack of concordance in these joint constraints at a quantitative level, cosmological
data still provide leading bounds for $10^{-31} \lesssim m_\varphi / \mathrm{eV} \lesssim 10^{-28}$, as shown in
\cref{fig:quadratic-coupling-constraints-Xi-1e-2}.

Furthermore, this work provides theoretical grounding for the substantial body of literature
constraining cosmological variations in fundamental constants on a phenomenological basis.
The framework we develop, in the style of effective field theory, enables a relatively general
comparison of the constraining power from cosmological data and independent laboratory and
astrophysical searches.
Prior work noted that linearly coupled scalars in principle do not redshift sufficiently after
recombination to satisfy atomic clock constraints and simultaneously yield observably large
variations in the CMB~\cite{Vacher:2024qiq}.
Our work shows that such arguments apply only to linearly coupled scalars that do not oscillate at
late times: the fundamental-constant variation (or oscillation amplitude thereof) due to
higher-order couplings redshifts faster, and marginalizing over the scalar's unknown oscillation
phase parametrically penalizes limits from terrestrial probes (see
\cref{sec:phase-marginalization}).
On the other hand, equivalence principle limits already entirely exclude any observable cosmological
signatures from linearly coupled scalars (\cref{sec:UFF}), largely independent of the scalar's
abundance and potential [in the regime of interest with $V(\phi) \ll \mathrm{eV}^4$].

To assess the implications of laboratory and astrophysical searches for cosmological signatures, we
may straightforwardly extrapolate the noncosmological limits presented in
\cref{fig:quadratic-coupling-constraints-Xi-1e-2} to limits on variations at recombination.
In the mass range for which the scalar remains frozen at recombination, the early-time variation in
fundamental constants is $\sim d_\lambda^{(2)} F_\varphi$ [\cref{eqn:dlambda-i-ito-Xi}].
The limits reported in \cref{fig:quadratic-coupling-constraints-Xi-1e-2}, taking fiducial
$F_\varphi = 10^{-2}$, thus correspond to the percent variation allowable at recombination.
Limits from quasars preclude subpermille early-time variations only for
$m_\varphi \lesssim 10^{-31}~\mathrm{eV}$ and quickly fall off at heavier masses.
The limits from nucleosynthesis [\cref{eqn:bbn-bounds-alpha-me}] directly apply at recombination, so
long as the scalar is static in the intervening epoch---i.e., if neither the bare nor in-medium
potentials are important.
The latter holds for scalar abundances near our fiducial choice, but for much less abundant scalars
(at fixed early-time variation in fundamental constants) the dimensionless coupling, and therefore
in-medium effects, are enhanced and nontrivially modify these conclusions.
In summary, while this work establishes minimal models for which early-time variations in
fundamental constants (including those motivated as early-recombination solutions to the Hubble
tension) are viable in light of both theory and independent searches, these considerations
essentially require it makes up a nonnegligible fraction of the total matter density today (the
regime studied in Ref.~\cite{Baryakhtar:2024rky}).

\subsection{Future directions}\label{sec:future}

In order to realize the phenomenological scenarios of prior study in which the change to the
fundamental constants is time independent through recombination, we restricted our analysis to
regimes in which the effect of matter on the scalar's dynamics is negligible.
The analysis of \cref{sec:cmb-matter-potential,app:thermal_field_evolution} shows that, though such
effects are only relevant for relatively large quadratic couplings, they lead the scalar to grow or
decay during the matter era, depending on the sign of the coupling.
That the field dynamically reacts to large matter couplings in particular complicates the
``phenomenological'' limit of varying-constant scenarios where time-independent shifts in
fundamental constants during recombination are in effect realized by a strongly coupled yet
negligibly abundant new scalar.
Our analysis of concrete models enables a quantitative assessment of the consistency of such
treatments, revealing that scalars which remain frozen during recombination must contribute
nonnegligibly to gravity.
A consistent treatment of the opposite regime, with strongly time-dependent fields leading
up to and during recombination (beyond existing studies of phenomenological toy
models~\cite{Hart:2017ndk, Lee:2022gzh}), is necessary to place genuine constraints on couplings
$\vert d_\lambda^{(2)} \vert \gtrsim 500$ [\cref{eq:variation_in_lambda_from_matter}] and scalar
abundances $\fphi \lesssim 10^{-4}$.
By the same token, scalars heavier than those we considered here would begin oscillating before last
scattering.
Such nontrivial dynamics before recombination are beyond the scope of this work and the cosmological
analysis of Ref.~\cite{Baryakhtar:2024rky} and are important targets for future work.

The possibility that the coupling to Standard Model matter invoked to change fundamental constants
itself explains the scalar's dynamics is also theoretically attractive.
The cosmological dynamics induced by SM matter were central to early studies of variations in
fundamental constants driven by massless scalar fields (``quintessence'')~\cite{Bekenstein:1982eu,Damour:1994ya}.
As discussed in \cref{sec:cmb-matter-potential,app:thermal_field_evolution}, such couplings to matter
induce nontrivial scalar dynamics starting only around matter-radiation equality, which could
potentially explain why the scalar begins diluting after recombination without invoking an
arbitrarily light mass, albeit without resolving the issue of large radiative corrections induced by the same matter coupling.
Models of early dark energy that attempt to address to Hubble tension are subject to a similar (but
more severe) coincidence problem in the timing of recombination and the start of the scalar's
dynamical evolution.
Indeed, proposed solutions invoke couplings to the dark matter~\cite{Karwal:2021vpk,
Lin:2022phm} or neutrinos~\cite{Sakstein:2019fmf, CarrilloGonzalez:2020oac, Kamionkowski:2024axz} that induce the field to
start rolling around matter-radiation equality, in exact analogy to the dynamics discussed here.

We treat only electron and photon interactions, but in a UV complete theory a scalar likely possesses further couplings with the SM and specific relationships between them.
Couplings to quarks, gluons, and the Higgs all alter the masses of nucleons, whose cosmological signatures (beyond
BBN) have not been studied in detail.
We also consider only the simplest models of scalars with bare masses, which nominally must be
extremely tuned against radiative corrections; our study of such prototype models may provide
inspiration for the construction of more complete theories.

To sum up, the phenomenology of scalar fields coupled to the SM is rich, and cosmological observations offer a
powerful probe of unique regimes of parameter space, both via precise sensitivity to altered microphysics in the early Universe and through the gravitational impact of the new fields.
Moreover, future opportunities to search for coupled scalars (hyperlight or otherwise) are
myriad~\cite{Antypas:2022asj}, including nuclear clocks~\cite{Flambaum:2006ak, Campbell:2012zzb,
Peik:2020cwm, Brzeminski:2022sde}, high-resolution CMB observations~\cite{CMB-S4:2016ple, SimonsObservatory:2018koc}, CMB spectral distortions~\cite{Hart:2022agu},
the Lyman-$\alpha$ forest~\cite{Hamaide:2022rwi}, and absorption spectra measurements from
quasars~\cite{ANDES:2023cif} and emission line galaxies~\cite{DESI:2024yok}.
In the near term, further data from DESI and other future galaxy surveys~\cite{DESI:2016fyo,
DESI:2022lza, Euclid:2024yrr, Euclid:2024imf, Eifler:2020vvg} and supernova
surveys~\cite{Rigault:2024kzb, LSSTDarkEnergyScience:2018jkl, YoungSupernovaExperiment:2020dcd}
should improve limits on electron couplings and clarify the discordant results from current
datasets.
Studying the interesting dynamical regimes of strong couplings and heavier masses, as well as couplings to the strong sector, would paint a complete picture of new hyperlight, interacting scalars, which may well be first revealed through their cosmological signatures in the wealth of upcoming data.

\begin{acknowledgments}
We thank Asimina Arvanitaki, Kimberly Boddy, Thomas Bouley, David Cyncynates, Thibault Damour, John Donoghue, Michael Fedderke, Subhajit Ghosh, Junwu Huang, Mikhail Ivanov, Marilena Loverde, Maxim Pospelov, Sergey Sibiryakov, Ken Van Tilburg, Neal Weiner, and Tien-Tien Yu for many useful discussions and correspondence and Marios Galanis for collaboration on initial stages of this work.
M.B.\ is supported by the U.S. Department of Energy Office of Science under Award Number DE-SC0024375.
M.B.\ and Z.J.W.\ are supported by the Department of Physics and College of Arts and Science at the University of Washington; Z.J.W. is also supported in part by the Dr. Ann Nelson Endowed Professorship.
O.S.\ is supported by a Princeton Center for Theoretical Science Fellowship at Princeton University, and the Department of Physics and was supported by a DARE Fellowship from the Office of the Vice Provost for Graduate Education at Stanford University at the time this research was initiated.
MB is grateful for the hospitality of Kavli Institute for Theoretical Physics (KITP) and Perimeter Institute where part of this work was carried out.
Research at Perimeter Institute is supported in part by the Government of Canada through the Department of Innovation, Science and Economic Development and by the Province of Ontario through the Ministry of Colleges and Universities.
This work was also supported by a grant from the Simons Foundation (1034867, Dittrich).
This research was supported in part by grant NSF PHY-2309135 to the Kavli Institute for Theoretical Physics (KITP).
This work made use of the software packages
\textsf{emcee}~\cite{Foreman-Mackey:2012any,Hogg:2017akh,Foreman-Mackey:2019},
\textsf{corner.py}~\cite{corner}, \textsf{NumPy}~\cite{Harris:2020xlr},
\textsf{SciPy}~\cite{Virtanen:2019joe}, \textsf{matplotlib}~\cite{Hunter:2007ouj},
\textsf{xarray}~\cite{hoyer2017xarray}, \textsf{ArviZ}~\cite{arviz_2019},
\textsf{SymPy}~\cite{Meurer:2017yhf}, and \textsf{CMasher}~\cite{cmasher}.
\end{acknowledgments}

\appendix

\section{Scalar dynamics with effective potentials sourced by matter}
\label{app:thermal_field_evolution}

In this appendix we describe solutions to the Klein-Gordon equation with an effective potential
sourced by matter, as discussed in \cref{sec:matter-potentials}.
The matter potentials are qualitatively different before and after electron-positron annihilation,
so we treat each regime separately.
In both cases we discuss what restrictions on parameter space ensure our treatment of the theory
remains self-consistent.

\subsection{After annihilation}\label{app:after-annihilation}

\begin{figure}[t!]
    \centering
    \includegraphics[width=\textwidth]{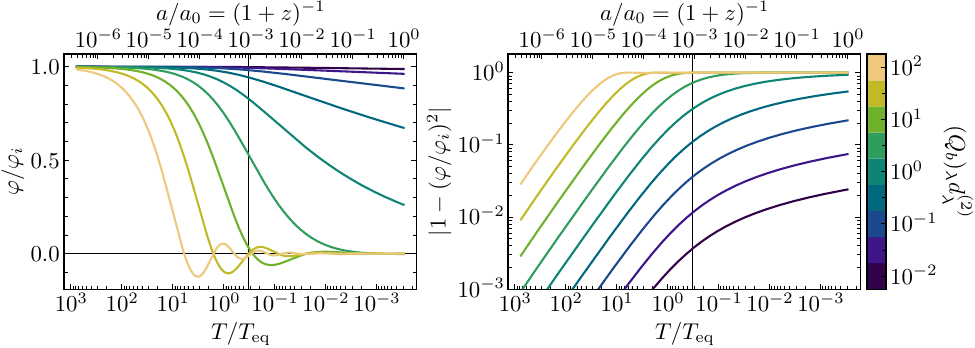}
    \caption{
        Solutions to the Klein-Gordon equation \cref{eqn:scalar-eom-nonrelativistic-potential} for a
        quadratically coupled scalar with a nonrelativistic matter potential, with values of
        $(Q_b)_\lambda d_\lambda^{(2)}$ varying by color.
        Vertical black lines indicate the moment of recombination (i.e., for the fiducial case that
        the fundamental constants do not change with time).
        The right panel depicts the fractional difference in the shift in $\lambda$ from the
        initial condition, i.e., $\vert \lambda / \lambda_i - 1 \vert$.
    }
    \label{fig:dynamics-nonrelativistic-thermal-mass}
\end{figure}

We first solve \cref{eq:scalar-eom} for a homogeneous scalar with
\cref{eqn:dVmatter-dphi} contributing to its dynamics.
Written in terms of $N \equiv \ln a / a_\mathrm{eq}$ where $a_\mathrm{eq}$ is the scale factor at
matter-radiation equality, the equation of motion is
\begin{align}
    \partial_N^2 \varphi
        + \left( 3 + \frac{\partial_N H}{H} \right) \partial_N \varphi
    &= - \left(
            \frac{m_\varphi^2}{H^2}
            + \frac{\mathcal{D}}{1 + e^{-N}}
        \right)
        \varphi
    \label{eqn:scalar-eom-nonrelativistic-potential}
\end{align}
where
\begin{align}
    \mathcal{D}
    &\equiv \frac{3}{2}
        \frac{
            (Q_b)_{e} d_{e}^{(2)} + (Q_b)_{m_e} d_{m_e}^{(2)}
        }{1 + \omega_c / \omega_b}
\end{align}
and $\partial_N H / H = - 3 (1 + w) / 2$ in a Universe with a constant equation of state $w$.
Taking the scalar's bare mass to be negligible, \cref{eqn:scalar-eom-nonrelativistic-potential}
admits analytic solutions in the limits of pure matter and radiation domination, $N \gg 0$ and
$N \ll 0$.
The radiation-era solution satisfying $\varphi \to \varphi_i$ and $\dot{\varphi} \to 0$
at early times $a \to 0$\footnote{
    To be completely clear, this initial condition is specified long before matter-radiation
    equality but after electron-positron annihilation, since
    the scalar is also in general dynamical during and before annihilation
    (see \cref{app:before-annihilation}).
    The specified initial conditions remain appropriate because the scalar refreezes after
    annihilation unless its coupling is so large ($\vert d_{\lambda}^{(2)} \vert \gtrsim 10^9$) that
    the two regimes cannot be considered separately.
} is
\begin{align}
    \varphi_\mathrm{RD}(a)
    &= \varphi_i
        \frac{
            J_1\left( 2 \sqrt{\mathcal{D} a / a_\mathrm{eq}} \right)
        }{
            \sqrt{\mathcal{D} a / a_\mathrm{eq}}
        }
    \label{eqn:radiation-era-nonrel-soln}
\end{align}
where $J_1$ is the order-one Bessel function of the first kind.
Since the scalar begins rolling (to some degree, at least) during the radiation era, we take general
initial conditions $\varphi_\mathrm{eq}$ and $\dot{\varphi}_\mathrm{eq}$ for the matter-era
solution, which is
\begin{align}
    \varphi_\mathrm{MD}(a)
    &= \frac{
            \left(
                - \beta_- \varphi_\mathrm{eq}
                + \partial_N \varphi_\mathrm{eq}
            \right)
            (a / a_\mathrm{eq})^{\beta_+}
            + \left(
                \beta_+ \varphi_\mathrm{eq}
                - \partial_N \varphi_\mathrm{eq}
            \right)
            (a / a_\mathrm{eq})^{\beta_-}
        }{
            \sqrt{9 / 4 - 4 \mathcal{D}}
        }
    \label{eqn:matter-era-nonrel-soln}
\end{align}
where
\begin{align}
    \beta_\pm
    &= - 3/4 \pm \sqrt{9/16 - \mathcal{D}}
    \label{eqn:matter-era-nonrel-soln-beta}
\end{align}
The matter-era solution can be matched onto the radiation-era one by setting $\varphi_\mathrm{eq}$
and $\partial_N \varphi_\mathrm{eq}$ according to \cref{eqn:radiation-era-nonrel-soln}; however, the
transition between the two epochs is gradual enough that matching only qualitatively reproduces
full solutions to \cref{eqn:scalar-eom-nonrelativistic-potential}, which are depicted in
\cref{fig:dynamics-nonrelativistic-thermal-mass} for positive couplings $d_\lambda^{(2)}$.
The dynamics well before and well after equality match the analytic solutions
\cref{eqn:radiation-era-nonrel-soln,eqn:matter-era-nonrel-soln} well, and clearly when
$(Q_b)_\lambda d_\lambda^{(2)} \gtrsim 1$ the scalar rolls substantially before recombination.
For negative couplings, the squared in-medium mass is negative, yielding growing rather than decaying
solutions; for $\vert (Q_b)_\lambda d_\lambda^{(2)} \vert \gtrsim 1$, the growth is exponential
[as captured by \cref{eqn:radiation-era-nonrel-soln} when evaluating the Bessel function with
complex argument] and clearly not viable cosmologically.
For $\vert (Q_b)_\lambda d_\lambda^{(2)} \vert \lesssim 1$, the growth remains a power law described
by \cref{eqn:matter-era-nonrel-soln}.

We now determine what maximum $\vert d_\lambda^{(2)} \vert$ permits the scalar to be
sufficiently stable that we may treat the variations in fundamental constants as themselves
time independent before recombination.
For $\vert (Q_b)_\lambda d_\lambda^{(2)} \vert \lesssim 1$, most of the scalar's evolution takes
place in the short period between matter-radiation equality and recombination and can be understood
with the matter-era solution; we empirically find
$\varphi / \varphi_i \approx (1 + 2 a / 3 a_\mathrm{eq})^{\beta_+}$ provides an excellent
approximation, yielding \cref{eqn:apx-soln-matter-effects-late} at small $\mathcal{D}$.
For \Planck{}'s preferred \LCDM{} parameters~\cite{Planck:2018vyg}, the scale factor of
recombination relative to that at equality is $a_\star / a_\mathrm{eq} \approx 3$
and $\omega_c / \omega_b \approx 5.4$.
In the small-$\mathcal{D}$ limit, $\beta_+ \approx - 2 \mathcal{D} / 3$
and the \emph{shift} in $\lambda$ varies by recombination by an amount given by \cref{eq:variation_in_lambda_from_matter}.

The scalar departs from its initial condition even more by the time it begins oscillating due to its
bare mass, i.e., as in \cref{eqn:apx-soln-matter-effects-late} with $a_\osc$ substituted for
$a_\star$.
Between $t_\osc$ and today, the field amplitude oscillates approximately at the bare mass
frequency $m_\varphi$ with an envelope well described by a Wentzel-Kramers-Brillouin (WKB)
approximation~\cite{Bouley:2022eer,Sibiryakov:2020eir,Olive:2001vz},
\begin{align}
    \varphi(a)
    &\propto \frac{1}{a^{3/2}} \frac{1}{\sqrt{m_\varphi^2 + \delta m_T(a)^2}},
\end{align}
with $\delta m_T(a)^2 = \mathcal{D} H^2$ in the matter era.
The scalar's energy density relative to that of CDM is therefore not constant after
$t_\osc$ when the in-medium mass is nonnegligible.
Combined with \cref{eqn:apx-soln-matter-effects-late}, at small $\vert \mathcal{D} \vert$ and
$a > a_\osc$ we have
\begin{align}
    \frac{\bar{\rho}_\varphi(t)}{\bar{\rho}_c(t)}
    &\approx \fphi \frac{1}{\left( 1 + 2 a_\osc / 3 a_\text{eq} \right)^{4 \mathcal{D} / 3}}
        \frac{m_\varphi^2 + \delta m_T(a_\osc)^2}{m_\varphi^2 + \delta m_T(a(t))^2}
    \approx \fphi \frac{
            1 + \mathcal{D} \left( 1 - [a(t) / a_\osc]^{-3} \right)
        }{
            \left( 1 + 2 a_\osc / 3 a_\text{eq} \right)^{4 \mathcal{D} / 3}
        }
    \label{eqn:Xi-generalization}
\end{align}
where $\fphi$ is the scalar-CDM ratio given by \cref{eqn:Xi-ito-varphi}, i.e., for an uncoupled scalar.
\Cref{eqn:Xi-generalization} encodes the correction to the relationship between the early-time
parameter shift $\Delta \lambda_i / \lambda(0)$, $d_{\lambda}^{(2)}$, and the late-time abundance
$\fphi$ for an uncoupled scalar, \cref{eqn:dlambda-i-ito-Xi}.
As explained in \cref{sec:discussion}, because these effects are relevant only at large
$d_{\lambda}^{(2)}$ but cosmological data place relatively strong constraints on
$\Delta \lambda_i / \lambda(0)$, they are only relevant in the regime in which the scalar's
abundance is negligibly small.
Any changes to the scalar's dynamics are then themselves negligible.
\Cref{eqn:Xi-generalization} still encodes the correction between $\fphi$ [taken to be
\emph{defined} as the combination in \cref{eqn:dlambda-i-ito-Xi}] and the present-day scalar
abundance relevant to, e.g., laboratory probes.
But these effects are again small in the parameter space we consider.

\subsection{Before annihilation}\label{app:before-annihilation}

In the early Universe, the Klein-Gordon equation with the in-medium mass
\cref{eq:thermal_mass_early} (in the relativistic limit) may be written as
\begin{align}
    \partial_N^2 \varphi
        + \partial_N \varphi
    &= - \left(
            \frac{m_\varphi^2}{H^2}
            + \frac{15 d_{m_e}^{(2)}}{2 \pi^2} e^{2 N}
            + \frac{25 \alpha(0) d_{e}^{(2)}}{8 \pi}
        \right)
        \varphi
    \label{eqn:scalar-eom-relativistic-potential}
\end{align}
where $N = \ln a / a_\mathrm{ann} \equiv \ln m_e / T$.
\Cref{eqn:scalar-eom-relativistic-potential} may be solved analytically for each coupling
individually; we also compare to numerical solutions using the full potential
\cref{eqn:matter-potential-early} which is valid through the end of annihilation.

Because the scalar's in-medium mass becomes exponentially suppressed at annihilation (for both
couplings), the scalar refreezes except for very large $d_{\lambda}^{(2)}$.
Though nontrivial dynamics before annihilation therefore have no direct effect on phenomenology at
later times, they do affect how the scalar's value at, e.g., recombination ($\varphi_\star$) is
related to its value at earlier times.
For large enough couplings and $\varphi_\star$, the scalar at earlier times could probe regimes that
are beyond the scope of the of the effective theory \cref{eq:DD_lagrangian}.
For instance, interactions at higher order in $\varphi$ can in principle be present in the full
theory, but such higher-dimension operators should be suppressed in the regime that the effective
field theory remains valid.
In the sense that the power series in \cref{eqn:coupling-function-series-expansion} remains
perturbative, the leading-order contribution should remain smaller than unity to avoid sensitivity
to unknown details of the theory's UV completion.
Moreover, our calculation of the effective potential in \cref{sec:matter-potentials} relied on an
expansion in small $g_\lambda(\varphi)$.
In our case, $g_\lambda(\varphi) = d_{\lambda}^{(2)} \varphi^2 / 2$ should then remain small at all
times.
One could also require the scalar to remain below the Planck scale ($\varphi \lesssim 1$), but this
condition is weaker than the former unless $d_{\lambda}^{(2)} < 1$.

Second, our treatment does not capture possible backreaction of the scalar on the SM plasma (nor the
expansion of the Universe).
Setting aside the current lack of a formalism to treat it, such backreaction would likely have a
significant effect on nucleosynthesis (which is sensitive to plasma dynamics during this epoch).
As a proxy, one might require that the effective potential \cref{eqn:matter-potential-early} be
negligible compared to the plasma energy density; backreaction would otherwise be important on the
grounds of energy conservation.
Since $\rho_\mathrm{SM}(T) = \pi^2 g_\star(T) T^4 / 30$, this condition amounts to requiring
that the quantity in brackets in \cref{eqn:matter-potential-early} remain smaller than unity.

\begin{figure}[t!]
\begin{centering}
    \includegraphics[width=\textwidth]{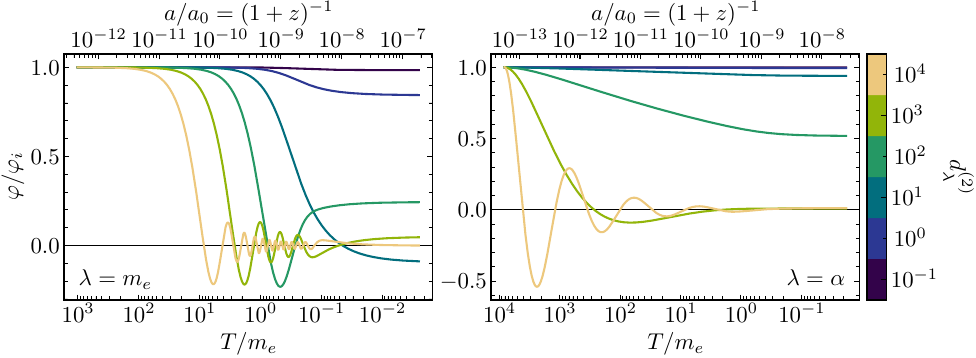}
    \caption{
        Solutions to the Klein-Gordon equation \cref{eqn:scalar-eom-relativistic-potential} for a
        scalar coupled quadratically to the electron (left) and photon (right), with couplings
        $d_\lambda^{(2)}$ varying by color.
    }
    \label{fig:dynamics-relativistic-thermal-mass}
\end{centering}
\end{figure}

To determine the limits imposed by these considerations, we solve
\cref{eqn:scalar-eom-relativistic-potential}.
We again compare approximate, analytic results to numerical results, the latter of which are
presented in \cref{fig:dynamics-relativistic-thermal-mass}.
Under the electron coupling, the solution to \cref{eqn:scalar-eom-relativistic-potential} with an
initially frozen value (say, whatever value $\varphi_\mathrm{RH}$ it took on at reheating) is
\begin{align}
    \varphi_{m_e}(a)
    &= \varphi_\mathrm{RH}
        \frac{
            \sin\left( \sqrt{\mathcal{D}_{m_e}} a / a_\mathrm{ann} \right)
        }{
            \sqrt{\mathcal{D}_{m_e}} a / a_\mathrm{ann}
        }.
    \label{eqn:electron-coupling-early-soln}
\end{align}
where $\mathcal{D}_{m_e} \equiv 15 d_{m_e}^{(2)} / 2 \pi^2 g_\star(T)$.
\Cref{eqn:electron-coupling-early-soln}, however, is only valid while electrons and positrons are
relativistic.
When they annihilate, the in-medium mass is exponentially suppressed and the scalar refreezes.

\begin{figure}[t!]
\begin{centering}
    \includegraphics[width=4in]{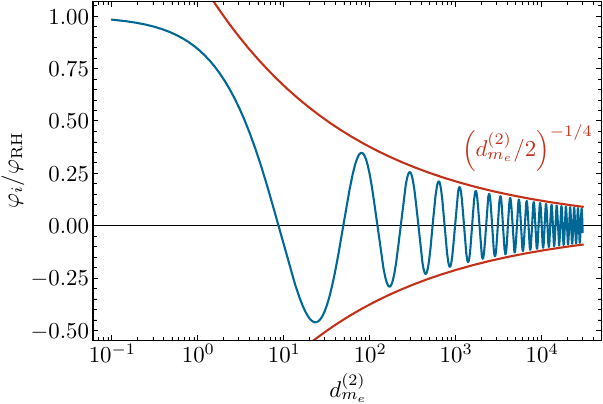}
    \caption{
        Value $\varphi_i$ at which a scalar refreezes after electron-positron annihilation as a
        function of its quadratic coupling to the electron, obtained by numerically solving the
        Klein-Gordon equation with effective potential \cref{eqn:matter-potential-early}, itself
        evaluated by numerical quadrature of \cref{eqn:rho-P-noninteracting-gas}.
    }
    \label{fig:refreeze-value}
\end{centering}
\end{figure}

From the form of \cref{eqn:electron-coupling-early-soln}, one would expect the scalar to refreeze
at a value $\sqrt{\mathcal{D}_{m_e}}$ times smaller than its initial value $\varphi_\mathrm{RH}$.
While numerical results match this scaling when comparing solutions evaluated precisely when
$T = m_e$, the subsequent dynamics are not captured by matching onto solutions to the massless
Klein-Gordon equation because annihilation is not instantaneous.
The dynamics during annihilation---in part that the in-medium mass becomes negligible later the
larger $d_{m_e}^{(2)}$ is---are nontrivial, and the scalar ultimately refreezes at values that are
instead $\sim \sqrt[4]{\mathcal{D}_{m_e}}$ times smaller than its initial value, as displayed in
\cref{fig:refreeze-value}.
Moreover, the precise timing of the scalar's oscillations as its in-medium mass shuts off allows the
scalar to ultimately refreeze near zero.
As a function of the refrozen shift in the electron mass, the constraint that
$g_{m_e}(\varphi_\mathrm{RH}) \lesssim 1$ requires
\begin{align}
    d_{m_e}^{(2)}
    &\lesssim 2 \left( \frac{\Delta m_{e, i}}{m_e(0)} \right)^{-2},
\end{align}
where $\Delta m_{e, i}$ is the change in the electron mass post-annihilation---i.e., that which
would be the initial condition for the dynamics during recombination.

Oscillatory solutions for positive couplings again switch to exponentially growing ones for negative
couplings.
For large negative couplings $- d_{m_e}^{(2)} \gtrsim 100$, numerical solutions determine that the
scalar refreezes at a value $\sim e^{3 \sqrt{\vert \mathcal{D}_{m_e} \vert}}$ times larger than
$\varphi_\mathrm{RH}$; the dependence on $\sqrt{\vert \mathcal{D}_{m_e} \vert}$ is as expected from
\cref{eqn:electron-coupling-early-soln}, while the numerical coefficient reflects the particular
dynamics of refreezing as the in-medium mass drops exponentially during annihilation.
At the classical level, there is no theoretical issue with the scalar having an arbitrarily small
value before reheating that is substantially amplified to its post-annihilation value.
However, during inflation, a quantized scalar scalar field with $m_\varphi < H_\text{inf}$ acquires an average amplitude of order $\sim H_\mathrm{inf}^2 / m_\varphi > H_\text{inf}$~\cite{Starobinsky:1994bd}.
Heuristically, this is because an inflationary Universe has a IR cutoff at the inflationary scale $H_\text{inf}$, meaning that the smallest physically meaningful scale is $\mathcal{O}(H_\text{inf})$.
At the very minimum, inflation had to occur in time for reheating to complete at a temperature
$T_\mathrm{RH} \approx 4~\MeV$~\cite{Giudice:2000ex, Kawasaki:2000en, Kawasaki:2004qu,
Hannestad:2004px, Ichikawa:2005vw, deSalas:2015glj, Hasegawa:2019jsa},
limiting $H_\mathrm{inf} \gtrsim 10^{-14}~\eV$.
Requiring that $\vert \varphi_\text{RH} \vert > H_\text{inf}/\sqrt{2}\Mpl$ implies
\begin{align}
    \sqrt{\frac{2 \vert \Delta m_{e,i}/m_e(0) \vert}{\vert d_{m_e}^{(2)} \vert}} e^{-3 \sqrt{\vert \mathcal{D}_{m_e} \vert}} \gtrsim \frac{H_\text{inf}}{\sqrt{2} \Mpl},
\end{align}
so
\begin{align}
    - d_{m_e}^{(2)}
    &\lesssim \frac{2\pi^2 g_\star(T_\text{RH})}{135}
        W^2\left(
            \sqrt{\frac{270}{\pi^2 g_\star(T_\text{RH})}}
            \sqrt{\frac{\Delta m_{e,i}}{m_e(0)}}
            \frac{\Mpl}{H_\text{inf}}
        \right)
\end{align}
where $W(z)$ is the Lambert $W$ function, which at large argument is $W(z) \approx \ln x - \ln \ln x$.

The scalar's effective mass induced by its coupling to photons is proportional to $H$ at all times after reheating.\footnote{
    Technically, the rest of the charged SM content would contribute to the interacting free energy
    at higher temperatures.
    The SM presumably also generates an effective potential for the scalar before it fully
    thermalizes, but it would not be captured by the thermal field theory calculations
    \cref{sec:plasma} relies on.
    We set aside these complex and uncertain details for the sake of simplicity.
}
Now defining $\mathcal{D}_e = 25 \alpha(0) d_e^{(2)} / 8 \pi g_\star$, the solution under the photon
coupling is
\begin{align}
    \varphi_{e}(a)
    &= \frac{
            \beta_+ \varphi_\mathrm{RH} (a / a_\mathrm{RH})^{\beta_-}
            - \beta_- \varphi_\mathrm{RH} (a / a_\mathrm{RH})^{\beta_+}
        }{
            \sqrt{1 - 4 \mathcal{D}_e}
        }
    \label{eqn:photon-coupling-early-soln}
\end{align}
where $\beta_\pm = - 1/2 \mp \sqrt{1/4 - \mathcal{D}_e}$,
where we have taken $\dot{\varphi}_e(a_\text{RH}) = 0$.
The solutions are oscillatory if $\mathcal{D}_e > 1/4$, i.e., when $d_e^{(2)} \gtrsim 370$.
To leading order in small $\mathcal{D}_e$, $\beta_- \approx -\mathcal{D}_e$ and $\beta_+ \approx -1$.
\Cref{eqn:photon-coupling-early-soln} precisely reproduces the numerical solutions displayed in
\cref{fig:dynamics-relativistic-thermal-mass} well before annihilation, at which point the scalar
again refreezes.

For positive couplings the dominant mode is $\propto a^{-\mathcal{D}_e}$, meaning the
scalar would have been roughly $(T_\mathrm{RH} / m_e)^{\mathcal{D}_e}$ times larger at
reheating than at annihilation.
Requiring $\vert g_e(\varphi_\mathrm{RH}) \vert \lesssim 1$ then amounts to
\begin{align}
    d_e^{(2)}
    &\lesssim
        - \frac{8 \pi g_\star}{50 \alpha(0)}
        \frac{
            \ln \left[ \Delta \alpha_i / \alpha(0) \right]
        }{
            \ln \left( T_\mathrm{RH} / m_e \right)
        }
\end{align}
where $\Delta \alpha_i$ is the change in $\alpha$ post-annihilation that is the initial condition
for the dynamics leading up to recombination.
Again taking the minimum reheat temperature $T_\mathrm{RH} \approx 4~\MeV$, this bound amounts to
$d_e^{(2)} \lesssim 360$; for instantaneous reheating after high-scale inflation, the bound tightens
by a factor $\sim 20$.

For negative photon couplings, the solutions are always growing power laws; for large and negative
$\mathcal{D}_e$, the growing mode is $\beta_- \approx \sqrt{-\mathcal{D}_e}$.
Again assuming a initial field value generated stochastically during inflation
$\sim H_\mathrm{inf}^2 / m_\varphi$
and assuming reheating was instantaneous ($H_\mathrm{RH} = H_\mathrm{inf}$),
$\vert \varphi_\mathrm{RH} \vert > H_\text{inf}/\sqrt{2}\Mpl$ if
\begin{align}
    \sqrt{\frac{2 \vert \Delta \alpha_i / \alpha(0) \vert}{\vert d_e^{(2)} \vert}}
    \left( \frac{m_e}{T_\text{RH}} \right)^{\sqrt{-\mathcal{D}_e}}
    &\gtrsim \frac{H_\text{inf}}{\sqrt{2} \Mpl},
\end{align}
so
\begin{align}
    -d_e^{(2)}
    &\lesssim \frac{8\pi g_\star(T_\text{RH})}{25 \alpha(0)}
        \frac{1}{\ln^2\left( T_\text{RH} / m_e \right)}
        W^2\left(
            \ln[ T_\text{RH} / m_e ]
            \sqrt{\frac{25\Delta \alpha_i}{2\pi g_\star(T_\text{RH})}}
            \frac{\Mpl}{H_\text{inf}}
        \right).
\end{align}

\section{Analysis variants for cosmological parameter inference}\label{app:analysis-variants}

\subsection{Impact of priors on marginalized constraints}\label{sec:priors}

As the discussion of \cref{sec:cosmo-constraints-couplings-and-Xi} suggests, the marginal posteriors
over both $\fphi$ and $d_{\lambda}^{(2)}$ are strongly sensitive to the choice of prior, which
impedes drawing robust constraints on the parameters of the underlying model. We now consider the
effect of priors in more detail.
In \cref{sec:cosmo-constraints-couplings-and-Xi} we selected uniform priors over $\fphi$ and the
early-time parameter values $\alpha_i$ and $m_{e, i}$ (as taken in all previous phenomenological
studies).
Such priors lack any theoretical input but allow the data to drive the shape of the posteriors, at
least in determining where the tails of the posterior are located.
The effective prior on the fundamental parameters $d_\lambda^{(2)}$ and $\bar{\varphi}_i$ induced by
taking uniform priors over $\fphi$ and $\lambda_i / \lambda_0$ is
\begin{align}
    \ud (\lambda_i / \lambda_0) \ud \fphi
    &= \frac{3 \bar{\varphi}_i^3 (1 + \omega_b / \omega_c)}{4}
        \ud d_\lambda^{(2)} \ud \bar{\varphi}_i
    .
    \label{eqn:effective-prior-from-pheno-parameters}
\end{align}
Notably, this prior rather steeply favors $\bar{\varphi}_i$ as large as possible, akin to the
implicit preference in models of axionlike early dark energy for Planckian initial
misalignments~\cite{Hill:2020osr}.

While any choice of prior on phenomenological parameters ultimately lacks direct theoretical
motivation, even that for fundamental parameters is ambiguous.
For instance, from the perspective of effective field theory one might consider the dimensionless parameter
$d_\lambda^{(2)}$ an order-unity Wilson coefficient suitable for a uniform prior.
On the other hand, $\Mpl / d_\lambda^{(2)}$ may be better described as a new-physics scale for a UV
completion of the model, and one might choose for it a log-uniform prior in order to avoid
arbitrarily selecting a preferred energy scale.
[That is, a uniform prior between $a$ and $b$ identifies $(b-a)/2$ as a ``special'' value, whereas a
log-uniform prior allocates equal weight to each decade within the allowed limits.]
A choice for $\bar{\varphi}_i$ could also be informed by the choice for $d_\lambda^{(2)}$---perhaps
$\bar{\varphi}_i$ is itself a characteristic energy scale in the theory, or derives from some other
physics that generates the initial conditions, or so on.
Posteriors, of course, are most useful when they are least sensitive to prior choices.

Given that the analysis of Ref.~\cite{Baryakhtar:2024rky} does not demonstratively ``detect'' an
early-time shift in $\alpha$ or $m_e$ nor the presence of a hyperlight scalar, such issues are of
less practical concern at the present.
To illustrate what effect an prior defined instead over the fundamental parameters can have on the
analysis in Ref.~\cite{Baryakhtar:2024rky}, we consider a uniform prior over $d_{\lambda}^{(2)}$ and
a log-uniform prior over $\bar{\varphi}_i$.
The resulting prior over phenomenological parameters is
\begin{align}
    \ud d_\lambda^{(2)} \ud \ln \bar{\varphi}_i
    &= \frac{3 (1 + \omega_b / \omega_c)}{4 \fphi^2}
        \ud (\lambda_i / \lambda_0) \ud \fphi.
    \label{eqn:lin-log-fundamental-prior}
\end{align}
Compared to a uniform prior over $\fphi$ and $\lambda_i / \lambda_0$, this prior penalizes
large $\fphi$ quadratically.
If the posteriors over $d_{\lambda}^{(2)}$ are largely driven by the likelihood's dependence on
$\lambda_i / \lambda_0$, we would expect such a prior to drive the marginal posteriors over the
quadratic couplings $d_{\lambda}^{(2)}$ to values even larger in magnitude.
Priors must be proper, and the chosen limits on, e.g., $\bar{\varphi}_i$ determine the relative
prior weight allocated to percent-level values of $\fphi$---i.e., those that were allowed in the
results taking uniform phenomenological priors.
Specifically, we impose the effective prior \cref{eqn:lin-log-fundamental-prior} with boundaries
corresponding to $d_{\lambda}^{(2)} \sim \mathcal{U}(-10^3, 10^3)$ and
$\log_{10} \bar{\varphi}_i \sim \mathcal{U}(-2, 0)$, where $\mathcal{U}$ denotes a uniform
distribution.
We additionally restrict $10^{-4} \leq \fphi \leq 0.3$ in order to roughly match the range of values
covered by the posteriors using uniform, phenomenological priors.

\begin{figure}[t!]
\begin{centering}
    \includegraphics[width=\textwidth]{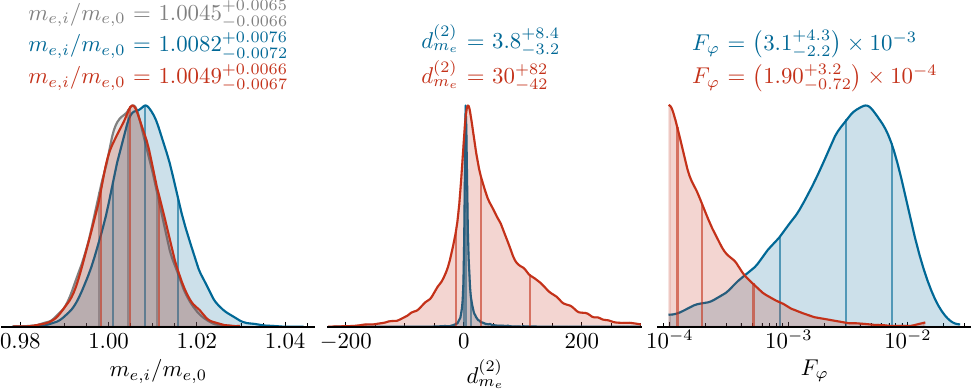}
    \caption{
        Comparison of marginalized posteriors over $m_{e, i}$, $d_{m_e}^{(2)}$, and $\fphi$ with
        uniform priors over both $m_{e, i}$ and $\fphi$ (blue) and a uniform prior over
        $d_{m_e}^{(2)}$ and a log-uniform prior over $\bar{\varphi}_i$ (red).
        The marginal posterior over $m_{e, i}$ for purely a phenomenological model (with no scalar
        field) appears in grey.
        All cases use \Planck{} 2018 data and BAO measurements and fix the scalar's mass to
        $10^{-30}~\eV$, where applicable.
        Vertical lines depict the median and $\pm 1 \sigma$ quantiles of the marginal posteriors,
        which are also reported above each panel.
        Results are obtained as described in Ref.~\cite{Baryakhtar:2024rky}.
        The evident prior sensitivity of the marginalized posteriors for $d_{m_e}^{(2)}$ and $\fphi$
        only slightly affects that of $m_{e, i}$, as interpreted in the text.
    }
    \label{fig:electron-coupling-prior-comparison}
\end{centering}
\end{figure}

\Cref{fig:electron-coupling-prior-comparison} compares marginalized posteriors for this prior with
those using the same uniform prior over phenomenological parameters applied in
Ref.~\cite{Baryakhtar:2024rky}.
The $\fphi$ dependence of the prior \cref{eqn:lin-log-fundamental-prior} clearly translates directly
to the marginalized posterior; the prior penalizes larger $\fphi$ so severely that the data's disfavor
for $\fphi \gtrsim 10^{-2}$ is not even apparent.
As a result, the posterior over $d_{m_e}^{(2)}$ broadens by about an order of magnitude.
Since the data only directly constrain $\fphi$ and $m_{e, i}$, we therefore cannot derive any
meaningful marginalized constraints on $d_{m_e}^{(2)}$ that are driven by the likelihood rather than
the prior.

However, the posterior over $m_{e, i}$ is much more robust, only shifting marginally toward lower
values.
The degeneracy discussed in Ref.~\cite{Baryakhtar:2024rky} allows for larger $m_{e, i}$ correlated
to a nonnegligible scalar abundance, $\fphi > 0$, but \cref{eqn:lin-log-fundamental-prior}
effectively precludes any gravitational impact of the scalar.
Given that \Planck{} data ultimately disfavor a substantial scalar
abundance~\cite{Baryakhtar:2024rky}, the impact of this effect on the marginalized posteriors is
minor.
We therefore conclude that constraints derived directly from the early-time values of fundamental
constants (and therefore the combination $d_{\lambda}^{(2)} \fphi$) are robust to the choice of
prior.
Bounds on $d_{\lambda}^{(2)}$ at particular values of $\fphi$ from cosmological data are thus on par
with quasar- and BBN-derived bounds, which are insensitive to gravitational effects of the scalar in
the first place.

\subsection{Impact of BBN consistency}\label{sec:bbn-consistency}

In the class of models we consider (\cref{sec:models}), shifts in the fundamental constants during
BBN are generally the same as during recombination (unless the matter effects of
\cref{sec:matter-potentials} are important, in which case for positive couplings the constants
differ from their present-day values even more at earlier times).
Following the discussion in Ref.~\cite{Baryakhtar:2024rky}, the primary effects of the helium yield
on the CMB anisotropies are on diffusion damping rate and the width of the visibility function.
The change to the helium yield predicted by BBN in particular introduces \emph{implicit} dependence
on $m_{e, i}$ per \cref{eqn:Yp-dependence-on-constants} that (in principle) breaks the degeneracy of
$m_{e, i}$ with \LCDM{} parameters.

\begin{figure}[t!]
    \centering
    \includegraphics[width=\textwidth]{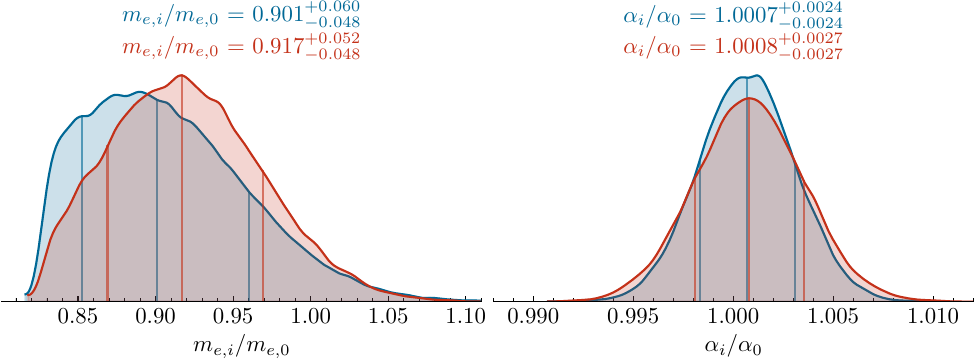}
    \caption{
        Marginalized posterior distributions over the early-time electron mass (left) and
        fine-structure constant (right), each varied independently, for the corresponding scalar
        field models with (red) and without (blue) the effect of the variation in fundamental
        constants on BBN accounted for.
        The posteriors only use the \Planck{} 2018 likelihoods, since the difference is negligible
        when including data that better constrain the models.
        Vertical lines mark the median and $16$th and $84$th percentiles; the median and
        corresponding $\pm 1 \sigma$ uncertainties for each parameter are reported above each panel.
    }
    \label{fig:vary-al-me-consistent-bbn}
\end{figure}

Here we use the approximate calculation of \cref{sec:BBN} to assess the impact of varying constants
at BBN on CMB-derived constraints thereof.
\Cref{fig:vary-al-me-consistent-bbn} demonstrates that the most extreme values of $m_{e, i}$ allowed
by \Planck{} data are more disfavored when this implicit dependence is accounted for.
Namely, the marginalized posterior over $m_{e, i}$ skews less toward values smaller than one when
treating BBN consistently.
The degeneracy is therefore only meaningfully broken for $\mathcal{O}(10\%)$ early-time shifts in
the electron mass, far outside of the strongest constraints obtained in
Ref.~\cite{Baryakhtar:2024rky} (that come from combining \Planck{} data with BAO or supernova
datasets).
The CMB independently constrains the fine-structure constant to so narrow a region that the effect
on $Y_\mathrm{He}$ is even more minor.
The diffusion damping rate already explicitly depends on $\alpha_i^3$, while the dependence of the
helium yield on $\alpha_i$ [using \cref{eqn:Yp-dependence-on-constants}] slightly decreases this
scaling to $\sim \alpha_i^{2.6}$~\cite{Baryakhtar:2024rky}, leading to the marginally broader
posterior evident in \cref{fig:vary-al-me-consistent-bbn}.

\bibliography{bib,manual}

\end{document}